\documentclass[12pt]{JHEP3}
\newcommand{\intd}{\mathrm{d}}
\newcommand{\ex}{\mathrm{e}}

\usepackage{amsmath,epsfig}
\usepackage{amssymb,amsfonts}
\usepackage{latexsym}
\usepackage[latin1]{inputenc}
\usepackage{slashed}
\usepackage{empheq}
\numberwithin{equation}{section}
\usepackage{dsfont}
\usepackage{cancel}
\usepackage{overpic}
\usepackage{subcaption}
\usepackage{subeqnarray}
\usepackage{xcolor}
\let\normalcolor\relax
\usepackage{graphicx}
\usepackage{amsmath}
\usepackage{longtable}
\usepackage{color}
\usepackage{bbm}

\allowdisplaybreaks

\newcommand{\exclude}[1]{}
\def\nn{\nonumber}

\def\L{\mathcal{L}}

\def\d{\mathrm{d}}

\def\a#1{\alpha_{#1}}

\def\tr{\mathrm{tr}}

\def\beq{\begin{equation}}
\def\eeq{\end{equation}}
\def\be{\begin{equation}}
\def\ee{\end{equation}}
\def\bea{\begin{eqnarray}}
\def\eea{\end{eqnarray}}
\def\bal{\begin{align}}
\def\eal{\end{align}}

\def\td{\tilde}
\def\frr{\mathcal{X}\ex^{2A}\D_{rr}}
\def\fxr{\mathcal{X}\ex^{2A}\D_{\xi r}}
\def\fxx{\mathcal{X}(1-\ex^{2A}\D_{\xi\xi})}
\def\cX{\mathcal{X}}

\def\2b2[#1,#2][#3,#4]{\left( \begin{array}{cc} #1 & #2 \\ #3 & #4 \end{array}
\right)}
\def\3b3[#1,#2,#3][#4,#5,#6][#7,#8,#9]{\left( \begin{array}{ccc} #1 & #2 #3 \\
#4 & #5 & #6\\#7&#8&#9\end{array} \right)}

\newcommand{\ka}{\kappa}

\newcommand\fverb{\setbox\pippobox=\hbox\bgroup\verb}
\newcommand\fverbdo{\egroup\medskip\noindent%
                        \fbox{\unhbox\pippobox}\ }
\newcommand\fverbit{\egroup\item[\fbox{\unhbox\pippobox}]}

\newcommand{\la}{\lambda}

\newcommand{\bear}{\begin{eqnarray}}

\newcommand{\eear}{\end{eqnarray}}
\newcommand{\de}{\partial}
\newcommand{\bsea}{\begin{subeqnarray}}
\newcommand{\esea}{\end{subeqnarray}}
\newbox\pippobox

\def\d{\delta}

\def\6{\partial}
\def\a{\alpha}
\def\nn{\nonumber}

\def\e{\epsilon}
\def\m{\mu}
\def\n{\nu}

\def\s{\sigma}

\def\sp{\;\;\;,\;\;\;}

\def\sq
\def\a{\alpha}
\def\b{\beta}
\def\l{\lambda}

\def\tr{{\rm Tr}}
\def\k{\chi}
\def\hri#1#2{\href{http://arxiv.org/abs/#1}{[ArXiv:#1]#2}}
\def\hre#1#2{\href{http://arxiv.org/abs/#1/#2}{[ArXiv:#1/#2]}}
\def\hrj#1#2{\href{https://doi.org/#1}{#2}}

\def\e{\epsilon}

\def\d{\delta}

\def\L{\Lambda}

\def\D{\Delta}

\title{The V-QCD baryon : numerical solution and baryon spectrum}
\author{
M. J\"arvinen$^\sharp$,  E. Kiritsis$^\natural$$^\flat$, F. Nitti$^\natural$, E. Pr\'eau$^\natural$
~\\
~\\
$^\natural$ \href{http://www.apc.univ-paris7.fr}{Universit\'e Paris Cit\'e, CNRS, Astroparticule et Cosmologie,  F-75006 Paris, France}\\
~\\
$^\flat$ \href{http://hep.physics.uoc.gr}{Crete Center for Theoretical Physics}, Institute for Theoretical and Computational Physics,
Department of Physics,  P.O. Box 2208,\\
University of Crete, 70013, Heraklion, Greece
~\\
~\\
$^\sharp$ \href{http://apctp.org}{Asia Pacific Center for Theoretical Physics}, Pohang 37673, Republic of Korea and Department of Physics, Pohang University of Science and Technology, Pohang 37673, Republic of Korea
}

\preprint{APCTP Pre2022 - 027\\
CCTP-2022-8\\
ITCP-2022/8}

\abstract{The single baryon solution of  V-QCD is numerically computed.
The spectrum of spin and isospin modes is also computed by quantizing the light fluctuations around the baryon.
It is shown that there is a partial restoration of chiral symmetry at the baryon center.}

\begin{document}
\maketitle

\section{Introduction}

The holographic correspondence, \cite{Malda,GKP,Witten98,review} has
offered from the start a new tool to approach the  strongly coupled
low energy physics of QCD or QCD-like theories in the large-$N$ limit
\cite{Witten98a,Cobi,COOT,RY,MW,B+M,SS,Casero}. This applies in
particular to the description of baryon bound states. These have been investigated both in top-down
string-theoretical  constructions \cite{WB,SS2,Kim,Sugimoto,Yi,Cherman,Hata:2007tn,Bolognesi} and in simple bottom-up
phenomenological models \cite{Pomarol07,Pomarol08,Gorsky2015}. In all these models, the baryon can
be understood as an instanton of the bulk non-abelian  gauge fields
which are holographically dual to the $U(N_f)_L \times U(N_f)_R$ flavor
currents.

Within the bottom-up approach, {one of} the most complete phenomenological
frameworks  is the   V-QCD setup \cite{ihqcd,gkmn,Jarvinen,JM,spectrum,fT,fD}, a five-dimensional holographic model
which aims at describing the Veneziano limit (large $N_c$, large $N_f$
with $N_c/N_f$ fixed) of QCD with $N_c$ colors and $N_f$ flavors.

The V-QCD  model has 22 parameters in its CP-even part and one extra parameter in the Chern-Simons term.
Although upon a more complete fit it may turn out that some parameters may not be as important, a substantial number of parameters is necessary because the model has the ambition to describe a rather complete set of observables that go well beyond other competing models.
The model can provide upon calculation, most $T=0$ mass spectra including baryons, thermodynamic functions and phase diagram in a multidimensional space of temperature, baryon and isospin chemical potentials, correlation functions of several local operators at finite temperature and density, including transport coefficients and quasinormal modes
and eventually way out-of equilibrium observables like quenches. It therefore can provide info that goes well beyond the parameter input to define the theory.

There are several approaches  and competing phenomenological models that we summarize below.

\begin{itemize}
\item Lattice Quantum Chromodynamics \cite{Lattice}. This is an ab-initio approach. Wherever it is applicable, and the numerics are reliable,  it gives the proper answer of QCD.  It has  however, computational limitations: finite density is out of direct reach as the probability density is not real and similar remarks apply to real time dynamical processes.
Clearly, phenomenological models like V-QCD are useful if they can address issues that cannot be studied by Lattice techniques. The problems mentioned above are in this class and as known, holographic models are well-tuned to address  them.

\item  Chiral Perturbation Theory (CPTh), \cite{CPTh1}-\cite{CPTh3}. This is an Effective Theory based on symmetry principles and as such reliable in the low-energy pion sector.
The addition of baryons is more tricky and although the formalism is well-studied,  it rests on shakier principles.
That being said, this approach is the workhorse for analysing low-energy phenomena and low-density phase diagrams of the hadronic phase. It is however unsuitable for the study of other phases where confinement and chiral symmetry are realized differently. Therefore V-QCD is more appropriate to address static and dynamical issues in the various plasma phases.

\item  FRG improved CPTh, \cite{FRG}.  This is  CPTh improved with the functional Renormalization Group. It allows CPTh to extend a bit up in energy but has similar limitations as CPTh.

\item PNJL models, \cite{PNJL,PNJL2}. These are extensions of the NJL model,\cite{NJL,NJL2}, by including the Polyakov loop as one extra order parameter. They are phenomenological, but well-motivated and have made predictions on the phase diagram of QCD. They are expected to work  less well in phases where chiral symmetry is unbroken, and quarks are not confined.
V-QCD is well tuned for dynamical questions in such phases.

\item Quark-Meson-Coupling models, \cite{QMC1,QMC2}. They are hybrids between MIT bags for nucleons and meson exchanges and they were developed as extensions of nuclear QFT Serot-Walecka models. They can describe well parts of nuclear phenomenology in the hadron phase, and possibly in medium quark condensates, but are not best suited for pure deconfined/quark-gluon plasma phases.

\item  HTL quasiparticle models, \cite{HTL}. This is a class of weakly-coupled models of massive quasi-particles with improved HTL propagators, used to extrapolate calculations of finite temperature static quantities to finite chemical potential. Their use so far is restricted to static properties and their success is localized.

\end{itemize}

Compared to all the approaches above, V-QCD is a model that can address a wider variety of problems than any of them.
 Moreover, unlike other approaches, it describes naturally  both phases and dynamics with broken or unbroken  chiral symmetry, as well as confined and deconfined phases.

Up to very recently, a description of the baryon state in V-QCD was
missing. This was due in part to the incomplete understanding of
Chern-Simons terms in this theory.  These terms are crucial for the
construction of the baryon as an instanton, since a) they contribute
to  stabilising  the instanton size and position in the bulk; b) they
provide the correct identification between bulk instanton number and
boundary baryon number.

Recently, this gap was filled   by the authors of the present work in
\cite{BaryonI}, where, in the limit of zero quark masses, a
systematic analysis of the allowed Chern-Simons terms in V-QCD was
performed and the general
construction of the baryon solution was presented. In that work, it
was shown that an axial $SU(2)$ instanton  ansatz satisfying appropriate boundary
conditions (normalizability close to the AdS boundary and regularity in
the interior) has indeed all the properties of a single localized
baryon state: finite boundary energy and unit boundary baryon
charge. The latter is indeed a topological quantity, which matches
both the bulk instanton number and the boundary Skyrmion number
(written in terms of an appropriate unitary pion matrix).

\subsection{Summary}

The present work is the direct continuation of \cite{BaryonI}. While that
work presented the general features of the solution, here we construct
the baryon in a specific model: we first obtain
  numerically the static instanton solution (which corresponds to the
  baryon ground state), then we analyse the instanton collective modes and their
quantization (which correspond to baryon excited states).

As important ingredient  of this work, we present a specific model in the
V-QCD class which offers a good quantitative match to low-energy QCD
parameters, including some parameters in the flavor sector (like the
pion decay constant) which were not correctly reproduced in previous
models.

We now give a summary of the results obtained in this work.  We refer the
reader to the introduction of \cite{BaryonI} for a more
detailed discussion and an extensive list of references on holographic
baryons.

\subsubsection*{Fit to QCD data}

We carry out an extensive comparison of the model predictions to experimental and lattice QCD data in order to pin down the parameters of the V-QCD action in section~\ref{Sec:VQCD_intro}. The comparison consists of two main steps:
\begin{enumerate}
 \item Qualitative comparison to QCD physics. The action both in the limit of weak and infinitely strong coupling, up to a few remaining parameters, can be determined by requiring that the model respects known properties of QCD such as confinement and asymptotic freedom. This work has been done in earlier literature~\cite{ihqcd,Jarvinen,spectrum,JM,Ishii}, and we simply review the results here.
 \item Quantitative comparison to QCD data. The details of the action at intermediate coupling, as well as the few remaining weak and strong coupling parameters, can be tuned so that the predictions of the model match with QCD data. In this step, the dependence of the predictions on the exact values of the model parameters is typically weak. {Despite this, the model has been able to describe various observables to a high precision~\cite{gkmn,Remes,Amorim}.}
\end{enumerate}
As for the second step, the work in this article extends the earlier work where the full V-QCD model was separately compared to data for thermodynamics~\cite{Remes} and to meson spectra~\cite{Amorim}:
\begin{itemize}
 \item We fit the model parameters to data for thermodynamics and spectra simultaneously.
 \item Unlike in~\cite{Amorim} (where the model was fitted to a large number of excited meson states), we stress the lowest lying meson states.
 \item Importantly, we require a good match of the model with the experimental value of the pion decay constant $f_\pi \approx 92~\mathrm{MeV}$, which was poorly reproduced in both  previous fits.
\end{itemize}

We now discuss the fit in more detail.
The V-QCD model contains two sectors, corresponding to gluons (improved holographic QCD~\cite{ihqcd}) and  quarks (tachyon Dirac-Born-Infeld actions for space filling branes~\cite{Bigazzi,Casero}). The former sector can be separately compared to data from lattice analysis of pure Yang-Mills theory~\cite{gkmn}. In this article, we use the fit of~\cite{Amorim} for the gluon sector, and check explicitly that it reproduces both the lattice data for the thermodynamics of Yang-Mills at $N_c =\infty$ (see Figure~\ref{fig:YMfit}) and for the glueball spectrum (see table~\ref{tab:YMratios}) to a good precision.

Most of the freedom in the full V-QCD model is however in the quark sector. In order to determine the model parameters in this sector, we compare the predictions to
\begin{itemize}
 \item Lattice data for the thermodynamics of full QCD with $N_f=2+1$. We use both the data for the equation of state at vanishing chemical potential (see Figure~\ref{fig:thermofit}) and for the first nontrivial cumulant of the pressure at nonzero chemical potential (see Figure~\ref{fig:suskisfit}).
 \item Experimental data for lowest lying meson masses and the pion decay constant. See table~\ref{table:masses}.
\end{itemize}
The final values of the model parameters are given in table~\ref{table:parameters}. Apart from a few exceptions, the model depends on these parameters through four different functions of the coupling, which are shown in Figure~\ref{fig:potentials} for the final fit.

The fit in Figures~\ref{fig:thermofit},~\ref{fig:suskisfit} and table~\ref{table:masses} has rather good quality. However, the agreement when fitting the thermodynamics~\cite{Remes} and the spectra~\cite{Amorim} separately was significantly better. This is the case because the combined fit is challenging: there is some clear tension between the fit to the properties of the finite temperature state and the zero temperature vacuum state. It is likely that this tension can be reduced by carrying out a simultaneous numerical fit of all parameters to all data. We do not attempt to do this technically demanding task here, but are planning to return to it in future work.
{Notice also that, at least to our knowledge, an overall fit to QCD data of the similar extent as presented in this article has not been attempted in any other model in earlier literature. }

\subsubsection*{Static baryon solution}

Starting from the formalism introduced in \cite{BaryonI}, we compute in section \ref{Sec:SB} the numerical bulk solution for a single static baryon, with the V-QCD potentials presented in Section \ref{Sec:VQCD_intro}.\footnote{In addition, the solutions with a different choice of potentials are discussed in Appendix~\ref{App:OldPot}.} In the Veneziano limit $N_c\to\infty,\ N_f\to\infty$, the baryon contribution
to the bulk action is of order $N_c$, which is negligible compared to $N_c^2$
and $N_c N_f$. This implies that the leading order baryon solution can be treated as a ``probe" on the background dual to the vacuum of the boundary theory.

The numerical baryon solution is computed both at leading order and including the first corrections to the background, which are of order $\mathcal{O}(1/N_c,1/N_f)$.
The leading order baryon solution is reliably calculated, by ignoring both the tachyon and glue backreactions.
The backreaction of the baryon solution to the tachyon is computed by neglecting the glue backreaction for simplicity.
It is expected that the back-reaction on the color sector, does  not affect the qualitative results for the tachyon backreaction.

For the leading order probe solution, the following results are obtained:
\begin{itemize}

\item The instanton number and bulk Lagrangian densities (Figure \ref{fig:rho_Ni_M_v8_probe}) are confined to a region of finite extent in the bulk, which confirms the solitonic nature of the baryon solution. The integrals of these densities give respectively the baryon number, which is confirmed to be equal to 1 numerically, and the classical contribution to the nucleon mass. The latter is found to be relatively close to the experimental nucleon mass for $N_c = 3$ colors
\be
\label{M0numI} M_0 \simeq \frac{N_c}{3} \times 1150\,\text{MeV} \, .
\ee
We recall however, that the full result for the V-QCD baryon mass
should also include quantum corrections: these should be computed  from the perturbations around
the baryon, together with an appropriate subtraction of the similar fluctuations
around the vacuum state. Computing these corrections goes beyond the
scope of this work.
Although these corrections are subleading at large $N_c$, starting at order $\mathcal{O}\big(N_c^0\big)$, they may give a sizeable contribution when $N_c$ is set to 3. Note that this state of affairs regarding quantum corrections is not particular to our model, and is true also for both the Skyrme model and other holographic models.

\item In \cite{BaryonI}, it was found that, for the baryon solution, the pion matrix at the boundary follows the Skyrmion hedgehog ansatz
\be
\label{UPI} U_P(\xi) = \exp{\left(i\theta(\xi)\frac{x\cdot\s}{\xi}\right)} \, ,
\ee
with $\xi$ the 3-dimensional radius. Also, the baryon number was shown to be equal to the skyrmion number for the pion matrix. This indicates that the baryon solution in V-QCD is qualitatively similar to the Skyrme model skyrmion solution\footnote{This should not come as a surprise, as we already know that the dual boundary theory can be understood in the confined phase as a chiral effective theory coupled to a tower of massive mesons.}. To measure the difference with the Skyrme model skyrmion, the pion phase $\theta(\xi)$ is compared with the Skyrme result in Figure \ref{fig:thx_probe_v8}. This indicates that the two solutions are quantitatively close.

\end{itemize}

At the next order in the large N expansion, the back-reacted solution provides the following information:
\begin{itemize}

\item The modulus of the chiral condensate $\left|\left<\bar{\psi}\psi\right>\right|$ is observed to decrease towards the baryon center, as shown in Figure \ref{fig:CC_New_v8}. This signals the expected partial restoration of the chiral symmetry inside the baryon.

\item The correction to the baryon Lagrangian density from the back-reaction is calculated numerically and presented in Figure \ref{fig:rhoM_x}. The observed behavior is well understood in terms of the chiral restoration, which has two effects. First, the negative contribution in the UV is mainly understood as a direct consequence of the decrease of the contribution from the chiral condensate to the Lagrangian density. Second, a positive correction is observed in the IR, which corresponds to a shift of the baryon towards the IR. This shift also contributes to the negative region in Figure \ref{fig:rhoM_x}, and is understood as a consequence of the weakening of the IR-repelling bulk force felt by the baryon in the chirally broken background. The results also indicate that the correction to the classical soliton mass is negative, and relatively small in absolute value.

\end{itemize}

\subsubsection*{Rotating baryon solution and spin-isospin spectrum}

The second part of this work is devoted to the quantization of the isospin collective coordinates of the bulk soliton dual to a baryon. In the large $N_c$ limit, the moment of inertia $\l$ of the baryon is of order $\mathcal{O}(N_c)$ and the quantization is that of a solid rotor. The result of this procedure is the derivation of the spin-isospin baryon spectrum for the V-QCD model considered in this work.

The starting point for the quantization of the collective coordinates is the classical bulk solution obtained by a time-dependent isospin rotation of the static soliton, parametrized by
\be
\nn V(t) \equiv \exp{\left( it\, \omega^a \l^a \right)} \in SU(N_f)_{L+R}
\ee
with $\omega^a$ the rotation velocity. In this work we restrict to the following regime
\begin{itemize}

\item Only an SU(2) subgroup (the same where the static soliton sits) of the full isospin subgroup is quantized. This means that we impose that $V(t)\in SU(2)_{L+R}$. By doing so, we compute only a subset of the full spin-isospin spectrum, corresponding to baryons composed of quarks with 2 flavors (or equivalently, the states with strong hypercharge $Y=1$).

\item The rotation is assumed to be stationary and slow. In terms of the rotation velocity $\omega^a$, this means that $\omega^a$ is assumed to be a constant and obey $\omega^2 \ll M_0/\l$. This regime describes well the baryon states with spins
\be
\nn s \ll N_c\ .
\ee

\end{itemize}

The quantization therefore requires the calculation of the bulk solution corresponding to a slowly rotating baryon. This calculation is done at linear order in $\omega$. Already at this order, it turns out that a simple rotation of the static soliton fields with $V(t)\in SU(2)_{L+R}$ is not a solution of the bulk equations of motion. Instead, as soon as the soliton is made to rotate, some new flavor fields are turned on in the bulk, at linear order in $\omega$ \cite{Panico08}. These are the flavor equivalents of the magnetic field sourced by a rotating charge.

The appropriate ansatz for the rotating fields is constructed in Section \ref{Sec:rot}, by imposing the same symmetries as for the static solution, apart from time-reversal. These include 3-dimensional rotations and parity. Once this ansatz is determined, the construction of the rotating soliton solution follows the same steps as in the static case:
\begin{itemize}

\item We derive the expression of the moment of inertia of the soliton in terms of the ansatz fields.

\item We derive the full equations of motion for the fields of the rotating ansatz.

\item We identify the boundary conditions such that the moment of inertia is finite. The boundaries here are 1) the near-AdS region UV boundary $r\to 0$, where the solution should satisfy vev-like boundary conditions for all the fields;
2) the boundary at spatial infinity $|\vec{x}|\to\infty$, where the fields have to vanish fast enough for the moment of inertia to be finite.

\item We identify suitable  regularity conditions in the IR region of the geometry.
\end{itemize}

The last part of this work presents the results of the numerical calculation of the solution to the equations of motion for the rotating ansatz fields. From this solution, the moment of inertia is calculated and the corresponding spin-isospin baryon spectrum in Table \ref{tab:bspecI}.
\begin{table}[h]
\centering
\begin{tabular}{|c|c|c|}
\hline
Spin & V-QCD mass & Experimental mass \\
\hline
$s = \frac{1}{2}$ & $M_N \simeq 1170 \,\text{MeV} $ & $M_N = 940 \,\text{MeV} $\\
\hline
$s = \frac{3}{2}$ & $M_\D \simeq 1260 \,\text{MeV}$ & $M_\D = 1234 \,\text{MeV}$
 \\
\hline
\end{tabular}
\caption{Baryon spin-isospin spectrum in the V-QCD model with the potentials of Section \ref{Sec:VQCD_intro}, compared with experimental data.}
\label{tab:bspecI}
\end{table}
We emphasize that these numbers where obtained by substituting $N_c = 3$ in the leading order large $N_c$ and $N_f$ result. In principle, at
small values of $N_c$ and $N_f$, one does not expect this result to be
quantitatively accurate.

\subsection{Discussion and outlook}

The V-QCD baryon solution we present here has several  advantages with
respect to similar constructions in the literature, as well as some
limitations. In order to discuss them, we shall compare the results of
this work with the two main models of holographic QCD in which
single-baryon  solutions were analyzed. These are the top-down Witten-Sakai-Sugimoto model (WSS) \cite{SS,Sugimoto} and the bottom-up Hard-Wall model (HW) \cite{hard,Gorsky2015}.

The main improvement with respect to both models mentioned above is
that the V-QCD background solution on which the baryon solution is
constructed is a more accurate description of the QCD vacuum. The
V-QCD vacuum possesses a rich structure, including the running of the
Yang-Mills coupling and the spontaneous breaking of the chiral
symmetry in the chiral limit. Moreover, it  incorporates the
back-reaction of the flavor sector onto the color sector, due to the
Veneziano limit. The model can have several parameters that can be adjusted
to experimental data if one wants to produce a precise
phenomenological model for strongly-coupled QCD
\cite{spectrum,fT} although generically, the dependence on these parameters is weak.

Let us  now focus on the comparison with the HW model. In the HW model, a
baryon state was constructed as a bulk axial instanton for the chiral
gauge fields, using the same kind of ansatz that is considered in this
work \cite{Pomarol08}. However, the main difference is that the bulk
geometry was arbitrarily fixed to AdS$_5$, where a hard wall was
placed in the IR for the boundary theory to be confining. Because of
the gravitational potential,  the bulk
soliton was found to fall on the IR wall. This indicates that the
  hard wall model is too crude to stabilize the baryon solution
  dynamically. Moreover, since its position is at  the very end of space, the
  properties of the  soliton will strongly depend on the IR boundary
  conditions.
On the contrary, the baryon solution that is constructed in the present
 work is well localized in the holographic direction. It stands at a
 value of the holographic coordinate of the order of the inverse of
 the soliton mass. This is due to the fact that, beyond the metric,
 the V-QCD vacuum contains another field under which the baryon is
 charged: the tachyon field\footnote{The baryon solution including a
   non-trivial  tachyon in the context of the HW model was considered
   in \cite{GorskyDyonic,Gorsky2015}. In that work it was also found,  as in our
   model, that considering a   non-trivial tachyon  resulted in a
   repulsive force on the baryon from the  IR, although the mechanism for this to happen is  different (in our case the baryon is a probe on the tachyon background). At large chiral condensate, this could eventually make the baryon detach from the IR wall, but only a finite distance from it.} dual to the quark bilinear operator. The combined effect of the baryon boundary conditions and the interaction of the gauge fields with the tachyon field, result in a force that balances the gravitational attraction towards the IR.

Let us now discuss the WSS model. There, the main drawback of the baryon solution that was constructed in \cite{Sugimoto} is that the baryon size was found to be parametrically small at large 't Hooft coupling.  Instead, the size of the baryon solution that we derived in V-QCD is set by the mass scale of the boundary theory, which roughly corresponds to $\L_{\text{QCD}}$. Note that this was not an obstacle to the calculation of meaningful baryon form factors in the WSS model, as the latter were found to be related to the scale set by the rho meson mass rather than the soliton mass \cite{Hashimoto:2008zw}. Nevertheless, the infinitesimal size of the soliton will be an issue for classical fields in the bulk, such as the chiral condensate.

Another aspect where our construction is an improvement, compared with
previous settings, lies in the tachyon dependence of the bulk
action. First, the DBI form of the kinetic action for the flavor
fields contains an infinite sum of corrections compared with the
quadratic action considered in the HW model. In vacuum, such a
square-root behavior was found to play an important role to reproduce
linear trajectories for the meson spectrum \cite{spectrum}. Although in the WSS
model the same kind of action was introduced in \cite{BSS}, the
present work is the first one in which a baryon solution is computed
by keeping  the full  DBI action for the tachyon\footnote{Note that
  the calculation is done here by expanding the DBI action at
  quadratic order in the non-abelian fields, but keeping the abelian part of the tachyon fully non-linear.}. Second, and most importantly, we consider for the fist time the tachyon dependence of the topological Chern-Simons (CS) term. In our bottom-up approach, this term was constructed in \cite{BaryonI} as the most general topological action compatible with QCD symmetries and chiral anomalies.

The approach we followed here presents also some limitations. Apart from
the usual drawbacks which are intrinsic in a bottom-up model (a
certain amount of indeterminacy in the action, no known embedding as a
low energy approximation  of string theory), the most
important limitation is that the  solution presented here is only valid in the exact
chiral limit. This is related to the CS term mentioned above: as was
explained  in \cite{BaryonI}, our construction only applies in the
limit of zero quark masses, as  turning on
non-zero quark masses requires modifying both the CS term
and the instanton ansatz.  We refer the reader to \cite{BaryonI} for a more
extended discussion of this point.

In this work,  we focused  on a small subsector of the baryon spectrum. This is due to the specific ansatz considered for the quantization of the soliton excitations:
\begin{itemize}

\item We considered only the zero-modes, which resulted in the spin-isospin spectrum of the baryons. However, the experimentally observed baryon spectrum contains higher excited states for each isospin eigenvalue. These states are understood in the holographic picture as non-zero modes of the soliton, that is modes that are associated with a non-trivial potential, for which the soliton solution sits at the minimum. These include for instance the dilation and oscillating modes considered in the WSS model \cite{Sugimoto}.

\item Within the isospin zero-modes, we focused on a subgroup containing $N_f=2$ flavors. For phenomenology, it is interesting to quantize higher subgroups, in particular $N_f=3$. When introducing asymmetric masses for the quarks, this will make it possible to discuss the properties of holographic hyperons.

\item We restricted to the case of a slowly rotating soliton, which is enough to compute the spin-isospin spectrum at $s \ll N_c$. In particular, at linear order in the rotation velocity $\omega$, there is no deformation of the static fields, such that the cylindrical symmetry is preserved. The consequence on the spectrum is that only states with equal spin and isospin $s=I$ are reproduced. States that do not obey $s=I$ are observed experimentally, such as N(1520) or $\D(1950)$. Reproducing such states will require a rotating ansatz that deviates from cylindrical symmetry. This can be obtained, for example, by computing the solution at next order in $\omega$.

\item The assumption of slow rotation also means that the linear Regge trajectories that are observed experimentally cannot be reproduced. Namely, the spin-isospin spectrum that we compute is that of the rigid rotor, for which the masses go as $M\sim s^2$ at large spin, instead of $M\sim s^{1/2}$ for a linear trajectory. It is expected that reaching the linear regime, if it can be reached in this framework, will require to consider states with $s \gtrsim N_c$. For such high spins, the relevant ansatz for the rotating soliton should be fully non-cylindrical. In particular, it will reproduce the linear Regge behavior if it turns out that the solution resembles a string at high rotation velocity \cite{Cobi16}.
\end{itemize}
All the  points above can be the subject of future improvements.

\section{The V-QCD model: comparison to data}

\label{Sec:VQCD_intro}

We start by briefly reviewing the V-QCD model~\cite{Jarvinen}. First, we discuss the definitions needed for the data comparison.
We only discuss the main points, see the companion article~\cite{BaryonI} and the review~\cite{JR} for more details.

We then go on and determine the parameters of the model by comparing to QCD data. This is done in two stages: In the first stage, we choose the asymptotics of the functions so that the model has the potential to resemble QCD. In the second stage, we fit the remaining parameters to QCD data.   In earlier work, this second step has been done for pure Yang-Mills~\cite{gkmn} (see also~\cite{Ballon-Bayona,Ballon-BayonaT}), as well as for full QCD by fitting the thermodynamics~\cite{Remes} and meson spectrum~\cite{Amorim} separately. Here we carry out a simultaneous fit to the thermodynamic and spectrum data for full QCD, with a slightly higher weight on the spectrum fit because it probes directly the vacuum phase which is relevant for the current study. Apart from just combining the earlier thermodynamics and spectra fits, we also choose a different set of observables for the latter fit as compared to~\cite{Amorim}. In this reference, the stress was on fitting a high number of meson masses including relatively heavy states, in order to obtain a good description of Regge trajectories. Here we focus on the mesons with lowest masses and also fit the pion decay constant.

\subsection{The action of V-QCD}

V-QCD is a bottom-up holographic model for QCD with $N_c$ colors and $N_f$ flavors.
The holographic description is obtained by working in the Veneziano large-$N$ limit~\cite{Veneziano}:
\be \label{v1}
N_c \to \infty \quad  \mathrm{and}\quad N_f\to \infty, \quad \mathrm{with}\quad x\equiv {N_f \over N_c} \quad \mathrm{fixed} \ .
\ee

The five-dimensional dynamical fields of the bulk theory are in one-to-one correspondence
with the lowest-dimensional gauge-invariant operators in QCD. They are
\begin{enumerate}
\item The  metric $g_{MN}$, dual to the
  stress-tensor of QCD;
\item The dilaton $\lambda$, dual to the operator
  $\mathrm{Tr}\, G_{\mu\nu}G^{\mu\nu}$  (where $G_{\mu\nu}$ is the Yang-Mills field strength);
\item  Non-Abelian gauge fields
  $\mathbf{L}_M$, $ \mathbf{R}_M$, of the bulk gauge group $U(N_f)_L \times U(N_f)_R$, dual to the currents
\be \label{v2}
 (J^{(L)}_\mu)^i_j =
\bar{q}_{L}^i\gamma_\mu  q_{L\,j}, \qquad  (J^{(R)}_\mu)^i_j =
\bar{q}_{R}^i\gamma_\mu  q_{R\,j} \qquad i,j = 1\ldots N_f
\ee
 where  $q_{L,R}$ are the left and right-handed quarks;
\item The tachyon, i.e., a $N_f\times N_f$ complex scalar field  ${T^i}_j$, dual  to the quark bilinear $\bar{q}_R^i q_{L\,j}$, which therefore transforms in the  bi-fundamental representation of  $U(N_f)_L \times U(N_f)_R$.
\end{enumerate}

The five-dimensional action
\be \label{v3}
S_{V-QCD}  = S_g + S_{DBI} + S_{CS}
\ee
is the sum of three terms\footnote{The action generically contains an additional CP-odd term that couples the flavor fields to the holographic axion, dual to the $\text{Tr}(G\wedge G)$ operator. This term was derived in \cite{Arean:2013tja,theta} based on the ideas of~\cite{Casero}. We do not write this term here because, as explained in Section \ref{Sec:rot1}, the coupling to the axion is ignored in the following.}. First is the glue term $S_g$, equivalent to the action of IHQCD~\cite{ihqcd,ihqcd1}. The second and third terms are the flavor terms inspired by a configuration of space-filling D$_4$-branes and $\overline{\rm D}_4$-branes. The flavor action separates into a Dirac-Born-Infeld action (second term) and a Chern-Simons action (third term).

We now discuss explicit expressions for the terms in the action. The first term associated to pure Yang-Mills, is given by the five-dimensional Einstein-dilaton gravity:
\be \label{v4}
S_g = M^3 N_c^2 \int d^5x \sqrt{-g}\left[R - {4\over
    3\lambda^2} g^{MN}\de_M \lambda \de_N \lambda + V_g(\lambda)\right] \ .
\ee
This action is that of IHQCD~\cite{ihqcd,ihqcd1,ihreview}: it can be obtained from noncritical five-dimensional string theory, which also gives a prediction for the potential $V_g$. However, in order to obtain a phenomenologically interesting model, we need to switch to a bottom-up approach and choose a potential which reproduces known features of Yang-Mills theory. We shall discuss this in more detail below.
For the vacuum background solution, the dilaton field will be running from $\l = 0$ at the UV boundary to $\l = \infty$ in the IR (see Appendix~\ref{app:vac}). Because the dilaton is dual to the $\mathrm{Tr}\, G_{\mu\nu}G^{\mu\nu}$ operator, the corresponding source is the 't Hooft coupling $g_\mathrm{YM}^2 N_c$. As the source term dominates the background solution near the boundary, the dilaton can be in practice identified with the 't Hooft coupling of YM theory, in this region.

As for the flavor terms, the simplified DBI action introduced in \cite{fD} is enough to discuss the properties of the background, assuming that it is homogeneous and that all quark flavors have the same mass. For the full DBI action see~\cite{BaryonI}. The simplified DBI action reads
\begin{align}
\label{v7}
S_{DBI,simpl}
= &- M^3 N_c N_f\times \\
\nn &\times \int d^5 x\ V_f(\lambda,\tau)
\sqrt{-\mathrm{det}\left(g_{MN}+\kappa(\lambda) \de_M\tau \de_N \tau + w(\lambda) F_{MN}\right)} \, ,
\end{align}
where $\tau$ is the scalar tachyon field defined through $T^i_j = \tau \delta^i_j$, with $i,j
= 1\ldots N_f$, and $F_{MN}$ is the field strength of the Abelian vectorial gauge field $v_M$, defined through
\be \label{v9}
\left(\mathbf{L}_M\right)^{i}_{j}
=\left(\mathbf{R}_M\right)^{i}_j=  v_M \delta^i_j \ .
\ee
This form of the DBI action follows the ideas of Sen~\cite{AS} for the decay of a pair of unstable D-branes. Accordingly, the tachyon potential should be chosen to be exponentially suppressed at large $\tau$, $V_f(\l,\tau)\sim \exp(-a \tau^2)$, which models the annihilation of the branes in the IR. As the tachyon is dual to the $\bar qq$ operator in QCD, the bulk tachyon condensate driven by the exponential potential gives rise to a chiral condensate and chiral symmetry breaking on the QCD side.

We do not present explicitly the CS term $S_{CS}$ here, as it is complicated and not needed for the background analysis -- please see~\cite{BaryonI} for details. It is however essential for the baryon solutions that we  construct below.

\subsection{Comparing V-QCD to data: asymptotics}

We now discuss in more detail how the potentials $V_g(\l)$, $V_f(\l,\tau)$, $\kappa(\l)$ and $w(\l)$ are determined. We start by considering the asymptotic constraints as $\l \to 0$ (UV) or $\l \to \infty$ (IR), which arise from comparison to QCD.

In the UV, the field theory becomes weakly coupled and it is far from obvious that gauge/gravity duality can provide useful predictions. In this region we follow the usual practice for bottom-up models and adjust the model by hand so that it mimics closely the behavior of QCD in the perturbative regime. Such choices are expected to give good boundary conditions for the more interesting, strongly coupled IR physics.

In order to write down the asymptotics as $\l \to 0$, we write down the effective potential
\be \label{v11}
V_{eff}(\lambda,\tau) = V_g(\lambda) - {N_f \over N_c} V_{f}(\lambda,\tau) \ ,
\ee
as well as the expansion of the flavor potential at small tachyon:
\be
 V_{f}(\lambda,\tau) = V_{f,0}(\l)\left[1-\hat a(\l) \tau^2+\mathcal{O}\left(\tau^4\right)\right] \ .
\ee
The UV behavior is in general chosen such that the geometry is asymptotically AdS$_5$ and the asymptotics of the bulk fields match with the expected (free field) UV dimensions of the dual operators. Moreover, one can require that the logarithmic running of the dilaton agrees with the two-loop perturbative running of the QCD coupling. This maps the subleading terms in the potential $V_g$ and the effective potential $V_{eff}$ to the two-loop $\beta$ functions of the YM theory and full QCD, respectively. We implement this constraint here for the gluon potential $V_g$ only, and choose to fit the other higher order coefficients to low energy data instead. The explicit definitions of the UV expansions take the form
\bea
&& V_g(\lambda) = V_{g,0}\left[1 + V_{g,1}\lambda + O(\lambda^2)\right]  \nonumber \\
&& V_{eff}(\lambda,\tau=0) = V_0 \left[1+  V_1\lambda +
  O(\lambda^2)\right] \nonumber \\
&& V_{f,0}(\lambda) = W_0 \left[1+  W_1\lambda + O(\lambda^2)\right],
\nonumber \\
&&\label{v12}\kappa(\lambda)  = \kappa_0 \left[1+ \kappa_1\lambda + O(\lambda^2)\right],
\qquad \qquad
\lambda \to 0, \\
&&w (\lambda) = w_0 \left[1+  w_1 \lambda + O(\lambda^2)\right], \nonumber \\
&& \nn \hat a(\lambda)  = 1 + a_1 \l + \mathcal{O}(\l^2) \, .
\eea
Here the coefficients $V_{g,0}$ and $V_0$ are linked to the radii of the UV AdS$_5$ geometry in the absence of flavors, and for the full background, respectively:
\be \label{v13}
V_{g,0} = \frac{12}{\ell_g^2} \ , \qquad V_0 = V_{g,0} - x W_0 = \frac{12}{\ell^2} \ .
\ee
We fixed the normalization of the tachyon field such that $\hat a(\l=0) = 1$\footnote{This is without loss of generality, as we have left the normalization of the tachyon kinetic term free for the moment.}.

The requirement that the asymptotic dimension of the $\bar q q$ operator equals 3 sets
\be
 \kappa_0 = \frac{2\ell^2}{3} \ .
\ee
The subleading coefficients $V_{g,1}$ and $V_{g,2}$ are determined by comparing to the YM  $\beta$ function, which sets
\be\label{Vgivals}
  V_{g,1} = \frac{11}{27 \pi^2} \ , \qquad V_{g,2} = \frac{4619}{46656\pi^4} \ .
\ee
The rest of the parameters are left free at this point.

The IR ($\l \to \infty$) asymptotics of the various functions is chosen such that the model agrees with various qualitative features. First, the asymptotics of $V_g(\lambda)$ is adjusted such that the model has a discrete glueball spectrum with Regge-like linear radial trajectories (i.e. that the masses behave as $m_n^2 \sim n$ as a function of the radial excitation number $n$). This requirement automatically implies also color
confinement and  magnetic screening, \cite{ihqcd}.

The geometry ends in a ``good'' IR singularity according to the classification by Gubser~\cite{Gubser}. Second, several requirements constrain the asymptotics of the functions in the DBI action:

\begin{itemize}
\item The IR singularity should remain fully repulsive, so that the boundary conditions for the background and fluctuations are properly determined~\cite{Jarvinen}.

\item The meson mass spectrum needs to be discrete, admit linear Regge-like radial trajectories, and all meson trajectories should have the same universal slope~\cite{spectrum}\footnote{There is a certain freedom in this direction.  Tachyon condensation induces a mass term for the axial gauge field in the bulk, and depending on bulk IR asymptotics, it may change the slope of the radial  trajectories for the axial and vector mesons, \cite{IKP}. It is not yet clear whether this is true in QCD, \cite{SV}.}. The mass gap of, say, vector mesons should grow linearly with quark mass at large values of the mass~\cite{JM}.

\item The flavor action should vanish in the IR for chirally broken backgrounds, corresponding to the annihilation of the flavor branes in the IR, \cite{AS}. Moreover, the phase diagram as a function of $x$, $T$, $\mu$ and the $\theta$-angle, should have the qualitatively correct structure.
\end{itemize}

These requirements can be met if we choose the flavor action which behaves as
\be \label{v10}
V_f(\lambda, \tau) \sim e^{-a(\lambda) \tau^2 }, \qquad  \tau \to + \infty
\ee
with $a(\l)>0$, and require that as $\l\to\infty$
\bea \label{v15}
&&V_g \sim V_{IR} \lambda^{4/3} (\log \lambda)^{1/2} , \quad
V_{f,0} \sim W_{IR} \lambda^{v_p} , \nonumber \\
&& \kappa \sim \kappa_{IR}
\lambda^{-4/3}(\log \lambda)^{1/2},
\quad a\sim a_{IR}, \quad   w \sim w_{IR} \lambda^{-4/3} (\log \lambda)^{w_l},
\eea
with $4/3<v_p< 10/3$ and $w_l>1/2$ (see~\cite{spectrum} for details).

\subsection{Comparing V-QCD to data: fit of parameters }

After the asymptotics of the potentials have been fixed, the remaining task is to determine the leftover freedom by doing a more precise comparison to QCD data. We shall be using lattice data for the thermodynamics of the YM theory and QCD as well as glueball masses, and experimental data for meson masses and decay constants.

The complete ansatz is given by
\begin{align}
& V_g(\lambda)=12\,\biggl[1+V_{g,1} \l+{V_{g,2}\lambda^2
\over 1+c_\l \l/\l_0}+V_\mathrm{IR} e^{-\l_0/(c_\l\l)}\left(\frac{c_\l\l}{\l_0}\right)^{4/3}\!\!\sqrt{\log(1+c_\l\lambda/\l_0)}\biggr]\  \\
\label{Vfan}
& V_{f}(\lambda,\tau) = V_{f0}(\l) \left(1+ \tau^4\right)^{\tau_p} \exp\left(-a(\l) \tau^2\right)   \\
&V_{f0}(\l)= W_0 + W_1 \l +\frac{W_2 \l^2}{1+c_f\l/\l_0} + W_\mathrm{IR}\left(1+\frac{\bar W_1 \l_0}{c_f\l} \right) e^{-\l_0/(c_f\l)}(c_f\l/\l_0)^{2}\\
\label{aan}
& a(\l) = 1+\frac{1}{2} a_\mathrm{st} \left[1+\tanh \left(-a_\mathrm{sh}+\log (\l/\l_0) \right)\right]\\
\label{kappa}
&\kappa(\l) = \kappa_0 \biggl[1 +
\bar \kappa_0 \left(1+\frac{\bar \kappa_1 \l_0}{c_\kappa\l} \right) e^{-\l_0/(c_\kappa\l) }\frac{(c_\kappa\l/\l_0)^{4/3}}{\sqrt{\log(1+c_\kappa\lambda/\l_0)}}\biggr]^{-1}  \\
\label{wan} &w(\l) =  w_0\biggl\{1 + \frac{w_1 c_w\l/\l_0}{1+c_w\l/\l_0} +
\\\nonumber &\qquad \qquad \quad +\
\bar w_0
e^{-\hat\l_0/(c_w\l)}\frac{(c_w\l/\l_0)^{4/3}}{\log(1+c_w\lambda/\l_0)}\left[1+ w_\mathrm{as} \left(\log \left(1+c_w\lambda/\lambda_0\right)\right)^4\right]\biggr\}^{-1}
\end{align}
where $\l_0 = 8\pi^2$.
 We have also chosen $v_p=2$ in the IR, and set $\ell_g=1$, so that the AdS radius is given by
\be
 \ell = \frac{1}{\sqrt{1-x W_0/12}}
\ee
and consequently
\be \label{kappa0}
 \kappa_0 =\frac{2}{3\left(1-x W_0/12\right)} \ .
\ee
Requiring agreement with the two-loop $\beta$-function of Yang-Mills fixes the subleading coefficients of $V_g$  in the UV to the values given in~\eqref{Vgivals}.

We now set $x=2/3$, roughly corresponding to QCD with $N_f=2$ light quarks and $N_c=3$. The rest of the parameters are fitted to data. In addition to the parameters of the potentials, we also fit the 5D Planck mass $M$ and the characteristic scale $\Lambda$ of the background solutions, which is roughly dual to the scale $\Lambda_{QCD}$ in field theory, see Appendix~\ref{app:vac}.

 The fit is carried out in steps. We  fit the various functions sequentially to appropriately  chosen observables as we  explain below.
Although the best strategy would be a global fit of all the parameters to all appropriate data, this is numerically very demanding and we are not yet able to do it.

\begin{figure}[ht]
\begin{center}
\includegraphics[width=0.7\textwidth]{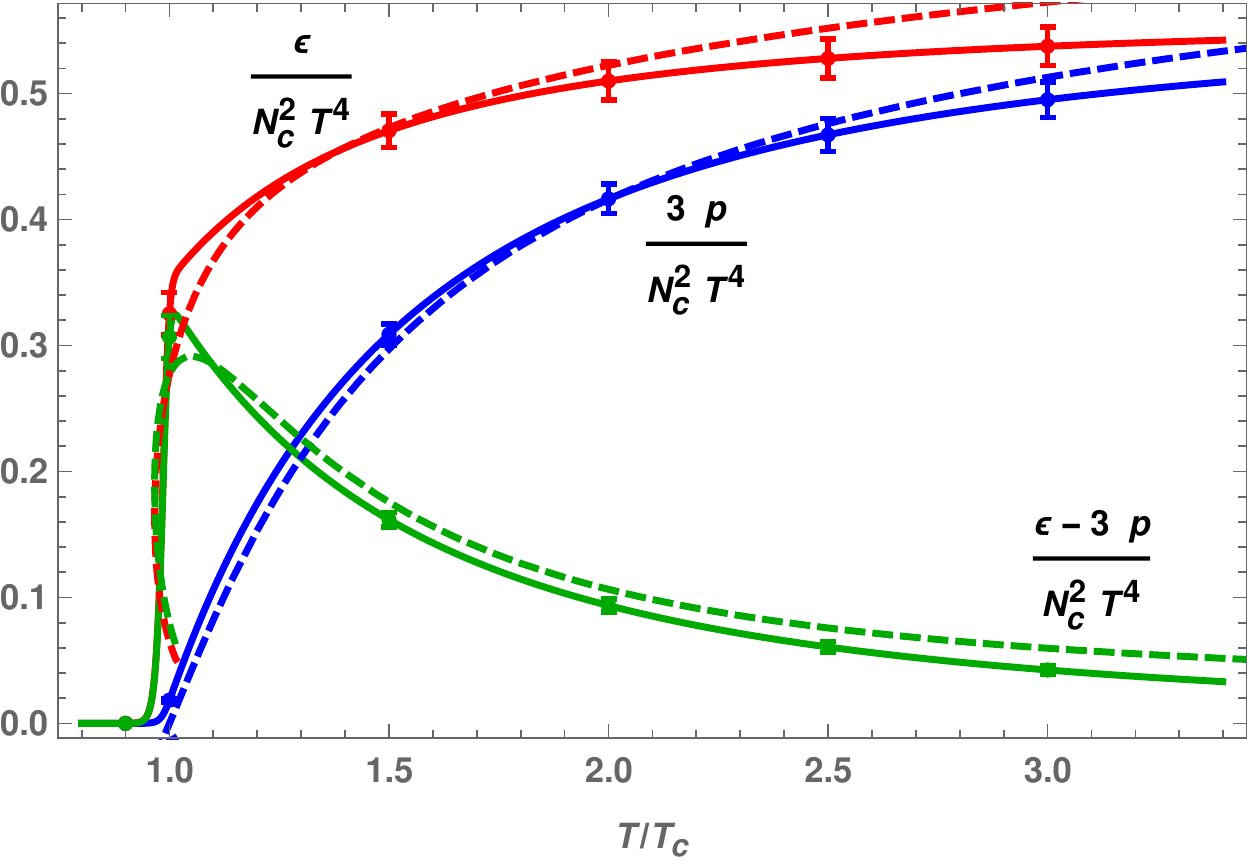}
\caption{Our result for the thermodynamics of pure Yang-Mills theory (dashed curves) compared to lattice data interpolated to $N_c \to \infty$~\cite{Panero} (solid curves and error bars). The red, blue, and green curves show the normalized energy density, pressure, and interaction measure, respectively. }
\label{fig:YMfit}
\end{center}
\end{figure}

\begin{table}
\centering
\begin{tabular}{ | c | c | c | c |}
\hline
Ratio & Model & Lattice ($N_c=3$) & Lattice ($N_c=\infty$) \\
\hline
\hline
$m_{0^{*++}}/m_{0^{++}}$ & 1.52 & $\begin{array}{c}1.603 \pm 0.042~\cite{Lucini} \\ 1.719 \pm 0.016~\cite{Athenodorou}  \end{array}$
  & $\begin{array}{c}1.835 \pm 0.032~\cite{Lucini} \\ 1.903 \pm 0.018~\cite{Athenodorou} \end{array} $ \\
\hline
$m_{2^{++}}/m_{0^{++}}$ & 1.29 & $\begin{array}{c}1.346\pm 0.037~\cite{Lucini} \\ 1.437\pm 0.011~\cite{Athenodorou} \end{array}$  & $\begin{array}{c}1.451\pm 0.048~\cite{Lucini} \\ 1.497 \pm 0.008~\cite{Athenodorou} \end{array}$ \\
\hline\hline
$T_c/m_{0^{++}}$ & 0.162 & $0.182 \pm 0.004~\cite{Lucini,LuciniT} $ & $0.181 \pm 0.003~\cite{Lucini,LuciniT} $ \\
\hline
\end{tabular}
\caption{Our results for glueball mass ratios and the ratio of the mass to the critical temperature compared to lattice results.}
\label{tab:YMratios}
\end{table}

The function $V_g$ is chosen to be the same as in the fit of~\cite{Amorim} where it was determined through a global fit to the QCD spectrum. Even if it was not directly fitted to the data from Yang-Mills theory, it does produce a good description of the lattice data for Yang-Mills thermodynamics~\cite{Panero} and the glueball spectra~\cite{Lucini,LuciniT,Athenodorou}, see Figure~\ref{fig:YMfit}\footnote{In this figure, the Planck mass $M$ and the scale parameter $\Lambda$ were fitted independently of the later fits to full QCD data. This reflects the dependence on $x$ of these parameters: the values for Yang-Mills are those with $x=0$, whereas for full QCD we take $x=2/3$.} and Table~\ref{tab:YMratios}.

\begin{figure}[h]
\begin{center}
\includegraphics[width=0.48\textwidth]{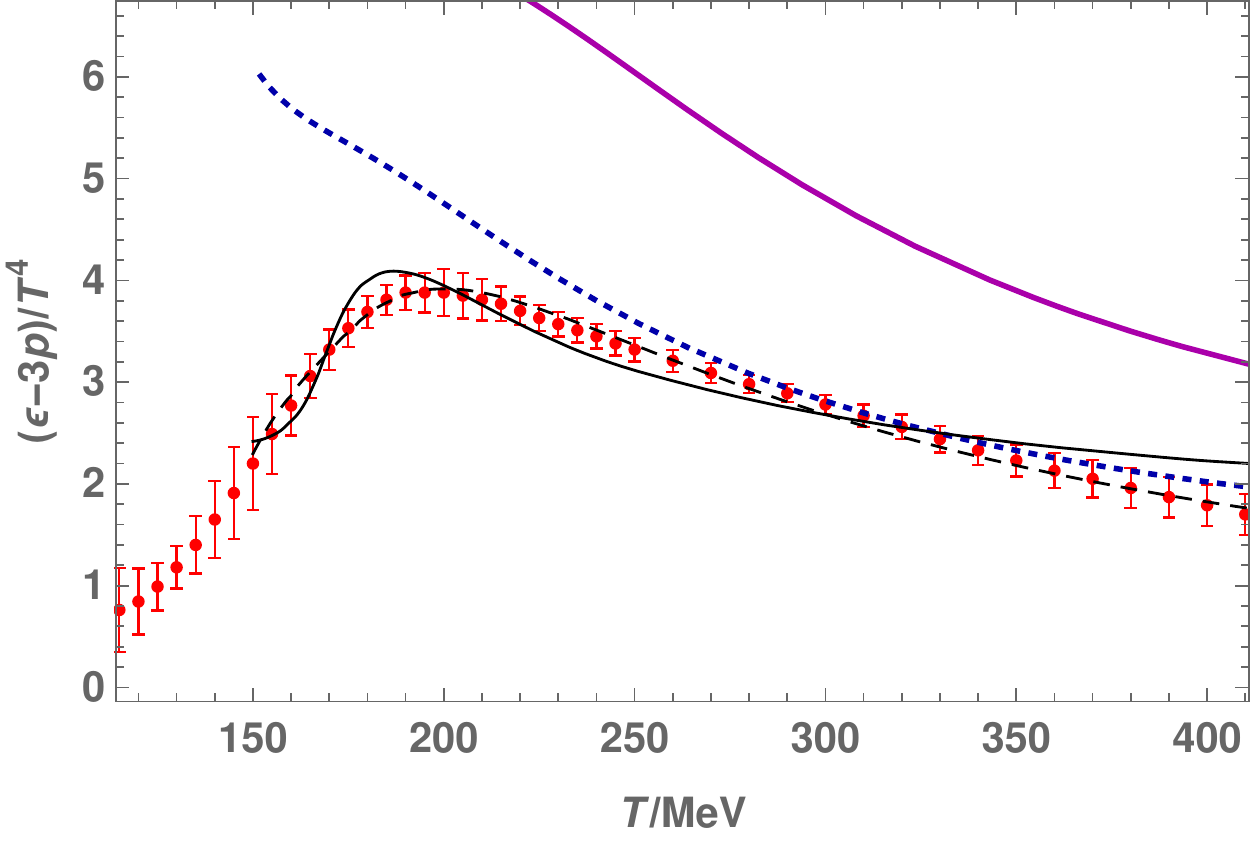}\hspace{5mm}%
\includegraphics[width=0.48\textwidth]{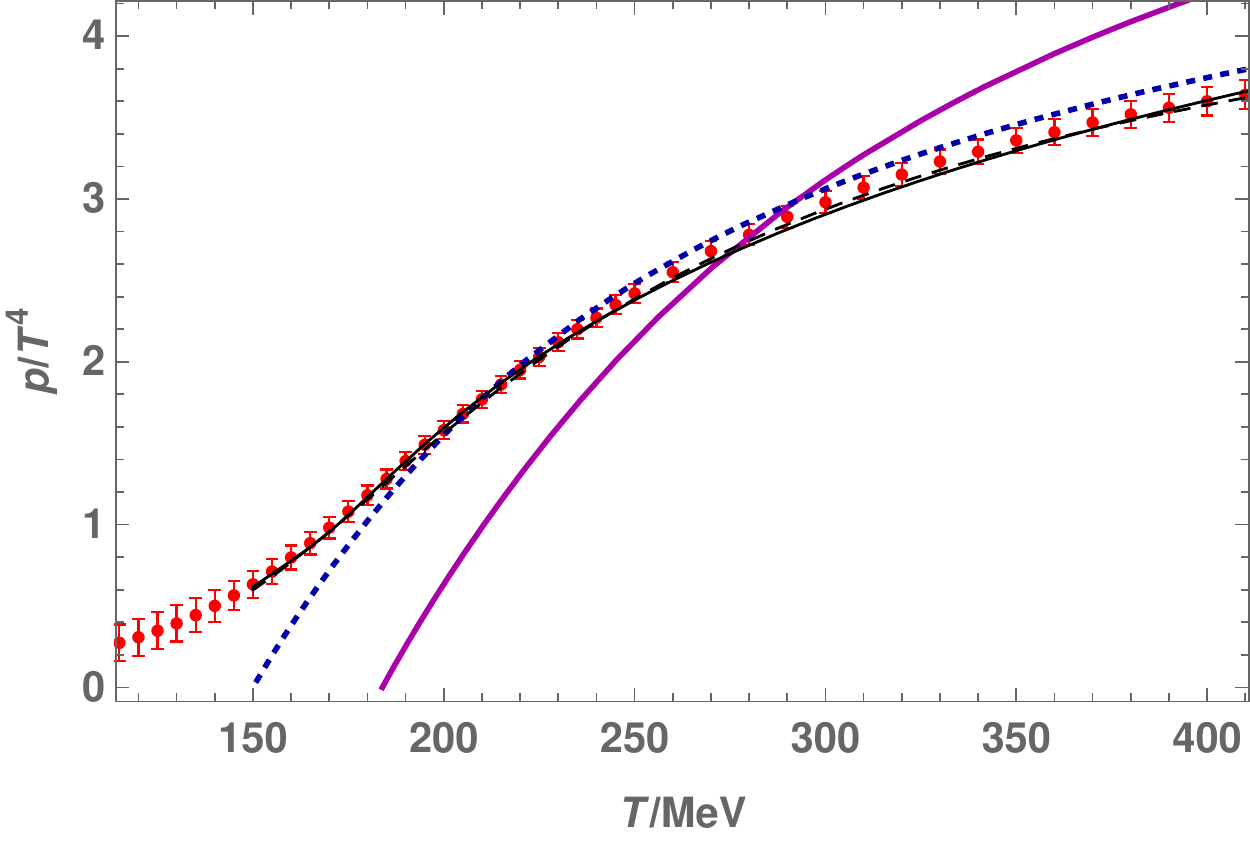}
\caption{Fits to the thermodynamics lattice data~\cite{Borsanyi} of QCD with 2+1 flavors. The thin black solid and dashed curves are direct fits of the effective potential $V_{eff}$. The dotted blue and solid magenta curves are final fits with scale parameters $M$ and $\Lambda$ optimized for thermodynamics and spectrum data, respectively, with fit parameters given in Table~\protect\ref{table:parameters}. These two fits have different transition temperature as the direct fits of the effective potential; see the  text for details.}
\label{fig:thermofit}
\end{center}
\end{figure}

The next function to be fitted is $V_f(\l,\tau)$ at zero tachyon, i.e., $V_{f0}(\l)$. For this function, we mostly use the lattice data for QCD thermodynamics at high temperatures and zero density in the chirally symmetric phase. As we are working at zero quark mass in the holographic model, chiral symmetry is fully restored in this phase, meaning that the tachyon is identically zero. This means that the thermodynamics only depends on $V_{f0}$ through the effective potential 
$$V_{eff}(\l,\tau=0) = V_g(\l) - x V_{f0}(\l)\;.$$
 In order to present our results, we introduce a reference value $M_{\text{UV}}$ for the Planck mass, defined by
\be
M_{\text{UV}}^3 \equiv \frac{1}{45\pi^2\ell^3}\left(1+\frac{7x}{4}\right) \sp x\equiv {N_f\over N_c}\, ,
\ee
where $x=2/3$.
This expression is obtained by requiring agreement of the pressure with the perturbative QCD result in the limit $T \to \infty$. We fitted this effective potential to latice data for various fixed values of the Planck mass $M$ while the other scale parameter $\Lambda$, which only affects the temperature scale in the plots, was allowed to vary freely. See Figure~\ref{fig:thermofit} for fits with $M^3/M_{\text{UV}}^3=1.5$ (dashed thin black curves) and $M^3/M_{\text{UV}}^3=4$ (solid thin black curves).\footnote{In the right hand plot disentangling the two curves is difficult because they almost overlap.}
Note, that while the fit results are determined mostly by $V_{eff}(\l,\tau=0)$, $M$, and $\Lambda$, they also  depend 
on other parameters of the theory via  the critical temperature of the phase transition between the chirally-unbroken and the chirally-broken vacuum.

 When fitting, we therefore took the critical temperature as an additional fit parameter. In the next step, the tachyon dependent functions were adjusted such that the critical temperature is indeed close to the fitted value. The two values will however be slightly different, and therefore the results for the thermodynamics for the final fit parameters, which are also shown in Figure~\ref{fig:thermofit} and will be discussed in detail below, differ from the direct fits of the effective potential.

The best fits were obtained at low $M^3/M_{\text{UV}}^3 \lesssim 2$. However, in the analysis that follows we chose a high value $M^3/M_{\text{UV}}^3 \simeq 4$. The reason is that such high values lead to a better description of the spectrum and in particular of the pion decay constant, which, as we remarked above, are stressed in the fit. We also note that there is a simple flat direction in this fit as the thermodynamics is unchanged under uniform rescalings of the coupling $\l \mapsto c \l$ in the effective potential. This happens because the kinetic term in~\eqref{v4} is invariant under such rescalings, so that a constant in $V_{eff}$ multiplying $\l$ can be eliminated through a field redefinition. We use this freedom to ensure that the constructed effective potential is consistent with the choice of $V_g$, meaning, in particular, that $V_{f0}$ is set to be positive and monotonic.

The most complicated step in the fit is the next step, where we choose the form of $V_f(\l,\tau)$ at nonzero tachyon, including the function $a(\l)$ in the exponential factor, and the tachyon kinetic term $\kappa(\l)$. These functions are probed by the chirally broken vacuum, which has nonzero bulk tachyon condensate. The main observables are the pion decay constant, the mass of the $\rho$ meson, and the mass of the lightest scalar flavor nonsinglet (i.e., isotriplet for $N_f=2$) state. When doing the fit, it is important to keep an eye on the ratio of the meson masses to the critical temperature.

As it turns out, fitting the thermodynamics and meson spectra simultaneously leads to tensions in the choices of potentials. The basic issue is that it is difficult to find a choice of function that would, at the same time, give high enough pion decay constant, heavy enough scalar states, and the experimentally observed meson mass to critical temperature ratio. In order to alleviate this tension we introduced an ansatz for the tachyon dependence of $V_f$ in~\eqref{Vfan} and~\eqref{aan}, which is somewhat more detailed than those used in previous studies. This ansatz has the following properties
\begin{itemize}

\item It includes a new parameter $\tau_p$ in (\ref{Vfan}),  controlling a term which depends on the tachyon only. We find that increasing $\tau_p$ leads to better fits, so we choose the value $\tau_p=1$ which is close to the maximal possible value. This maximum arises because $V_f(\l,\tau)$ needs to be monotonic in $\tau$ at small $\l$, otherwise no appropriate background solutions exist.

\item The exponent in the tachyon exponential $a$ in (\ref{Vfan}), is taken to be a function of the dilaton $a(\l)$. $a(\l)$ needs to be constant at both large and small $\l$, but may have a step in the middle. The fit result is that the IR value of $a(\l)$ should be significantly higher than the UV value (that was normalized to one).

\end{itemize}
In addition, the function $\kappa(\l)$ has three free parameters (notice that $\kappa_0$ in the ansatz~\eqref{kappa} is given in~\eqref{kappa0}). One of these (in practice $\bar \kappa_1$) is fitted such that the final critical temperature is close to that obtained from the fit of $V_{eff}$ discussed above. Due to tension with the fit to thermodynamics, we however choose a value that is a bit higher than obtained in the fit. The other two  parameters $\bar \kappa_0$ and $c_\kappa$, are used to adjust the function such that the pion decay constant and the scalar masses are optimal. This means, in practice, taking $\bar \kappa_0$ to be close to the critical value beyond which the model stops to be confining for mesons (see~\cite{spectrum}), and adjusting $c_\kappa$ according to the fit of the masses and the decay constant.

The remaining task is to fit the function $w(\l)$, parametrized in (\ref{wan}). The spectra, in particular the vector and axial meson masses, do depend on this parameter. But as it turns out, the dependence is rather weak. Therefore, we fit this function to lattice data on the baryon number susceptibility
\be
 \chi_B(T) = \frac{1}{N_c^2}\frac{\partial^2 p(T,\mu)}{\partial \mu^2}\Big|_{\mu=0}
\ee
following~\cite{Remes}. Here $\mu$ is the quark chemical potential. For the parameter $w_\mathrm{as}$ in (\ref{wan}), we choose a small value\footnote{This value can also be chosen to be zero without affecting the quality of the fits.}
\be
w_\mathrm{as} = 2\times 10^{-5}
\ee
so that this fit is essentially independent of the IR modification (i.e., the factor in the square brackets in~\eqref{wan}). Due to the weak dependence of the spectrum on $w(\l)$ the last two steps of the fitting procedure need to be done in part in parallel: in practice we determine first $V_f(\l,\tau)$ and $\kappa(\l)$ using a choice of $w(\l)$ that produces a rough fit to the lattice data for the susceptibility. When $V_f(\l,\tau)$ and $\kappa(\l)$ are known, we then tune $w(\l)$ to obtain a good fit as the last step.

\begin{table}
\centering
\begin{tabular}{ | c | c | }
\hline
Parameter & Value \\
\hline
$M^3/M_\mathrm{UV}^3$ & 4.929 (3.8) \\
\hline
$\Lambda$ & 4107~MeV (3350~MeV)\\
\hline
$M^3/M_\mathrm{UV}^3|_{x=0}$ & 1.4 \\
\hline
\hline
\hline
$V_\mathrm{IR}$ &  1.804\\
\hline
$c_\l$ &  2.833 \\
\hline
\hline
\hline
$W_0$ & 2.376 \\
\hline
$W_1$ & 0.04603 \\
\hline
$W_2$ & 0.02546 \\
\hline
$W_\mathrm{IR}$ & 1.783 \\
\hline
$\bar W_1$ & 4.357 \\
\hline
$c_f$ &  1.463 \\
\hline
\end{tabular}\hspace{8mm}
\begin{tabular}{ | c | c | }
\hline
Parameter & Value \\
\hline
$\tau_p$ & 1 \\
\hline
$a_\mathrm{st}$ & 2.5 \\
\hline
$a_\mathrm{sh}$ & 1.5 \\
\hline
\hline\hline
$\bar\kappa_0$ & 2.429 \\
\hline
$\bar\kappa_1$ & 0.32 \\
\hline
$c_\kappa$ & 2.0 \\
\hline
\hline
\hline
$w_0$ & 1.17 \\
\hline
$w_1$ & 52.5 \\
\hline
$\bar w_0$ & 200 \\
\hline
$c_w$ & 0.18 \\
\hline
$w_\mathrm{as}$ & $2\times 10^{-5}$ \\
\hline
\end{tabular}
\caption{Choices of model parameters, split in groups. For the first two parameters, the Planck mass $M$ and the scale $\Lambda$, the first set of values are determined by the $\rho$ mass and $f_\pi$, whereas the values in parentheses are the values preferred by the fit to thermodynamics. We also set $x=2/3$ and $b=10$ for the additional parameter appearing in the CS action~\protect\cite{BaryonI}.}
\label{table:parameters}
\end{table}

\begin{figure}[h]
\begin{center}
\includegraphics[width=0.6\textwidth]{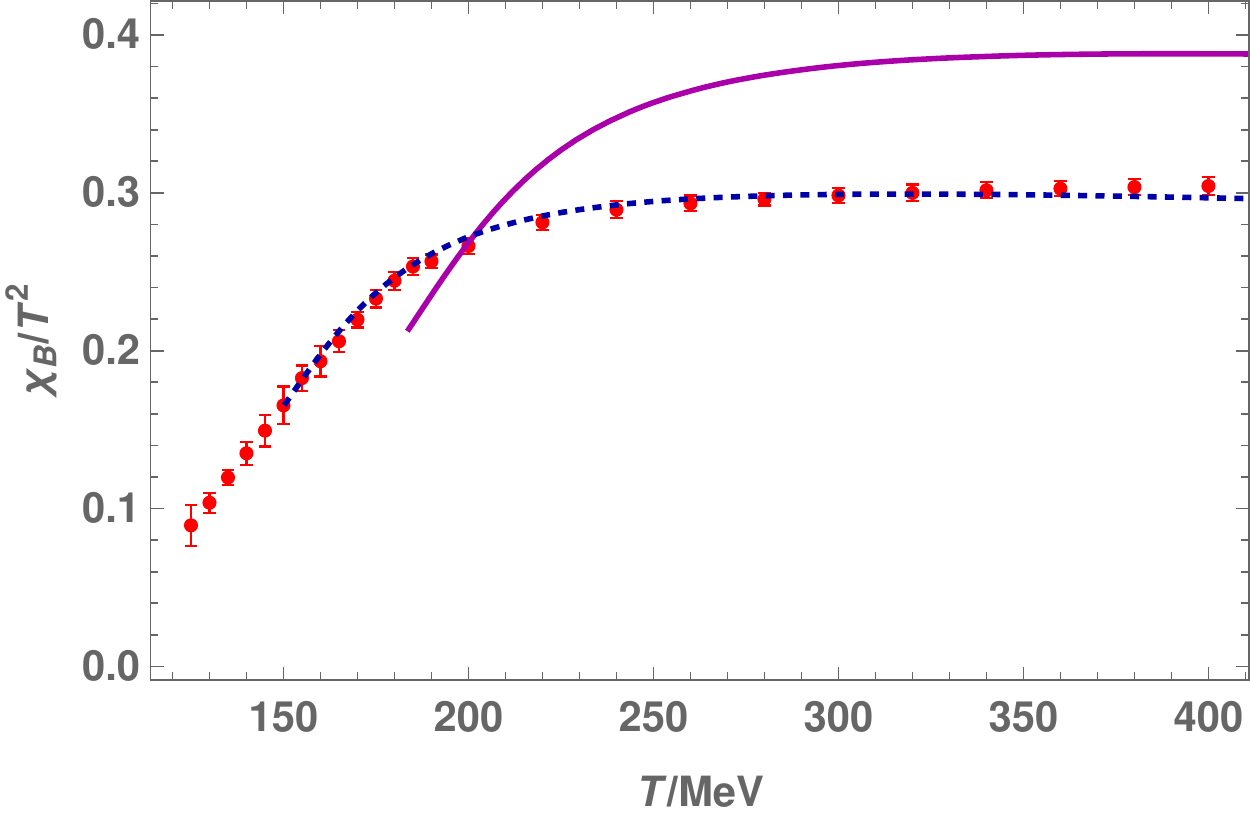}%
\caption{Fits to the lattice data (with 2+1 flavors)~\cite{Borsanyichi} for the light quark susceptibility of QCD. The dotted blue and solid magenta curves are final fits with scale parameters optimized for thermodynamics and spectrum data, respectively.}
\label{fig:suskisfit}
\end{center}
\end{figure}

\begin{table}
\centering
\begin{tabular}{ | c | c | c |}
\hline
Quantity & Model & Experiment~\cite{PDG} \\
\hline
\hline
$f_\pi$ & 92~MeV$^*$ & 92~MeV \\
\hline
$m_\rho$ & 775~MeV$^*$ & 775~MeV \\
\hline
$m_{\rho^*}$ & 886~MeV & $1465\pm 25$~MeV \\
\hline
$m_{a_1}$ & 1240~MeV & $1230\pm 40$~MeV \\
\hline
$m_{\pi^*}$ & 1260~MeV & $1300\pm 100$~MeV \\
\hline
$m_{a_0}$ & 639~MeV & -- \\
\hline
\end{tabular}
\caption{Fitted low lying (flavor nonsinglet) meson masses and the pion decay constant. The values of the scale parameters $M$ and $\Lambda$ were chosen such that the values of $f_\pi$ and $m_\rho$, marked with asterisks, match exactly the experimental values.}
\label{table:masses}
\end{table}

The final parameter values are collected in Table~\ref{table:parameters}. The fits for the QCD thermodynamics are shown in Figures~\ref{fig:thermofit} and~\ref{fig:suskisfit}. The results for the most important meson masses and the $f_\pi$ are shown in Table~\ref{table:masses}.

Before discussing the details of the fit results, we give a summary of the observables the various parameters were fitted to. Notice that the parameters in Table~\ref{table:parameters} were grouped in six groups. In the first group (top left), we show the final values fixed to $f_\pi$ and $m_\rho$ for the mass scales $M$ and $\Lambda$, as well as the values preferred by the fit to the lattice results for thermodynamics (values in parentheses). The value of $M$ at $x=0$ is the one used in Figure~\ref{fig:YMfit} for the thermodynamics of pure Yang-Mills.

 The parameters of $V_g$ (middle left group) were not fitted here but we used the values from~\cite{Amorim}. The thermodynamics and glueball masses from this choice were however seen to agree well with lattice data (Figure~\ref{fig:YMfit} and Table~\ref{tab:YMratios}).

 The parameters of the $V_{f0}$ potential (the bottom left group) are fitted to the thermodynamic data of Figure~\ref{fig:thermofit} and the parameters of the $w$ potential (the bottom right group) are fitted to the susceptibility in Figure~\ref{fig:suskisfit}. The remaining parameters of the tachyon potential $V_f$ (top right group) were adjusted to obtain a spectrum that mimics that of QCD, as shown in Table~\ref{table:masses}. The parameters of the $\kappa$ potential (middle right group) were fitted in part to the thermodynamics and in part to the spectrum.

Apart from the parameters fitted to data here, there is a single parameter arising from the CS sector,  which was discussed in the companion article~\cite{BaryonI}.  In~\cite{BaryonI} we derived the most general CS term which is compatible
with known constraints, and which contains four functions $f_i(\tau)$ which are only known at $\tau=0$ and at $\tau =\infty$. We will use here the functions derived in~\cite{Casero} arising from  flat space string theory, up to the parameter $b$ which corresponds to a rescaling of the tachyon field in the CS term. The precise functions are given in equation (3.17) of \cite{BaryonI}
\begin{align}
 f_1(\tau) &= -\frac{1}{6}e^{-b \tau^2}\,,& f_2(\tau) &= \frac{i}{12}(1+b\tau^2)e^{-b\tau^2}\,,\nn\\
 f_3(\tau) &= -\frac{1}{12}e^{-b \tau^2}\,,& f_4(\tau) &= \frac{1}{120}(2+2b \tau^2+b^2\tau^4)e^{-b\tau^2} \, .
\label{CKPfgen}
\end{align}
Our results here turn out to have little dependence on different finite values  of this additional parameter; we set $b=10$ following~\cite{Ishii}.

We now discuss some additional details of the fit.
 Figures~\ref{fig:thermofit} and~\ref{fig:suskisfit}, show the final fit to the thermodynamic data, with the values of $M$ and $\Lambda$ given in the parentheses in Table~\ref{table:parameters}. 
 
  The dotted blue curves  in Figure~\ref{fig:thermofit}, show the results for the thermodynamic fit to the equation of state. They  differ from the direct fit of $V_{eff}$, the solid black curve,
  because the transition temperature was adjusted differently in the final fit, in order to obtain a better agreement with the experimental meson spectrum. As we pointed out above, the final transition temperature is determined, apart from the effective potential, by the values of $\bar\kappa_0$, $\bar\kappa_1$ and $c_\kappa$ in Table~\ref{table:parameters}.
   
Finally, the solid magenta curves show the fit for the values of $M$ and $\Lambda$ that reproduce the values  of  $f_\pi$ and the $\rho$ mass, i.e., the values in Table~\ref{table:parameters} which are not in parentheses. Their difference to the dashed blue curves therefore demonstrates the remaining tension between the fits to thermodynamic and spectrum data. Because we are interested in baryons in the zero temperature vacuum state in this article, we chose to use this latter fit in the analysis of the properties of the baryon solution in the rest of this paper.

Notice that we only fit the lattice data above the QCD crossover, and in the deconfined phase of the holographic model, where the phases are separated by a first order phase transition.
The phase transition in the model, is at the same time a deconfining transition as well as a chiral restoration transition. Since we are in the massless quark case this is  in agreement with universality arguments, \cite{fT}. It is possible to obtain higher order phase transitions by tuning the potentials~\cite{Gursoy}, but such tuning would contradict the other constraints we have set, in particular the requirement of linear radial glueball trajectories.
However, it is expected that stringy loop corrections, which map to the pressure of pions and other light hadrons on the QCD side, can make the transition continuous also in the current setup for the holographic model~\cite{fT}. Such corrections are neglected in the holographic model, but may be added as in the second reference in \cite{fT}. After the holographic model has been fitted to lattice data, even simple hadron resonance gas models for the confined phase equation of state match almost continuously with the model in the deconfined phase~\cite{Demircik}.

Regarding the meson spectrum, we compare in Table~\ref{table:masses} the masses of the flavor nonsinglet mesons (i.e. the fluctuation modes with vanishing trace in flavor space) to the experimental values of isospin $I=1$ mesons from the particle data group tables~\cite{PDG}. We also include the pion decay constant, and its value as well as the $\rho$ meson mass are used to determine the final values of $M$ and $\Lambda$, as mentioned above.

Of the remaining mesons, the mass of the lowest  axial vector and the mass of the first pion excitation agree very well with the experimental values. The mass of the excitation of the $\rho$ meson is however too low. As it turns out, requiring the scalar mass to be high with respect to the $\rho$ mass leads to a situation where excited states in all sectors are rather close to the ground states. One should however also notice that the state in QCD that we are comparing is suspected not to be a clean radial excitation of the $\rho$ but to contain a significant hybrid component~\cite{PDG}. This may in part explain the difference in the numbers.

 As we remarked above, the mass of the lowest scalar $m_{a_0}$ is low: it is slightly less than $m_\rho$. We do not attempt to compare this mass directly to the experimental data as the scalar sector in QCD has a rather involved structure with several states that resemble pion and kaon molecules. Nevertheless our result is too low to be identified with any known state in the spectrum. This is perhaps not surprising since similar issues often appear in simple potential quark models.
We do note, however, that the model of~\cite{IKP}, which is closely related to V-QCD, does produce a significantly heavier flavor non-singlet scalar state. We also remark that we did not try to fit the pion mass as we carried out the fit at zero quark mass. It would be simple to fit the pion mass accurately by turning on a nonzero quark mass.

\begin{figure}[h]
\begin{center}
\includegraphics[width=\textwidth]{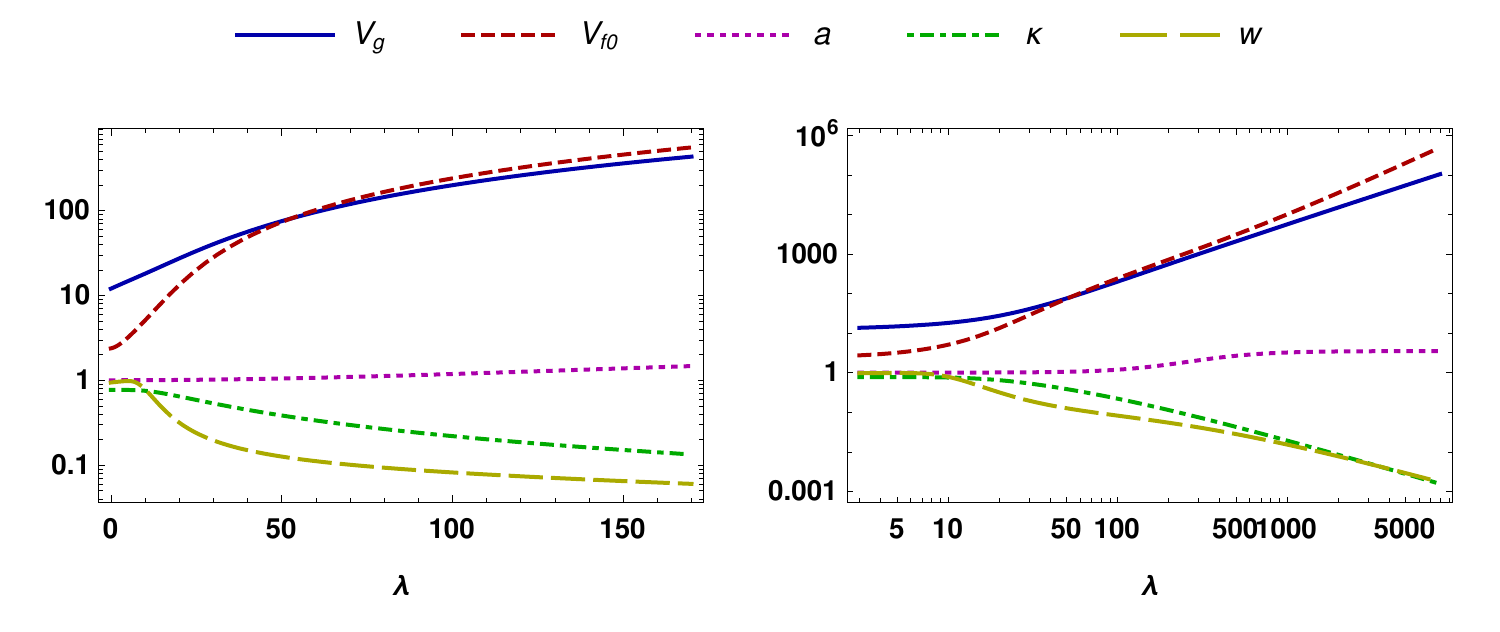}%
\caption{Potentials for the choice of parameters given in Table~\protect\ref{table:parameters}. Notice that the left (right) plot uses linear (logarithmic) scale for the dilaton $\la$.}
\label{fig:potentials}
\end{center}
\end{figure}

Finally we show the potentials after the fit in Figure~\ref{fig:potentials}. We remark that all the functions are simple, i.e. monotonic functions with no rapid changes in behavior. Notice also that even though we list a high number of parameters in Table~\ref{table:parameters}, almost all of these parameters only appear through the functional form of the functions shown in these plots. As the asymptotic form of the functions is fixed by comparison of QCD properties independently of the values of the parameters, they only affect the details of the functions in the middle, i.e., for $\l/\l_0 = \mathcal{O}(1)$. That is, despite the high number of parameters, the details of the model and therefore predictions for the observables are tightly constrained from the beginning, and the fit basically amounts to small tuning of the final results.

\section{The static soliton}

\label{Sec:SB}

We discuss in this section the bulk soliton dual to a static baryon state at the boundary.
A single static baryon is realised in the bulk as a  Euclidean instanton of
the non-abelian bulk gauge fields extended in the three
spatial directions plus the holographic  direction.
We start by reviewing the main ingredients of the formalism described in \cite{BaryonI} that are necessary for the present discussion.
In particular, we describe the ansatz that is used to compute the instanton solution in the bulk. We then present the numerical results for the static soliton solution.

\subsection{Ansatz for the instanton solution\label{31}}

The ansatz that is relevant for the instanton solution is obtained by requiring that the solution is invariant (up to a global chiral transformation) under a maximal set of symmetries of the V-QCD action (and QCD) compatible with a finite baryon number.  These symmetries are cylindrical symmetry (rotations in the 3 spatial directions of the boundary), parity and time-reversal. The parity and time-reversal symmetries act non-trivially on the fields in addition to their usual action on space-time. The explicit transformations can be found in \cite{BaryonI}.

\subsection*{Ansatz for the glue sector}

As explained in \cite{BaryonI} and reviewed in the next subsection, at leading order in $N_c$, the glue sector composed of the metric and dilaton is not affected by the presence of the baryon, and is identical to the vacuum solution. The latter depends only on the holographic coordinate $r$
\begin{equation}
\label{ds2}\mathrm{d}s^2 = \ex^{2A(r)}(-\mathrm{d}t^2 + \mathrm{d}\textbf{x}^2 + \mathrm{d}r^2)\, ,
\end{equation}
\be
\label{ansl} \lambda = \lambda(r) \, .
\ee

\subsection*{Gauge fields ansatz}

The left and right handed gauge fields are denoted
\begin{equation}
\label{defLR}\mathbf{L} = L + \hat{L} \mathbb{I}_{N_f} \sp \mathbf{R} = R + \hat{R} \mathbb{I}_{N_f} \, ,
\end{equation}
where $L$ and $R$ correspond to the $SU(N_f)$ part of the gauge fields, and $\hat{L}$ and $\hat{R}$ to the $U(1)$ part.

Because for any $N_f > 1$, the homotopy groups of $U(N_f)$ and $SU(2)$ are equal
\be
\label{p3SUN} \pi_3\left(U(N_f)\right) = \pi_3\left(SU(2)\right) = \mathbb{Z} \, ,
\ee
a $U(N_f)$ instanton can be constructed by embedding an $SU(2)$ instanton in $U(N_f)$. The $SU(2)$ subgroup couples to a $U(1)$ subgroup via the CS term, in such a way that the baryon ansatz belongs to a $U(2)$ subgroup of $U(N_f)$
\be
\label{U2gf} \mathbf{L} = L^a \frac{\s^a_L}{2} + L^{\text{T}} \mathbb{I}_{2,L} \sp \mathbf{R} = R^a \frac{\s^a_R}{2} + R^{\text{T}} \mathbb{I}_{2,R} \, ,
\ee
where $\s^a_{L/R}$ are the Pauli matrices and $\mathbb{I}_{2,L/R}$ are the unit matrices in the subgroups.

By imposing invariance under the cylindrical symmetry, time-reversal and parity, the $U(2)$ instanton ansatz for the gauge fields is
\begin{equation}
\label{ansatzSU2iL} L_i^a = -\frac{1+\phi_2(\xi,r)}{\xi^2}\e_{iak}x_k + \frac{\phi_1(\xi,r)}{\xi^3}(\xi^2\d_{ia}-x_ix_a) + \frac{A_\xi(\xi,r)}{\xi^2}x_ix_a \, ,
\end{equation}
\begin{equation}
\label{ansatzSU2iR} R_i^a = -\frac{1+\phi_2(\xi,r)}{\xi^2}\e_{iak}x_k - \frac{\phi_1(\xi,r)}{\xi^3}(\xi^2\d_{ia}-x_ix_a) - \frac{A_\xi(\xi,r)}{\xi^2}x_ix_a \, ,
\end{equation}
\begin{equation}
\label{ansatzSU2r} L_r^a = \frac{A_r(\xi,r)}{\xi} x_a \sp R_r^a = -\frac{A_r(\xi,r)}{\xi} x_a \, ,
\end{equation}
\begin{equation}
\label{ansatzU1} L^\text{T}_0 = \Phi(r,\xi) \sp R^\text{T}_0 = \Phi(r,\xi) \, ,
\end{equation}
where $i,k=1,2,3$ refer to spatial indices, $\xi \equiv \sqrt{x_1^2+x_2^2+x_3^2}$ is the 3-dimensional spatial radius and $a=1,2,3$ is the index for the components in the $SU(2)$ basis. Note that the gauge field ansatz is fully specified by 5 real functions
\be
\label{def2D} A_{\bar{\mu}} \equiv
(A_\xi,A_r) \sp \phi \equiv \phi_1 + i \phi_2 \;\;\; \text{and} \;\;\; \Phi \, ,
\ee
depending on
the two variables
$x^{\bar{\mu}}\equiv (\xi,r)$, that are used as coordinates on a 2D
space.

The choice of the ansatz partially fixes the gauge but there is still a residual
(axial) $U(1)$ invariance,  corresponding to the $SU(2)$
transformation
\be
\label{resg1}
 g_L = g_R^{\dagger} = \exp\left(i \a(\xi, r){x\cdot\s
  \over 2\xi}\right) \, ,
\ee
where
\be
\label{defxds} x\cdot \s \equiv x^a \s^a \, .
\ee
Under this gauge transformation, $A_{\bar{\mu}}$ is the gauge field, $\phi$ has charge +1 and $\Phi$ is neutral
\be
\label{gtgf} A_{\bar{\mu}} \to A_{\bar{\mu}} + \partial_{\bar{\mu}}\a \sp \phi \to \ex^{i\a} \phi \sp \Phi \to \Phi \, .
\ee

\subsection*{Tachyon ansatz}

For the tachyon matrix restricted to the $SU(2)$ subgroup $T^{SU(2)}$, the ansatz compatible with the symmetries of the V-QCD action is of the form
\be
\label{TU2} T^{SU(2)} = \tau(r,\xi) \exp{\left(i\theta(r,\xi)\frac{x\cdot\s}{\xi}\right)} \, ,
\ee
which reproduces the Skyrmion ansatz for the unitary part of the tachyon. This last point is more than a coincidence. Indeed, from the near-boundary behavior at $r\to 0$ of the tachyon field at zero quark mass, the pion matrix in the boundary theory is identified to be
\be
\label{Pmat} U_P(\xi)^{\dagger} \equiv \exp{\left(i\theta(0,\xi)\frac{x.\s}{\xi}\right)} \, .
\ee
Also, the baryon number in the boundary theory can be shown to be equal to the Skyrmion number for $U_P$ \cite{BaryonI}. Note that under the residual $U(1)$ gauge freedom \eqref{resg1}, the tachyon phase in \eqref{TU2} transforms as
\be
\label{gtt} \theta\to \theta - \a \, .
\ee

To summarize the content of this subsection, the ansatz for the instanton solution \eqref{ansatzSU2iL}-\eqref{ansatzU1}, and \eqref{TU2} contains 7 real dynamical fields
\be
\label{sumans} \Phi(r,\xi) \sp \phi(r,\xi)\equiv \phi_1(r,\xi) + i \phi_2(r,\xi) \sp A_{\bar{\mu}}(r,\xi) \equiv  \left(A_\xi(r,\xi), A_r(r,\xi)\right) \, ,
\ee
\be
\nonumber \tau(r,\xi) \sp \theta(r,\xi) \, ,
\ee
that depend on the two coordinates $x^{\bar{\mu}}=(\xi,r)$ and have a U(1) gauge redundancy under which the fields transform as
\be
\label{gtsum} \phi \to \ex^{i\a} \phi \sp A_{\bar{\mu}} \to A_{\bar{\mu}} + \partial_{\bar{\mu}}\a \sp \theta \to \theta - \a \, .
\ee

At this point, a useful observation is that there exists a redefinition of the fields \eqref{sumans} such that, in the equations of motion for the flavor fields, the phase $\theta$ in the tachyon ansatz \eqref{TU2} can be absorbed into the gauge field. By doing so, the dynamical field content is reduced to a set of 6 fields invariant under the residual gauge freedom. In practice, if we define
\be
\label{defgth} g(\theta) \equiv \exp{\left(i\theta \frac{x\cdot\s}{2\xi}\right)}\, ,
\ee
then we consider the following redefinition of the gauge fields
\be
\label{LtoLt} \mathbf{L}_M \to \tilde{\mathbf{L}}_M \equiv g(\theta)\mathbf{L}_M g(\theta)^\dagger + ig(\theta)\partial_M g(\theta)^\dagger  \, ,
\ee
\be
\label{RtoRt} \mathbf{R}_M \to \tilde{\mathbf{R}}_M \equiv g(\theta)^\dagger\mathbf{R}_M g(\theta) + ig(\theta)^\dagger\partial_M g(\theta) \, ,
\ee
which for the ansatz \eqref{ansatzSU2iL}-\eqref{ansatzU1} is equivalent to
\be
\label{LtoLtcyl} A_{\bar{\mu}} \to \tilde{A}_{\bar{\mu}} \equiv  A_{\bar{\mu}} + \partial_{\bar{\mu}}\theta \sp \phi \to \tilde{\phi} \equiv \ex^{i\theta}\phi \sp \Phi \to \Phi \, .
\ee
From  \eqref{gtsum}, we see that the gauge fields thus redefined are indeed invariant under the residual gauge transformation  \eqref{resg1}.

In the following, we find it convenient to use the redefined fields in several places. When they appear, we  always write the tildes so that it is clear that we are using the gauge-invariant fields.

\subsubsection*{Lorenz gauge}

To compute the baryon solution, we  need to solve the bulk equations of motion with the ansatz \eqref{ansatzSU2iL}-\eqref{ansatzU1}. The equations of motion can be written in terms of the redefined gauge fields \eqref{LtoLtcyl}. In particular, when solving the equations of motion numerically, we find convenient to work with $\td{\phi}$ instead of $\phi$ as a dynamical field.

However, as far as the 2-dimensional gauge field $A_{\bar{\mu}}$ is concerned, the equations of motion written in terms of $\td{A}_{\bar{\mu}}$ are not elliptic. While this is not problematic per se, the heat diffusion method that we use to solve the equations numerically requires that they are in elliptic form.

The equations of motion can  be recast in  elliptic form if we write
  them instead in terms of the gauge variant fields
  \eqref{ansatzSU2iL}-\eqref{ansatzSU2r}, and then  fix the gauge with the Lorenz condition
\be
\label{Lorenzc1} \partial_r A_r + \partial_{\xi} A_{\xi} = 0 \, .
\ee
This is the gauge that we shall use in the following.

\subsection{Probe and back-reacting solutions}

\label{Sec:AS}

The instanton dual to a baryon state is a configuration of the ansatz of equations \eqref{ansatzSU2iL}-\eqref{ansatzU1} and \eqref{TU2} that obeys the bulk equations of motion. These are written in Appendix F of \cite{BaryonI}.

{As explained in the previous subsection, we shall consider a baryon whose flavor quantum numbers are a $U(2)$ subgroup of the $U(N_f)$ flavor group. This implies that the flavor
action (composed of the DBI and CS actions in} \eqref{v3})  {for the baryon ansatz does not depend on $N_f$ and is of order $N_c$. On the other hand, the glue action is of order $N_c^2$.  Likewise, the tachyon modulus background contributes a factor $N_f$ more than the baryon fields to the bulk action, which can be seen explicitly from} \eqref{M0} below. {So, at leading order in the Veneziano limit, both the glue sector (metric and dilaton) and the tachyon modulus $\tau$ are not affected by the presence of the baryon, and remain identical to the vacuum solution.}

We start by computing the numerical baryon solution in this leading order probe regime. In this case, the dynamical fields are those listed in \eqref{sumans} with the exception of the tachyon modulus $\tau$, which is fixed to its background value. The equations of motion obeyed by those fields are written in Appendix F.1 of \cite{BaryonI}.

In the Veneziano limit, the back-reaction on the background starts at order $\mathcal{O}(1/N_c)$. At this order, the correction to the glue sector (metric and dilaton) and tachyon modulus $\tau$ can be computed by solving the linearized Einstein-dilaton equations sourced by the probe baryon solution, together with the linearized equations for $\tau$. Qualitatively, we do not expect a dramatic effect on the glue sector from the presence of the baryon. Correspondingly, the back-reaction on the glue sector is not expected to affect much the flavor structure of the baryon, which is its most important dynamical property.
This motivates the approximation that we consider in the following, where the baryon is assumed to back-react only on the tachyon background. The equations of motion in this case are written in Appendix F.2 of \cite{BaryonI}.

In summary, we  consider two different regimes for the baryon solution, with different treatments of the tachyon modulus:
\begin{itemize}
\item The \emph{probe baryon} solution, where the tachyon modulus $\tau$ is fixed to its vacuum value. This corresponds to the solution at leading order in the Veneziano expansion. At this order, the chiral condensate profile around the baryon is trivial, but the other flavor properties of the baryon are expected to be qualitatively correct.
\item The \emph{back-reacted tachyon} regime, where the equations of motion for $\tau$ are solved, with the gauge fields and tachyon phase $\theta$ fixed to the probe baryon solution. In the Veneziano limit, this will reproduce the leading order $\mathcal{O}(1/N_f)$ correction to the tachyon background, assuming no back-reaction on the glue sector. In this case, the solution obtained for the tachyon modulus should give a good idea of the qualitative behavior of the chiral condensate in presence of the baryon.
\end{itemize}

\subsection{Boundary conditions}

\begin{table}[h]
\centering
\begin{tabular}{|c|c|c|c|}
\hline
$\xi \to 0$ & $\xi \to \infty$ & $r \to 0$ & $r \to r_{\text{IR}} \to \infty $  \\ \hline
$\partial_{\xi}\tilde{\phi}_1 - (A_{\xi} + \partial_{\xi}\theta) \to 0$ & $\xi^{1/2} \tilde{\phi}_1 \to 0$ & $\tilde{\phi}_1 \to \sin{\theta}$ & $\tilde{\phi}_1 \to 0 $         \\
 $\frac{1+\tilde{\phi}_2}{\xi}\to 0$ & $\xi^{1/2}\left(\tilde{\phi}_2 - 1 \right) \to 0$ & $\tilde{\phi}_2 \to -\cos{\theta}$ & $\partial_r\tilde{\phi}_2 \to 0 $     \\
$\partial_{\xi}A_{\xi} \to 0$ & $\partial_{\xi}A_{\xi} \to 0$ & $A_{\xi} \to 0 $ & $A_{\xi} \to 0 $      \\
 $A_r \to 0$  & $\xi^{3/2} \left(A_r-\frac{\pi}{r_{\text{IR}}} \right) \to 0$ & $ \partial_r A_r \to 0$ & $ \partial_r A_r \to 0$       \\
$\partial_{\xi}\Phi \to 0$  & $\Phi \to 0$ & $\Phi \to 0$ &  $\partial_r\Phi \to 0$      \\
$\theta \to 0$ & $\theta \to \pi \left(1-\frac{r}{r_{\text{IR}}} \right) $ & $\partial_r\theta + A_r  \to 0$ & $ \theta \to 0$ \\
$\partial_\xi\tau \to 0$ & $\tau \to \tau_b(r)$ & $\tau \to 0$ & $\tau\to\tau_b(r_\text{IR})$ \\ \hline
\end{tabular}
\caption{Boundary conditions in Lorenz gauge. Here, $\tau_b(r)$ is the vacuum profile for the tachyon modulus. $L$ and $r_{\text{IR}}$ are the spatial and IR cut-offs, respectively.  These are introduced to solve the equations of motion numerically.\label{tab:bcsLg}}
\end{table}

The equations of motion for the fields of the instanton ansatz can be derived in the form presented in Appendix F of \cite{BaryonI}. These equations must be subject
to appropriate boundary conditions both at spatial infinity $\xi \to
+\infty$ and at the UV boundary $r\to 0$. The appropriate conditions were derived in \cite{BaryonI} from the requirement that the baryon mass be finite and the baryon number equal to 1. Moreover, it was observed that certain
(generalized) regularity
conditions must be imposed at the center of the instanton $\xi=0$ and
in the bulk interior. We list in Table \ref{tab:bcsLg} the conditions that
are imposed on the fields of the ansatz \eqref{sumans} in Lorenz gauge and refer to \cite{BaryonI} for more details about the derivation. The conditions are the same for the probe and back-reacted cases.

\subsection{Baryon mass}

Once the soliton solution is found, several static properties of baryons can be computed \cite{Pomarol08, Hashimoto:2008zw}. The most elementary of these properties is the nucleon mass. This mass is the sum of a classical contribution and quantum corrections
\be
\label{MB} M_{\text{nucleon}} = M_0 + \d M_Q \, .
\ee
The classical contribution is computed from the bulk on-shell action evaluated on the soliton solution \cite{BaryonI}. In terms of the ansatz fields \eqref{sumans}, its expression in the approximation where the back-reaction on the glue sector is neglected is given by
\be
\label{M0} M_0 = N_c\Big( N_f S_\tau + S_B \Big) - E_{DBI,vac} \, ,
\ee
where $E_{\text{DBI,vac}}$ is the DBI contribution to the vacuum energy 
\be
\label{EDBIvac} E_{DBI,vac} =  \int \mathrm{d}r\mathrm{d}\xi\,4\pi\xi^2 \rho_{DBI,vac} \sp \rho_{DBI,vac} \equiv M^3 \sqrt{1 + \ex^{-2A}\ka(\partial_r\tau_0)^2} \, V_f(\l,\tau_0)\ex^{5A}  ,
\ee
with $\tau_0(r)$ the vacuum profile of the tachyon field. The baryon contribution to the bulk action is split into two pieces
\be
\label{StB1} S_\tau = \int \mathrm{d}r\mathrm{d}\xi\,4\pi\xi^2 \rho_\tau \sp S_B = \int \mathrm{d}r\mathrm{d}\xi\,4\pi\xi^2 \rho_B \, ,
\ee
\be
\label{rtau} \rho_\tau \equiv M^3 \sqrt{1 + \ex^{-2A}\ka\left((\partial_r\tau)^2+(\partial_{\xi}\tau)^2\right)} \, V_f(\l,\tau)\ex^{5A} \, ,
\ee
\begin{align}
\nn \rho_B & \equiv   M^3 \sqrt{1 + \ex^{-2A}\ka\left((\partial_r\tau)^2+(\partial_{\xi}\tau)^2\right)}\, V_f(\l,\tau)\,\ex^A \times \\
\nonumber &\hphantom{= +4\pi M^3N_c } \times \left(\ex^{2A}\ka(\l)\tau^2\left(\ex^{2A}\D_{rr}\td{A}_r^2 + \left(1 - \ex^{2A}\D_{\xi\xi}  \right)\td{A}_{\xi}^2 + \frac{(\td{\phi} + \td{\phi}^*)^2}{2\xi^2} - \right. \right. \\
\nn &\hphantom{= +4\pi M^3N_c \times \left(\ex^{2A}\xi^2\ka(\l)\tau^2\left(\ex^{2A}\D_{rr}\td{A}_r^2 + \left(1 - \ex^{2A}\D_{\xi\xi}  \right)\td{A}_{\xi}^2 \right.\right.}  - 2\ex^{2A}\D_{\xi r}\td{A}_r\td{A}_{\xi} \Bigg) + \\
\nonumber &\hphantom{= +4\pi M^3N_c \times \Bigg( }  + w(\l)^2\left(\frac{1}{8}\ex^{2A}\left[\D_{rr}\left(1 - \ex^{2A}\D_{\xi\xi}  \right) - \ex^{2A}\D_{\xi r}^2\right](F_{\bar{\mu} \bar{\nu}})^2 + \right. \\
\nn &\hphantom{= +4\pi M^3N_c \times \Bigg( + w(\l)^2 \bigg( }
 + \frac{1}{2\xi^2}\left(\left(1 - \ex^{2A}\D_{\xi\xi}  \right)\left|\mathrm{D}_{\xi}\phi\right|^2 + \ex^{2A}\D_{rr} \left|\mathrm{D}_{r}\phi\right|^2\right) + \\
\nn & \hphantom{= +4\pi M^3N_c \times \Bigg( + w(\l)^2 \bigg(}
 + \frac{\left(1 - |\phi|^2\right)^2}{4\xi^4} - \frac{1}{2\xi^2}\ex^{2A}\D_{\xi r} (D_r\phi^*D_{\xi}\phi + h.c.) - \\
\nn & \hphantom{= +4\pi M^3N_c \times \Bigg( + w(\l)^2 \bigg(}
 - \Big(\ex^{2A}\D_{rr}(\partial_r\Phi)^2 + \left(1 - \ex^{2A}\D_{\xi\xi}  \right)(\partial_{\xi}\Phi)^2  - \\
\nn &\hphantom{= +4\pi M^3N_c \times \Bigg( + w(\l)^2 \bigg(  - \frac{N_f}{2} \xi^2 \Big(\ex^{2A}\D_{rr}(\partial_r\Phi)^2 }
 - 2\ex^{2A}\D_{\xi r}\partial_{\xi}\Phi\partial_r\Phi \Big)\bigg)\Bigg)+\\
\nn &\hphantom{=} + \frac{1}{\pi^2 \xi^2} \e^{\bar{\mu}\bar{\nu}}\partial_{\bar{\mu}}\Phi\times \\
\nn &\hphantom{=+}\times \left[(f_1(\tau)+f_3(\tau))\left(\td{A}_{\bar{\nu}} + \frac{1}{2}(-i\phi^*D_{\bar{\nu}}\phi + h.c.)+\frac{1}{4i}\partial_{\bar{\nu}}(\td{\phi}^2-(\td{\phi}^*)^2)\right)+ \right. \\
\label{rhoB} &\hphantom{=+\times\bigg[} \left. + \frac{1}{2}(3if_2(\tau)-f_1(\tau)-f_3(\tau))(\td{\phi} + \td{\phi}^*)^2 \td{A}_{\bar{\nu}} \right] \, .
\end{align} 
The covariant quantities $F_{\bar{\mu}\bar{\nu}}$ and $D_{\bar{\mu}}\phi$ are defined in Appendix \ref{Sec:2Dconv} and the symbol $\D_{\bar{\mu}\bar{\nu}}$ in \eqref{defDxx}-\eqref{defDrr}. The $f_i(\tau)$ are the Chern-Simons potentials, whose expressions are given in \eqref{CKPfgen}. Note that, as far as flavor fields are concerned, $S_\tau$ depends only on the tachyon modulus $\tau$, whereas
$S_B$ contains the dependence on the baryon fields. In particular, in the leading order probe baryon regime, only $S_B$ contributes to $M_0$. The total bulk Lagrangian density will be denoted by $\rho_M$
\be
\label{defrM} \rho_M \equiv N_c(N_f\rho_\tau + \rho_B) - \rho_{DBI,vac} \, .
\ee

Computing the quantum corrections $\d M_Q$ requires to take
the sum of the ground state energies for the infinite set of bulk
excitations on the instanton background, and subtract the vacuum
energy. It is not known how to do this calculation, so we can only
assume that the classical mass gives the dominant contribution. Note that, in terms of the expansion in $N_c$, the classical mass is of order
$\mathcal{O}(N_c)$, whereas the quantum corrections start at order
$\mathcal{O}(1)$. So the statement that the classical contribution dominates is correct at least in the large $N_c$ limit.

The experimental spectrum of baryons contains the nucleons, but also
many excited states, such as the isobar $\D$. The calculation of the spin dependence of the baryon mass spectrum is the subject of Section \ref{Sec:QI}.

\subsection{Numerical results}

We present in this subsection the numerical solution for the static baryon configuration. The equations of motion written in Appendix F of \cite{BaryonI} are solved with the gradient descent method\footnote{The name heat diffusion method also appears in the literature.}, imposing the boundary conditions of Table \ref{tab:bcsLg}. The same kind of method was used in \cite{Bolognesi,Gorsky2015} to compute baryon solutions in other holographic models. We focus here on the results and give more details about the numerical method in Appendix \ref{Num}. We start by presenting the leading order probe baryon solution and then discuss the back-reaction. We recall that the back-reacting solution is computed assuming no back-reaction on the color sector (metric and dilaton).

\subsubsection{Probe baryon solution}

\label{Sec:pb}

We start with the numerical results obtained for the probe baryon solution. In this case the modulus of the tachyon field $\tau$ is fixed to its background value, and the equations of motion take the form presented in Appendix F.1.3 of \cite{BaryonI}.

\begin{figure}[h]
\begin{center}
\begin{overpic}
[scale=0.49]{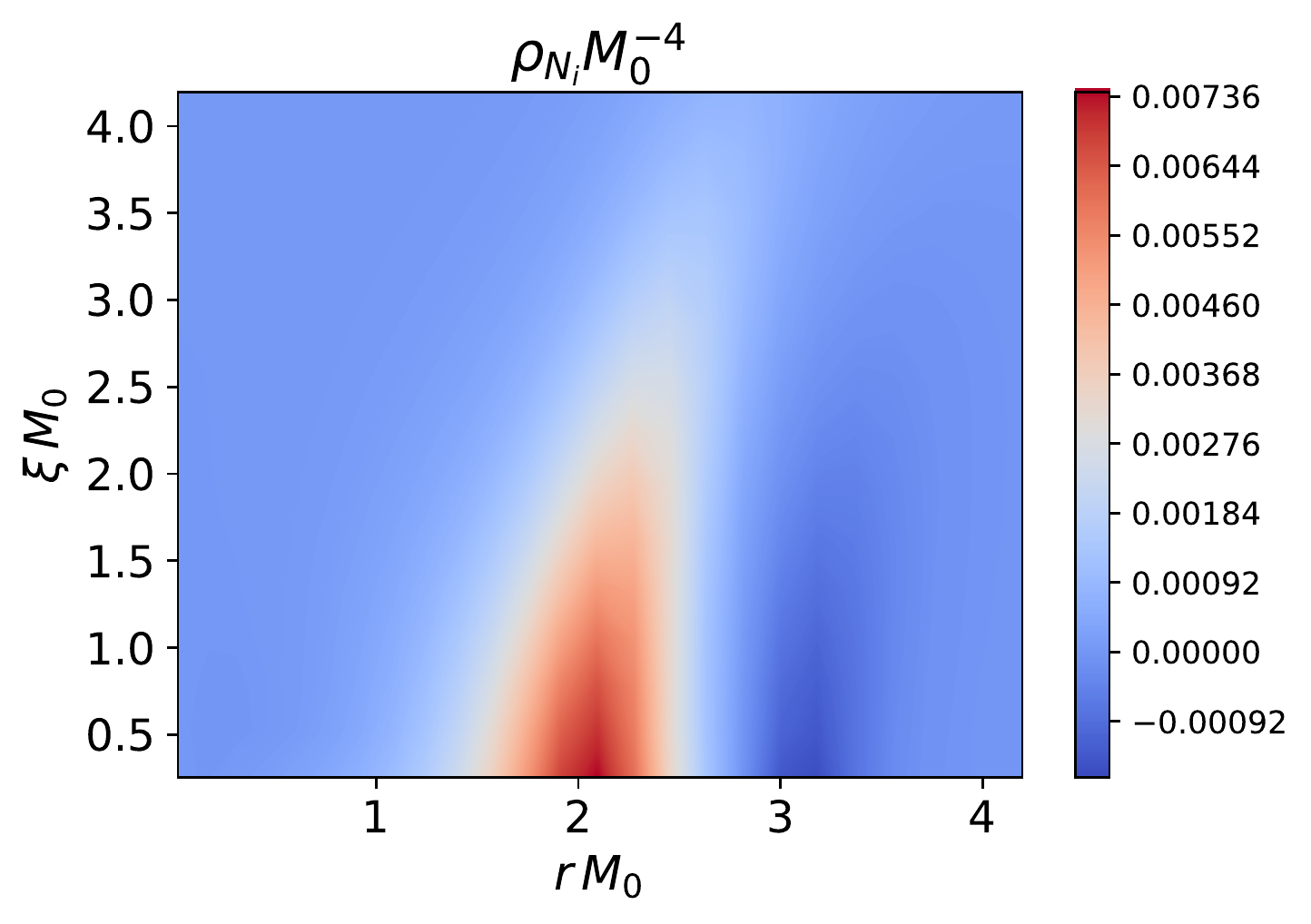}
\end{overpic}
\begin{overpic}
[scale=0.49]{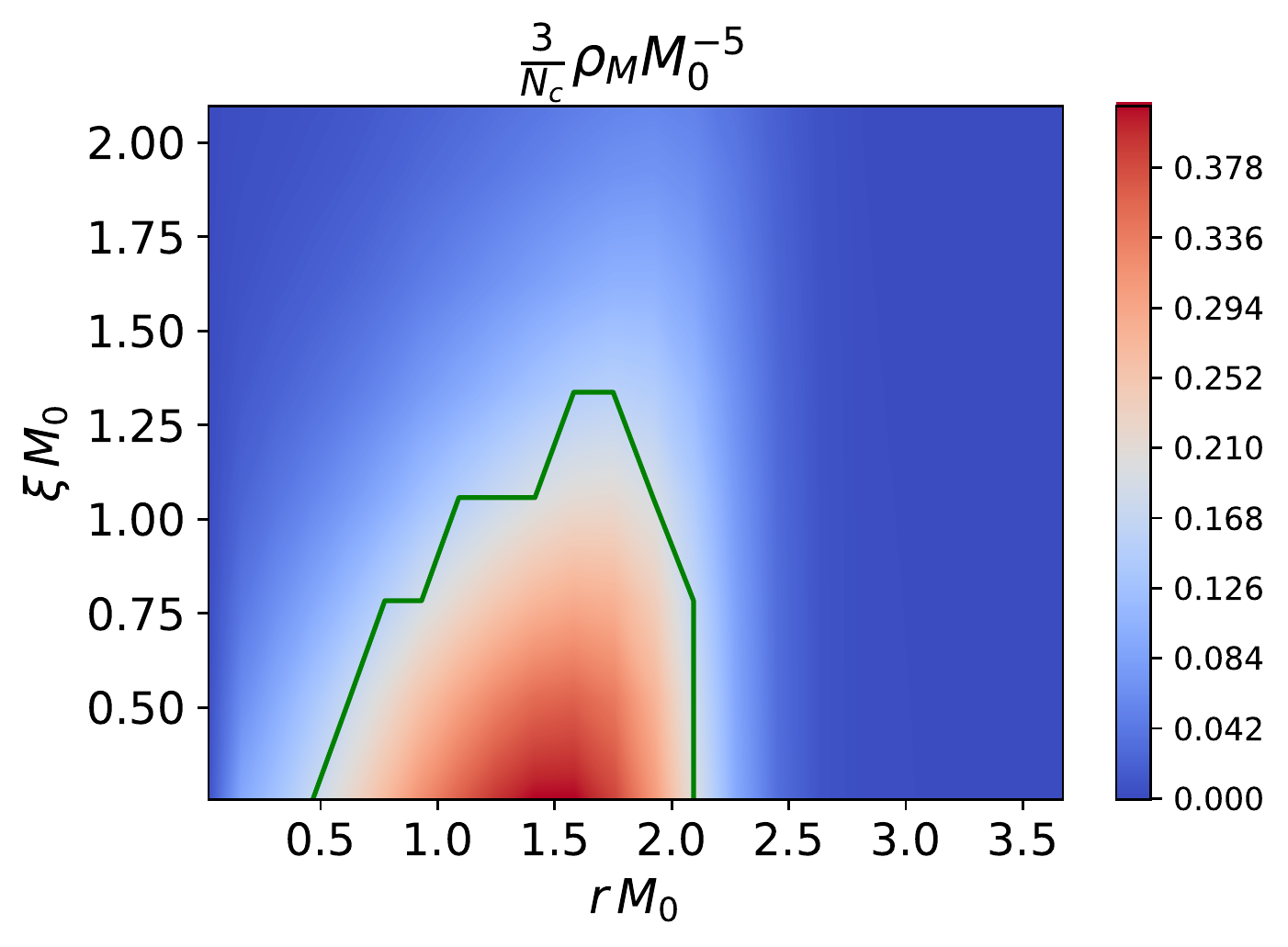}
\end{overpic}
\caption{Instanton number \eqref{rNi} (left) and bulk Lagrangian \eqref{defrM} (right) density for the static soliton solution in the probe baryon regime. All quantities are expressed in units of the classical mass of the soliton \eqref{M0}. The center of the soliton is located at $\xi=0$ where the density diverges as $\xi^{-1}$. The UV boundary is located at $r=0$. The green line in the right plot indicates the boundary of the region over which the mean value is computed to define the relative difference in Figure \ref{fig:rhoM_x}.}
\label{fig:rho_Ni_M_v8_probe}
\end{center}
\end{figure}

The instanton number and bulk Lagrangian density in the $(\xi,r)$-plane are presented in Figure \ref{fig:rho_Ni_M_v8_probe}, where all dimensionful quantities are expressed in units of the classical soliton mass \eqref{M0}. The bulk Lagrangian density is given by \eqref{defrM} , whereas the expression for the instanton number density can be obtained by dividing equation (6.7) of \cite{BaryonI}  by $4\pi\xi^2$
\be
\label{rNi} \rho_{N_i} \equiv \frac{1}{8\pi^2\xi^2} \e^{\bar{\m}\bar{\n}} \Big[ F_{\bar{\m}\bar{\n}} + \partial_{\bar{\m}}\big( -i\phi^*D_{\bar{\n}}\phi + h.c. \big) \Big] \, .
\ee 
Figure \ref{fig:rho_Ni_M_v8_probe} shows the expected behavior for a solitonic configuration, that is the densities are confined to a region of finite extent in the bulk. The size of this lump in the $\xi$ direction gives an estimate of the baryon size, which is of the order of $M_0^{-1}$. The numerical value for the classical soliton mass $M_0$ is obtained by integrating the Lagrangian density in Figure \ref{fig:rho_Ni_M_v8_probe}
\be
\label{M0num} M_0 \simeq \frac{N_c}{3}\times 1150\, \text{MeV} \, .
\ee
This number is expected to give the leading contribution to the nucleon mass in the V-QCD model with the parameters of Table \ref{table:parameters}. As discussed above, the full result for the nucleon mass also receives quantum corrections \eqref{MB} whose evaluation is an unsolved problem.

\begin{figure}[h]
\begin{center}
\begin{overpic}
[scale=0.75]{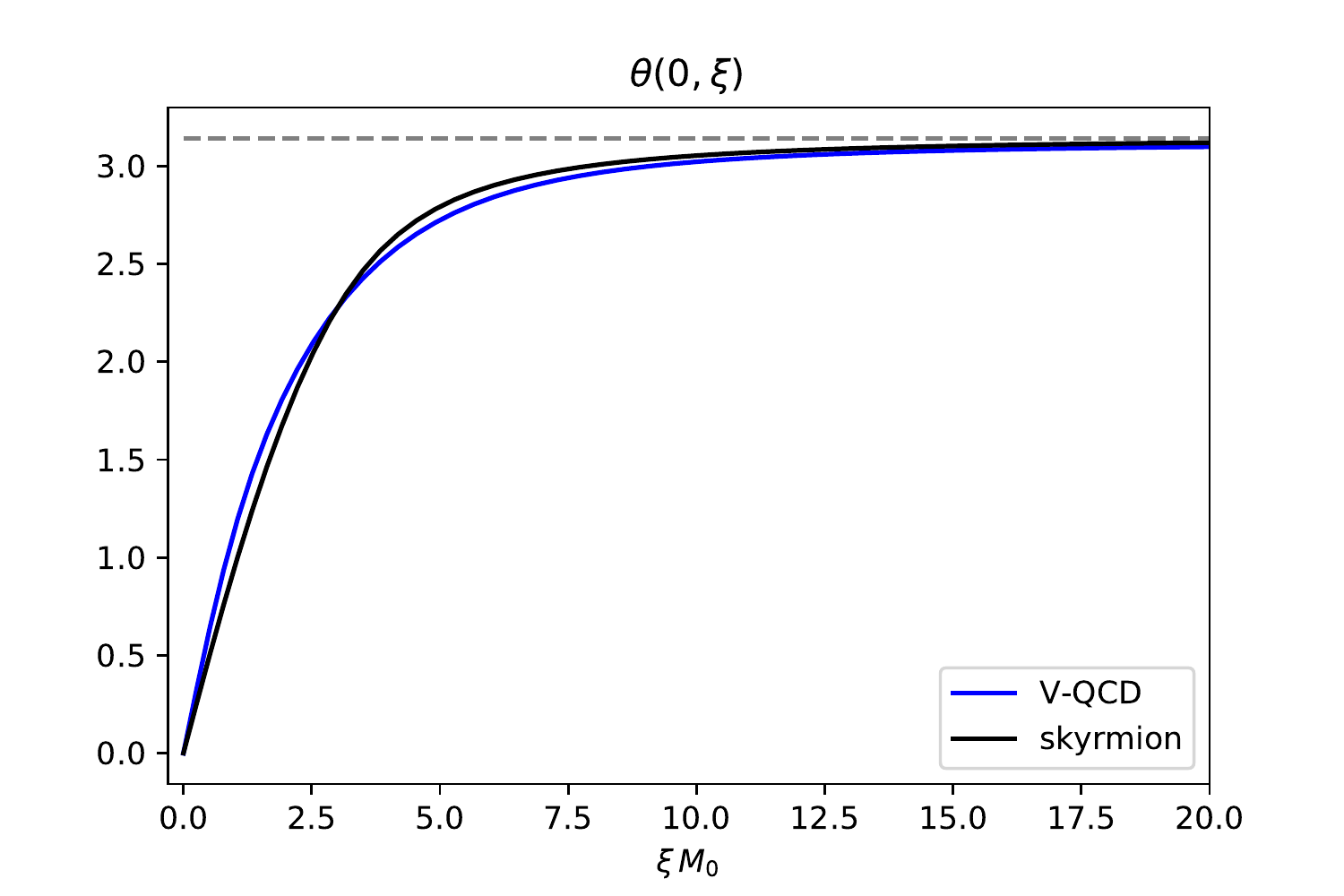}
\end{overpic}
\caption{Radial profile of the non-abelian phase of the tachyon field \eqref{TU2} at the UV boundary (blue line). The dashed gray line indicates the asymptotic value $\pi$. For comparison, we also plotted in black the profile for the pion fields in the skyrmion solution of the Skyrme model. The parameters of the Skyrme model were chosen such that $f_\pi$ and the soliton mass are equal to those of the V-QCD model, with the parameters of Table \ref{table:parameters}.}
\label{fig:thx_probe_v8}
\end{center}
\end{figure}

In Figure \ref{fig:thx_probe_v8}, we also plot the profile at the boundary ($r=0$) for the non-abelian phase $\theta$ of the tachyon field \eqref{TU2}. As stated above, the pion matrix in the boundary theory \eqref{Pmat} reproduces\footnote{This comparison is well defined even though the phase $\theta$ transforms under the residual $U(1)$ gauge freedom: the boundary gauge transformations of $U_P$ match exactly those of the pion matrix in the Skyrme model, and the gauge is fixed in both cases by requiring the absence of sources for the gauge fields.} the skyrmion hedgehog ansatz, and the associated skyrmion number is equal to the baryon number. Figure \ref{fig:thx_probe_v8} should therefore be compared with the corresponding plot of the pion field in the Skyrme model skyrmion solution in the chiral limit \cite{Adkins1983}. This plot is reproduced\footnote{Notice that $\theta$ goes from 0 to $\pi$, instead of $\pi$ to 0 for the skyrmion, because the boundary value of the tachyon field is the conjugate of the pion matrix \eqref{Pmat}. So it is actually $\pi-F_{\text{skyrmion}}$ that is plotted in Figure \ref{fig:thx_probe_v8}.} in Figure \ref{fig:thx_probe_v8}. It makes it clear that the shape of the boundary skyrmion is close to that of the Skyrme model. Note, in particular, that the asymptotic behavior is the same:
\be
\label{th0_asymp} \theta(0,\xi) \underset{\xi\to 0}{\sim} \xi \sp \theta(0,\xi)-\pi \underset{\xi\to\infty}{\sim} \frac{1}{\xi} \, .
\ee
This can be seen from the asymptotic analysis in Appendix G of \cite{BaryonI}.

\subsubsection{Back-reacted tachyon}

\label{Sec:br}

We now discuss the numerical results obtained when taking into account the back-reaction on the tachyon field. In this case, the gauge fields and tachyon phase $\theta$ are fixed to the probe baryon solution, and the equation of motion for the tachyon modulus $\tau$ is solved on this background. The corresponding equation of motion for $\tau$ is written in Appendix F.2 of \cite{BaryonI}.

\begin{figure}[h]
\begin{center}
\begin{overpic}
[scale=0.75]{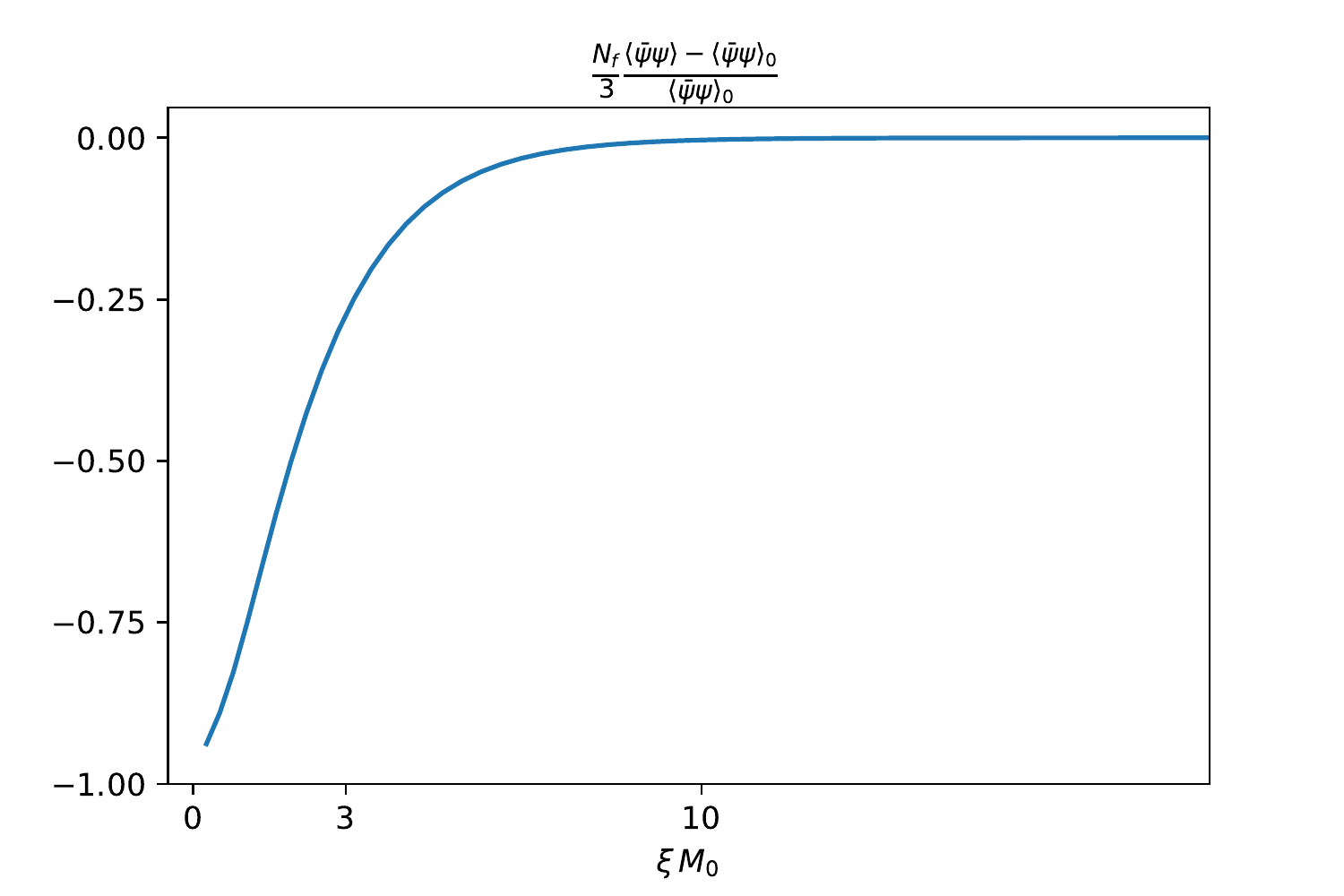}
\end{overpic}
\caption{Chiral condensate profile around the static soliton solution in the back-reacted tachyon regime, where the baryon center is at $\xi=0$. The plotted quantity is the relative difference between the modulus of the chiral condensate and the chiral condensate in vacuum. In the Veneziano limit $N_c,N_f \to \infty$, the difference is of order $\mathcal{O}(1/N_f)$, so that it has to be multiplied by $N_f$ in order to obtain a finite result. We also divide by 3, which means that what the figure shows is the large $N_f$ result, where the value $N_f = 3$ is substituted for the number of flavors. As discussed in the text, there is no reason for this result to be quantitatively accurate for $N_f=3$, but it gives an indication of the qualitative behavior.}
\label{fig:CC_New_v8}
\end{center}
\end{figure}

For a back-reacted tachyon, the chiral condensate profile around the baryon can be computed from the near-boundary behavior of the tachyon modulus
\be
\label{tauUV} \tau(r,\xi) \underset{r\to 0}{\sim} \Sigma(\xi)\, r^3 \, ,
\ee
where $\Sigma$ is proportional to the modulus of the chiral condensate $\left|\left<\bar{\psi}\psi\right>\right|$. The relative difference of $\Sigma(\xi)$ with the vacuum value $\Sigma(\infty)$ is plotted in Figure \ref{fig:CC_New_v8}.
This shows the expected behavior, where the chiral symmetry tends to be restored inside the baryon. {Note that the result that is shown is valid in the limit of large $N_c$ and large $N_f$. There is a priori no guarantee for it to be a quantitatively accurate approximation when a small number of flavors (for example $N_f = 2$ or 3) is substituted in the leading large $N_f$ result. So, at small $N_f$ and $N_c$, Figure} \ref{fig:CC_New_v8} {should not be considered as more than an indication of the qualitative behavior of the chiral condensate in presence of the baryon.}

Another interesting information that can be extracted from this back-reacted tachyon solution, is the effect of the back-reaction on the soliton mass \eqref{M0}. The leading order correction to the on-shell Lagrangian density due to the back-reaction on $\tau$ can be expressed as
\be
\label{delM0} N_c^{-1}\d \rho_M = \frac{1}{2}N_f\d\tau^2\frac{\d^2 S_\tau}{\d \tau^2}\bigg|_{\text{probe}} + \d\tau \frac{\d S_B}{\d\tau}\bigg|_{\text{probe}} + \mathcal{O}\big(N_f^{-2}\big) \, ,
\ee
where $\d\tau$ refers to the order $\mathcal{O}\big(N_f^{-1}\big)$ correction to $\tau$, and $S_\tau$ and $S_B$ are defined in \eqref{StB1}. Note that we dropped the terms that vanish on-shell
\be
\label{Sos} \frac{\d S_\tau}{\d\tau}\bigg|_{\text{probe}} = \frac{\d S_B}{\d\varphi_B}\bigg|_{\text{probe}} = 0 \sp \varphi_B \in \{ \Phi, \phi, A_{\bar{\mu}}, \theta \} \, .
\ee
Equation \eqref{delM0} can be simplified by using the back-reacted equations of motion for the tachyon modulus
\be
\label{brEOM} \frac{\d(N_f S_\tau + S_B)}{\d\tau}\bigg|_{\text{back-react}} = 0 \, ,
\ee
which, at leading order in $N_f$, implies that
\be
\label{brEOMO1} \frac{\d S_B}{\d\tau}\bigg|_{\text{probe}} + N_f \d\tau  \frac{\d^2 S_\tau}{\d\tau^2}\bigg|_{\text{probe}} = \mathcal{O}\big(N_f^{-1}\big) \, ,
\ee
and finally
\be
\label{delM02} N_c^{-1}\d \rho_M = -\frac{1}{2}N_f\d\tau^2\frac{\d^2 S_\tau}{\d \tau^2}\bigg|_{\text{probe}} + \mathcal{O}\big(N_f^{-2}\big) \, .
\ee

\begin{figure}[h]
\begin{center}
\begin{overpic}
[scale=0.75]{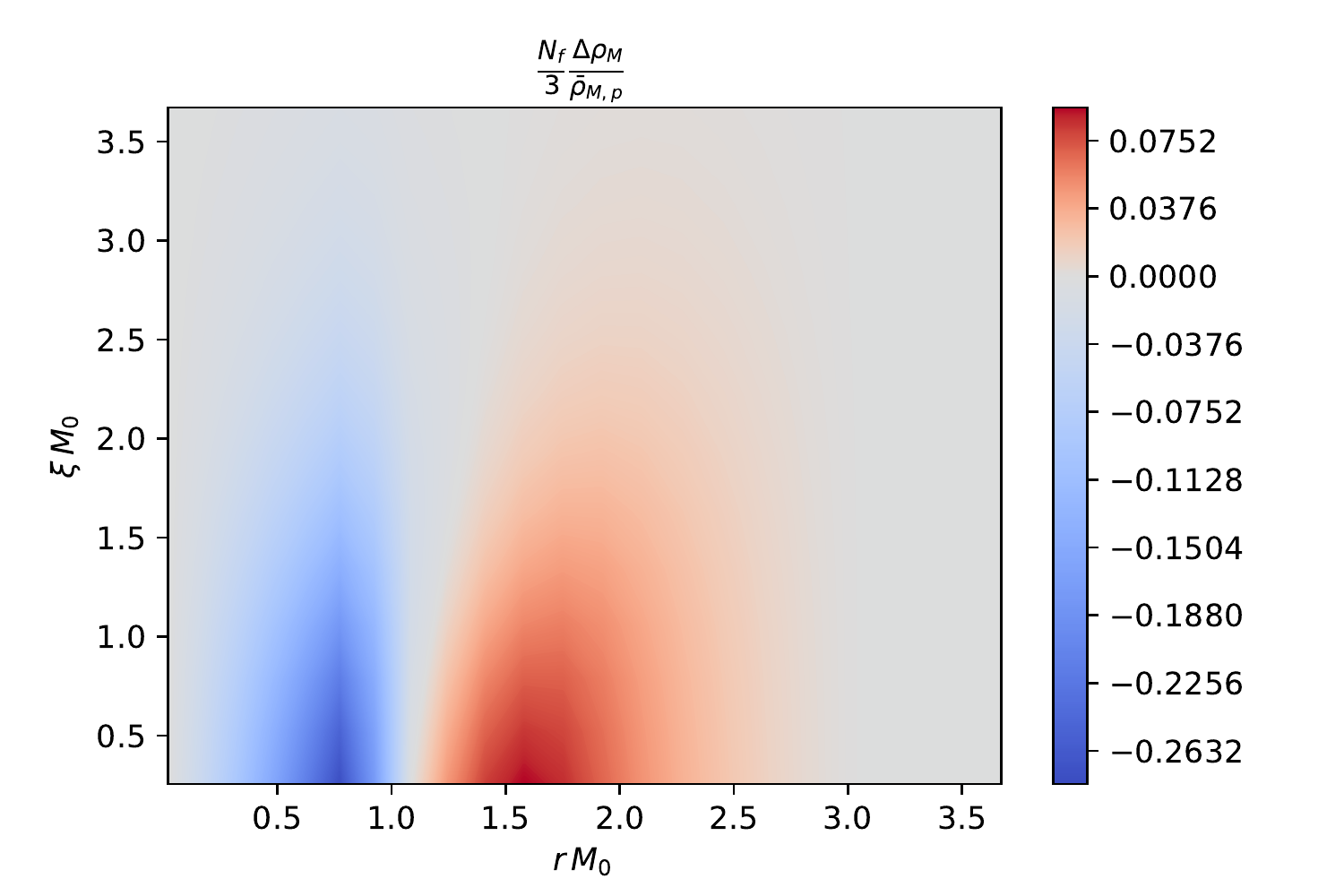}
\end{overpic}
\caption{Relative difference of bulk Lagrangian density for the static soliton solution in the
probe baryon and the back-reacted tachyon regime. The relative difference is defined as  the difference of the two densities divided by the mean value of the probe density \eqref{mean}.
The mean value is taken over the area delimited by the green line in the right of Figure \ref{fig:rho_Ni_M_v8_probe}, which is the region where the density is substantially different from zero. The ratio is multiplied by $N_f$ in order to obtain something finite in the Veneziano limit. The UV boundary is located at $r=0$ and the baryon center at $\xi = 0$. }
\label{fig:rhoM_x}
\end{center}
\end{figure}

Although the correction to the mass \eqref{delM02} is suppressed by a factor $\mathcal{O}(1/N_f)$ in the Veneziano limit, it can be sizeable when a realistic value is substituted for $N_f$. Figure \ref{fig:rhoM_x} shows the relative difference between the bulk Lagrangian density for the back-reacted solution and the probe baryon solution, when setting $N_f=3$ in the leading large N result. As for the chiral condensate, there is no reason for the result to be quantitatively accurate at $N_f=3$, but it gives an indication of the qualitative behavior.

We should also emphasize that the definition of the relative difference which is shown in Figure \ref{fig:rhoM_x} is not the standard one, where the difference of the two quantities that are compared is divided by the quantity of reference (as in \eqref{rdM1} for instance). The usual definition of the relative difference is not appropriate to compare the densities over the $(\xi,r)$ plane, since the place where the densities go to zero is not exactly the same for the probe and back-reacted solutions. Instead, we define the relative difference by dividing the difference of the two densities by a reference value $\bar{\rho}_M$
\be
\label{Drel} \D_{\text{rel}}\rho_M \equiv \frac{\rho_{M,\text{back-reacted}} - \rho_{M,\text{probe}}}{\bar{\rho}_M} \, .
\ee
$\bar{\rho}_M$ is defined as the mean value of the probe density over the region of the bulk where most of the density is contained. To be more precise, the criterion that we used to define the relevant region is given by
\be
\label{cA} \frac{3}{N_c} \rho_M \geq 0.7 M_0^4 \, ,
\ee
whose boundary is shown by the green line in the right of Figure \ref{fig:rho_Ni_M_v8_probe}. In practice, the mean value is then computed numerically by averaging over the cells contained in the given region, denoted $\mathcal{A}$ here
\begin{equation}
\label{mean} \bar{\rho}_M = \frac{1}{N_{\mathrm{cells}}} \sum_{i \in \mathcal{A}} \rho_M(i) \, ,
\end{equation}
where $N_{\text{cells}}$ is the number of grid cells contained in $\mathcal{A}$.

Figure \ref{fig:rhoM_x} indicates that there is a region near the UV boundary where the back-reacted Lagrangian density decreases with respect to the probe solution. This is understood easily as coming from the decrease of the tachyon modulus in presence of the baryon, which is dual to the decrease of the chiral condensate observed in Figure \ref{fig:CC_New_v8}. Another noticeable feature of Figure \ref{fig:rhoM_x} is the shift of the baryon Lagrangian density towards the IR. This can be understood as another consequence of the partial chiral restoration at the baryon center. Indeed, the interaction of the baryon with the tachyon modulus results in a repulsive force from the IR. So a decrease of the tachyon modulus weakens this force, and implies the observed shift towards the IR.

Even at small values of $N_f$, the relative difference between the probe and back-reacted solutions is observed to be relatively small numerically, of the order of a few $10\%$. This is also the case at the level of the soliton masses
\be
\label{rdM1} \frac{N_f}{3}\frac{M_{0,\text{back-reacted}}-M_{0,\text{probe}}}{M_{0,\text{probe}}} \simeq -10\% \, .
\ee
{Here again, the number in} \eqref{rdM1} {is the leading large $N_f$ result, that cannot precisely be trusted for small $N_f$ and should be considered as indicative. }

\section{Quantization of the isospin collective modes}

\label{Sec:QI}

The baryon states of equal half-integer spin and isospin are found by quantizing the soliton collective coordinates (or zero modes) around the static solution \cite{Adkins1983}. These modes are
\begin{itemize}
\item The spatial position of the soliton $\vec{X} = (X^1,X^2,X^3)$ .
\item The isospin orientation of the soliton, encoded in an $SU(N_f)_V$ matrix\footnote{The relevant collective coordinates are actually only a subgroup of the isospin group $SU(N_f)_V$. For instance, for $N_f = 2$ they are the elements of $SU(2)_V/\mathbb{Z}_2$ and for $N_f = 3$, the elements of $SU(3)_V/U(1)_Y$, where $U(1)_Y$ refers to the strong hypercharge subgroup. } $V$.
\end{itemize}
The baryon solution can be deformed  in many ways in addition to these modes, such as changing the position of the soliton in the holographic direction, or the size of the soliton \cite{Sugimoto}. Quantizing such modes leads to a tower of excited baryon states in each spin sector. In the following, we focus on the lowest states of these towers and consider only the quantization of the zero modes.

There is no guarantee in principle that the rotation modes can be studied as those of a rigid rotor, independently from the dilation mode of the soliton. In \cite{Sugimoto}, it was actually found that the geometry of the collective modes manifold for dilation and isospin rotation had to be such that the modes are rather quantized as a 4D harmonic oscillator, with energy levels given by equation (5.24) in this reference. The rigid rotor is a good approximation to the harmonic oscillator only when its fundamental frequency is very large compared with its inverse moment of inertia. In QCD, this condition is fulfilled in the large $N_c$ limit. This property is reproduced in the holographic QCD model of \cite{Sugimoto}, as is manifest from the large $N_c$ limit of the energy levels, equation (5.31) in~\cite{Sugimoto}. Their equation (5.24) also indicates that the rigid rotor (large $N_c$) approximation is better for lower spins. In the following, we assume that the rotation modes can be treated as those of a rigid rotor. According to the previous discussion, once we set the number of colors to its physical value $N_c=3$, this should not induce too large errors for the low spin modes that exist in real QCD ($s=1/2$ and $3/2$). Note that the same rigid rotor approximation was considered in the context of the hard wall model \cite{Panico08} .

The spatial position of the baryon is actually irrelevant to the study of baryon states, so it will be kept fixed at $\vec{X} = 0$. We are therefore left with the problem of quantizing the isospin rotation mode of the soliton. To do so, we consider a configuration where the soliton isospin orientation $V(t)$ evolves with time, but sufficiently slowly to be approximated by the ansatz
\be
\label{ansrotL} \mathbf{L}(t) = V(t) \mathbf{L}^{\text{(sol)}}\left(r, \xi \right) V(t)^{\dagger} - i\mathrm{d}V(t)V(t)^{\dagger} \, ,
\ee
\be
\label{ansrotR} \mathbf{R}(t) = V(t) \mathbf{R}^{\text{(sol)}}\left(r, \xi \right) V(t)^{\dagger} - i\mathrm{d}V(t)V(t)^{\dagger} \, ,
\ee
\be
\label{ansrotT} T(t) = V(t) T^{\text{(sol)}}\left(r, \xi \right) V(t)^{\dagger} \, ,
\ee
where the superscript (sol) refers to the field evaluated in the static soliton solution. $V(t)$ is parametrized as
\be
\label{defVo} V(t) \equiv \exp{\left( it\, \omega^a \l^a \right)} \, ,
\ee
the $\l^a$'s being the generators of $SU(N_f)$, and the angular velocity is identified to be
\be
\label{defo} \omega^a \l^a = -iV(t)^\dagger \frac{\mathrm{d}V(t)}{\mathrm{d}t} \, .
\ee
We assume that the rotation is stationary
\be
\label{dto0} \frac{\mathrm{d}\omega^a}{\mathrm{d}t} = 0 \, ,
\ee
and slow, so that $\omega$ can be treated as a perturbation on top of the static soliton background. Also, \emph{from now on we restrict to the case of $N_f=2$}.
This means that we shall quantize only a subset of the full isospin rotations. Specifically, we restrict to $V(t)\in SU(2)_V$, in the same subgroup as the baryon solution.

The starting point for the quantization of the isospin rotation modes is the classical Lagrangian that controls their dynamics. The latter is obtained in the next section by substituting the slowly rotating ansatz \eqref{ansrotL}-\eqref{ansrotT} into the bulk action \eqref{v3} and evaluating it on-shell for the slowly rotating soliton solution
\be
\label{Lrot1} L_{\text{rot}} = -M_0 + \frac{1}{2}\l\vec{\omega}^2 \, ,
\ee
where $M_0$ is the mass of the static soliton and $\l$ its moment of inertia. The classical Hamiltonian is then computed, and quantized in the canonical way described in Appendix \ref{App:QI}. As a result, the eigenstates of the Hamiltonian are shown to have same spin and isospin and its eigenvalues are given by
\be
\label{Es1} E_s = M_0 + \frac{1}{2\l}s(s+1) \, ,
\ee
where $s$ refers to the spin. In particular, the nucleon states correspond to $s=1/2$ and the isobar $\D$ to $s=3/2$.

\section{The rotating soliton}

\label{Sec:rot}

This section is dedicated to the calculation of the rotating soliton solution, from which can be computed the moment of inertia $\l$ that controls the splitting of the baryon energy levels \eqref{Es1}. We start by determining the ansatz relevant to the solution, before deriving the equations of motion for the fields of the ansatz as well as the boundary conditions they should obey. We finally describe the numerical solution for the rotating solution. We work with a slowly rotating soliton, at first order in the rotation velocity $\omega$. Also, we recall that we only consider the quantization of the subsector of the chiral group that contains the 2 flavors of the soliton solution.

\subsection{Ansatz for the rotating instanton}

\label{Sec:rot1}

Substituting the naive ansatz \eqref{ansrotL}-\eqref{ansrotT} with $V(t)\in SU(2)$ into the equations of motion, reveals that this ansatz in itself cannot solve the time-dependent equations of motion. The reason is that the components $(L/R)^a_0$ and $(L^\text{T}/R^\text{T})_{r,i}$ in \eqref{U2gf} are turned on at linear order in $\omega$ \cite{Panico08}, as is the abelian phase of the tachyon $(\det T-1)$. Accordingly, the ansatz \eqref{ansrotL} for the gauge fields should be supplemented by
\be
\label{ansrotL0} L_0(t) = V(t) L_0^{\text{(rot)}}\left(r, \xi ; \vec{\omega} \right) V(t)^{\dagger} - i\partial_0V(t)V(t)^{\dagger} \, ,
\ee
\be
\label{ansrotLhri} L^\text{T}_{r,i}(t) = L^{\text{T,(rot)}}_{r,i}(r, \xi ; \vec{\omega}) \, ,
\ee
and likewise for the right-handed fields. Also, the ansatz for the tachyon field \eqref{ansrotT} should be modified to
\be
\label{ansrotdetT} T(t) = V(t)T^{\text{(rot)}}(r,\xi ; \vec{\omega})V(t)^\dagger \, ,
\ee
where $\mathbf{L}^{\text{(rot)}}$ and $(\det T^{\text{(rot)}} -1)$ start at linear order in $\omega$.

To determine relevant ans\"atze for $\mathbf{L}^{\text{(rot)}}$ and $T^{\text{(rot)}}$, we proceed as in the case of the static soliton \cite{BaryonI} and impose the maximal number of symmetries of the bulk action. In the rotating case, this includes the cylindrical symmetry and parity. For $N_f=2$ flavors, the cylindrical symmetry of the static soliton solution \eqref{ansatzSU2iL}-\eqref{ansatzU1} implies that a constant isospin rotation of the soliton is equivalent to a constant spatial rotation. So the soliton rotating in isospin space can be seen as rotating instead in physical space, with angular momentum $\vec{\omega}$. In particular, $\vec{\omega}$ transforms as a pseudo-vector in 3-dimensional space\footnote{Strictly speaking, $\vec{\omega}$ is a definite 3-dimensional vector, and the rotation breaks the cylindrical symmetry. However, as is standard for broken symmetries, the appropriate ansatz can be derived by assuming that $\vec{\omega}$ transforms as a pseudo-vector (in that case, $\vec{\omega}$ is regarded as a field, called a spurion).}. At linear order in $\vec{\omega}$, the cylindrically symmetric ansatz for the gauge fields of the rotating soliton is then \cite{Cherman,Panico08}
\be
\label{ansLs} (L/R)_i = V(t)(L/R)_i^{\text{(sol)}}V(t)^\dagger \sp (L/R)_r = V(t)(L/R)_r^{\text{(sol)}}V(t)^\dagger \, ,
\ee
\be
\nn (L/R)^\text{T}_0 = (L/R)_0^{\text{T,(sol)}} \, ,
\ee
\begin{align}
\label{ansL0s} (L/R)_0 &= V(t)\bigg(\omega_b \left[\k_1^{(L/R)}(r,\xi)\e^{abc}\frac{x^c}{\xi} + \k_2^{(L/R)}(r,\xi)\left(\frac{x^a x^b}{\xi^2} -\d^{ab} \right) \right]+\\
\nn &\hphantom{= V(t)\bigg(} + \frac{v^{(L/R)}(r,\xi)}{\xi^2} (\vec{\omega} . \vec{x}) x^a \,+ \omega^a\bigg)\frac{\s^a}{2}V(t)^\dagger \, ,
\end{align}
\be
\label{ansLhis} (L/R)^\text{T}_i = \frac{\rho^{(L/R)}(r,\xi)}{\xi}\left( \omega_i - (\vec{\omega} \cdot \vec{x}) \frac{x_i}{\xi^2} \right) \!+\! B_{\xi}^{(L/R)}(r,\xi) (\vec{\omega} \cdot \vec{x}) \frac{x_i}{\xi^2} + Q^{(L/R)}(r,\xi) \e_{ibc} \omega^b \frac{x^c}{\xi}  ,
\ee
\be
\label{ansLhrs} (L/R)^\text{T}_r = B_r^{(L/R)}(r,\xi) \frac{\vec{\omega} \cdot \vec{x}}{\xi} \, ,
\ee
where the superscript $\text{(sol)}$ refers to the field in the static soliton configuration\footnote{At order $\mathcal{O}(\omega^2)$, the static fields will receive corrections from the rotation. These could in principle contribute to the moment of inertia in \eqref{Lrot}. It is not the case because the static fields sit at a saddle point of the static action. The leading contribution to the Lagrangian \eqref{Lrot} from the $\mathcal{O}(\omega^2)$ correction to the static fields therefore starts at order $\mathcal{O}(\omega^4)$, corresponding to an $\mathcal{O}(\omega^2)$ correction to the moment of inertia.}, and we introduced the new 2-dimensional fields
\be
\label{rotf1} \k_{1,2}^{(L/R)} \sp v^{(L/R)} \sp \rho^{(L/R)} \sp B_{\bar{\mu}}^{(L/R)} \sp Q^{(L/R)} \, .
\ee
Imposing symmetry under the parity transformation (the full transformation $P =P_1 \cdot P_2$ in terms of the definitions in Appendix~\ref{app: conventions})
\be
\label{Prot} P
\, :\quad \vec{x} \to -\vec{x} \sp L \leftrightarrow R \sp \vec{\omega} \to \vec{\omega} \, ,
\ee
relates the right-handed and left-handed fields as
\be
\nn \k_1 \equiv \k_1^L = -\k_1^R \sp \k_2 \equiv \k_2^L = \k_2^R \, ,
\ee
\be
\nn v \equiv v^L = v^R \, ,
\ee
\be
\label{parrot} \rho \equiv \rho^L = -\rho^R \, ,
\ee
\be
\nn B_{\bar{\mu}} \equiv B_{\bar{\mu}}^L = - B_{\bar{\mu}}^R \, ,
\ee
\be
\nn Q \equiv Q^L = Q^R \, .
\ee
For the tachyon field, the ansatz that has the right transformation properties under 3-dimensional rotations and parity takes the form
\be
\label{ansTs} T = V(t)\exp{\left( i\zeta(r,\xi) \frac{\vec{\omega}\cdot\vec{x}}{\xi} \right)} T^{(sol)} V(t)^\dagger \sp \zeta(r,\xi)\in\mathbb{R} \, .
\ee

The ansatz thus defined is invariant under a $U(1)_s\times U(1)_r$ residual gauge freedom, where the first factor was already present in the static case \eqref{resg1}, as denoted by the subscript ``$s$'', and the second factor appears in the rotating solution, as denoted by the subscript ``$r$''. This new factor is an axial $U(1)$ gauge freedom that is a subgroup of the chiral $U(1)_A$
\be
\label{resg2} \hat{g}_L = \hat{g}_R^{\dagger} = \exp{\left(i\b(r,\xi)\frac{\vec{\omega}\cdot\vec{x}}{\xi}\right)} \, ,
\ee
under which only $B_{\bar{\mu}}$, $\rho$ and $\zeta$ transform, with transformation rules
\be
\label{trB} B_{\bar{\mu}} \to B_{\bar{\mu}} + \partial_{\bar{\mu}}\b \sp \rho \to \rho + \b \sp \zeta \to \zeta -2 \b\, .
\ee
Also, the complex scalar field
\be
\label{defchi} \k \equiv \k_1 + i \k_2 \, ,
\ee
transforms as a charge 1 complex scalar field under $U(1)_s$ in  \eqref{resg1}.

In the case of a rotating soliton, one should in principle consider the coupling to the holographic axion $\mathfrak{A}$, dual to the boundary Yang-Mills instanton density operator $\text{Tr}\left( G\wedge G \right)$~\cite{ihqcd,Arean:2013tja,theta}. This coupling appears because of the additional residual $U(1)_A$ gauge freedom \eqref{resg2}, which turns on the abelian phase of the tachyon $\zeta$ and the axial part of the abelian gauge field $B_{\bar{\mu}}$ and $\rho$. However, as for the other color fields, the action of the baryon on the axion is suppressed by a factor $\mathcal{O}\big(N_c^{-1}\big)$ in the large $N_c$ limit. The rotating baryon solution will therefore decouple from the axion field at leading order in $N_c$. At next-to-leading order, the axion will contribute to the order $\mathcal{O}\big(N_c^{-1}\big)$ correction to the moment of inertia\footnote{The $g_{0i}$ component of the metric will also be turned on by rotation, and contribute at order $\mathcal{O}\big(N_c^{-1}\big)$}. In the following, we consider the same approximation as in the static case and ignore the action on the color sector. This implies in particular that we set the axion to 0.

As in the case of the static soliton, there exists a redefinition of the ansatz fields such that the tachyon phases $\theta$ and $\zeta$ are absorbed into the gauge fields
\be
\label{defkBrt} \tilde{\k} \equiv \ex^{i\theta} \k \sp \tilde{B}_{\bar{\mu}} \equiv B_{\bar{\mu}} + \frac{1}{2}\partial_{\bar{\mu}}\zeta \sp \tilde{\rho} \equiv \rho + \frac{1}{2} \zeta \, .
\ee
The resulting fields are invariant under the residual gauge freedom \eqref{resg1} and \eqref{resg2}. In addition to this field redefinition, for later use it is also convenient to define the field strength for the $B_{\bar{\mu}}$ gauge field
\be
\label{defBmn} B_{\bar{\mu}\bar{\nu}} \equiv \partial_{\bar{\mu}} B_{\bar{\nu}} - \partial_{\bar{\nu}}B_{\bar{\mu}} \, ,
\ee
and the covariant derivative for the $\rho$ and $\k$ field
\be
\label{defDrk} D_{\bar{\mu}} \rho = \partial_{\bar{\mu}}\rho - B_{\bar{\mu}}  \sp D_{\bar{\mu}} \k = (\partial_{\bar{\mu}} - i A_{\bar{\mu}})\k \, .
\ee

\subsubsection*{Lorenz gauge}

Because we obtain the static soliton solution by fixing the residual gauge freedom \eqref{resg1} to the Lorenz gauge
\be
\label{LgcAbis} \partial_rA_r + \partial_{\xi}A_{\xi} = 0 \, ,
\ee
we work in the same gauge for the rotating soliton. Also, we fix the additional gauge freedom  \eqref{resg2} by imposing a similar Lorenz condition for $B_{\bar{\mu}}$
\be
\label{LgcB} \partial_r B_r + \partial_{\xi}B_{\xi} = 0 \, .
\ee
For this choice, the equations of motion for $B_{\bar{\mu}}$  \eqref{Eabr} and \eqref{Eabi1} are elliptic and can be solved numerically by a heat diffusion method upon setting the right boundary conditions.

Note that the Lorenz condition \eqref{LgcB} leaves a residual gauge freedom of the form
\be
\label{resg_Lg} B_{\bar{\mu}} \to B_{\bar{\mu}} + \partial_{\bar{\mu}}\mathfrak{f} \sp \partial_r^2 \mathfrak{f} + \partial_\xi^2 \mathfrak{f} = 0 \sp \mathfrak{f}(0,\xi)=0 \, .
\ee
Part of the choice of boundary conditions will correspond to the choice of residual gauge freedom \eqref{resg_Lg}. The explicit choice that we make is discussed below, in Section \ref{Sec:bcrot}.

\subsection{Moment of inertia and equations of motion}

The Lagrangian for the rotational collective modes of the soliton is  obtained by substituting the ansatz \eqref{ansLs}-\eqref{ansLhrs} and \eqref{ansTs} into the bulk action \eqref{v3}. This yields the Lagrangian of a rigid rotor
\be
\label{Lrot} L_{rot} = -M_0 + \frac{1}{2} \l \vec{\omega}^2 \, ,
\ee
where $M_0$ is the mass of the static soliton \eqref{M0}, and we defined the moment of inertia
\be
\label{deflamb} \l \equiv \l_{DBI} + \l_{CS} \, ,
\ee
\be
\label{lDC} \l_{DBI} = \int \intd r \intd \xi\, 4\pi\xi^2 \rho_{\l,DBI} \sp \l_{CS} = \int \intd r \intd \xi\, 4\pi\xi^2 \rho_{\l,CS} \, ,
\ee
\begin{align}
\nonumber \rho_{\l,DBI} &\equiv - \frac{2}{3} M^3N_c V_f(\l,\tau)\ex^A\sqrt{1+\ex^{-2A}\ka \left((\partial_r\tau)^2 + (\partial_\xi\tau)^2\right)}\times  \\
\nonumber &\hphantom{=} \times \bigg(\ex^{2A}\ka(\l)\tau^2\bigg(4 \ex^{2A}\D_{rr}\td{B}_r^2 + 4(1-\ex^{2A}\D_{\xi\xi})\left[ \td{B}_{\xi}^2  + \frac{2}{\xi^2}\td{\rho}^2\right]- \\
\nn &\hphantom{=\times \bigg(\ex^{2A}\xi^2\ka(\l)\tau^2\bigg(}
- 8\ex^{2A}\D_{\xi r}\td{B}_\xi \td{B}_r - 2\td{\k}_1^2 \bigg) - \\
\nonumber &\hphantom{=\times\bigg(}
- w(\l)^2\bigg(\frac{1}{2}(1-\ex^{2A}\D_{\xi\xi})|D_{\xi}\k|^2 + \frac{1}{2} \ex^{2A}\D_{rr}|D_{r}\k|^2 -\\
\nn&\hphantom{=\times\bigg( -w(\l)^2}
- \frac{1}{2}\ex^{2A}\D_{\xi r}\left( D_\xi \k^* D_r \k + h.c. \right)+  \frac{1}{4}(1-\ex^{2A}\D_{\xi\xi}) (\partial_{\xi}v)^2 + \\
\nn &\hphantom{=\times\bigg( -w(\l)^2}
  + \frac{1}{4} \ex^{2A}\D_{rr} (\partial_r v)^2 - \frac{1}{2} \ex^{2A}\D_{\xi r} \partial_r v \partial_\xi v + \\
 \nn&\hphantom{=\times\bigg( -w(\l)^2}
 + \frac{1}{2\xi^2}(v^2+|\k|^2)(1+|\phi|^2) - v (\k\phi^* + h.c.) - \\
\nn &\hphantom{=\times\bigg( -w(\l)^2}
 - 2 \ex^{2A}\D_{rr} \xi^{-2}(D_r\rho)^2 - 2(1-\ex^{2A}\D_{\xi\xi})\xi^{-2}(D_{\xi}\rho)^2+\\
\nn&\hphantom{=\times\bigg( -w(\l)^2}
+ 4 \ex^{2A}\D_{\xi r} \xi^{-2}D_r\rho D_\xi\rho - 2 \ex^{2A}\D_{rr} (\partial_r Q)^2 -  \\
\nn &\hphantom{=\times\bigg( -w(\l)^2}
 - 2(1-\ex^{2A}\D_{\xi\xi})(\partial_{\xi}Q)^2 + 4 \ex^{2A}\D_{\xi r} \partial_r Q \partial_\xi Q- \\
\nn &\hphantom{=\times\bigg( -w(\l)^2}
- 2 \xi^{-2}Q^2\Big[2 - \ex^{2A}\xi \left( \partial_\xi\D_{\xi\xi} + \ex^{2A}\partial_r(\ex^{-2A}\D_{\xi r}) \right)  \Big]-  \\
\label{deflambDBI} &\hphantom{=\times\bigg( -w(\l)^2}
- \frac{1}{2} \Big[\ex^{2A}\D_{rr}(1-\ex^{2A}\D_{\xi\xi})-\ex^{4A}\D_{\xi r}^2\Big] (B_{\bar{\mu}\bar{\nu}})^2  \bigg)\bigg) \, ,
\end{align}
where the $\D$ symbol is defined in Appendix \ref{App:EOMrot}, 
\begin{align}
\nonumber \rho_{\l,CS} &\equiv \frac{2 N_c}{3\pi^2\xi^2} \e^{\bar{\mu}\bar{\nu}} \bigg(D_{\bar{\mu}}\rho\Big( (f_1(\tau)+f_3(\tau))(D_{\bar{\nu}}\phi\, \k^* + h.c.)-\\
\nn&\hphantom{=\times \bigg(D_{\bar{\mu}}\rho\Big(} - 2(2f_3(\tau) - f_1(\tau))\left( D_{\bar{\nu}}\td{\phi}+h.c.\right) \td{\k}_1 -\\
\nn &\hphantom{=\times \bigg( D_{\bar{\mu}}\rho\Big(}  -4 (f_1(\tau)-f_3(\tau)-if_2(\tau))\td{A}_{\bar{\nu}}\td{\phi}_1\td{\k}_2 \Big) + \\
\nn &\hphantom{=\times\bigg(} + \td{\rho} \Big( 2(f_3(\tau) - if_2(\tau))\big(-F_{\bar{\mu}\bar{\nu}}\,\td{\phi}_1\td{\k}_2 + \td{A}_{\bar{\mu}}(i\partial_{\bar{\nu}}\phi\k^* + h.c.)\,\big)- \\
\nn &\hphantom{=\times\bigg(+\rho} -\partial_{\bar{\nu}}\tau f_1'(\tau)(D_{\bar{\mu}}\phi\k^*+h.c.) +\partial_{\bar{\nu}}\tau f_3'(\tau)(D_{\bar{\mu}}\td{\phi}\td{\k} + h.c.)- \\
\nn&\hphantom{=\times\bigg(+\rho} - 2i\partial_{\bar{\nu}}\tau f_2'(\tau) \td{A}_{\bar{\mu}}\td{\phi}_1 \td{\k}_2\, \Big)+    \\
\nn &\hphantom{=\times\bigg(} + 2\partial_{\bar{\nu}}\tau (f_1'(\tau) - f_3'(\tau))D_{\bar{\mu}}\rho\, \td{\k}_1 \td{\phi}_1 + \\
\nn &\hphantom{=\times\bigg(} + \partial_{\bar{\mu}}(\xi Q)\Big( (f_1(\tau)+f_3(\tau))\left( i\k^*D_{\bar{\nu}}\phi + h.c. \right)  +\\
\nn&\hphantom{=\times\bigg(+ \partial_{\bar{\mu}}(\xi Q)\Big(}+ 4(f_1(\tau) + f_3(\tau) -3if_2(\tau))\td{A}_{\bar{\nu}}\, \td{\phi}_1\td{\chi}_1 \Big)+  \\
\nn &\hphantom{=\times\bigg(} +2\partial_{\bar{\nu}}\tau (f_1'(\tau) + f_3'(\tau))\partial_{\bar{\mu}}(\xi Q) \td{\phi}_1 \td{\k}_2- \\
\nn &\hphantom{=\times\bigg(} - 8(f_1(\tau)+f_3(\tau))\xi Q\,  D_{\bar{\mu}}\rho\partial_{\bar{\nu}}\Phi  + 8\partial_{\bar{\nu}}\tau (f_1'(\tau)+f_3'(\tau))\xi Q \td{\rho} \partial_{\bar{\mu}}\Phi +\\
\nn &\hphantom{=\times\bigg(} + v  \Big( B_{\bar{\mu}\bar{\nu}}\Big[\frac{1}{2}(f_1(\tau) + f_3(\tau))\, (|\phi|^2 - 1) + (f_1(\tau)-f_3(\tau)-if_2(\tau))\td{\phi}_1^2  \Big]+ \\
\nn &\hphantom{=\times\bigg(+ v} + (f_1(\tau)+f_3(\tau))\xi QF_{\bar{\mu}\bar{\nu}} - 2(f_3(\tau)-if_2(\tau))\td{B}_{\bar{\mu}} (D_{\bar{\nu}}\td{\phi} + h.c.)\td{\phi}_1 \Big)+ \\
\label{deflambCS} &\hphantom{=\times\bigg(}  +v \partial_{\bar{\nu}}\tau \Big( (f_1'(\tau) + f_3'(\tau)) \big( \td{B}_{\bar{\mu}}(1-|\phi|^2) - 2\xi Q \td{A}_{\bar{\mu}} \big) + 2i f_2'(\tau)\td{B}_{\bar{\mu}} \td{\phi}_1^2 \Big) \, \bigg) .
\end{align} 
We recall that the tildes on the static fields refer to the redefined fields that contain the non-abelian tachyon phase $\theta$ \eqref{LtoLtcyl}. The total bulk Lagrangian density for the rotating fields is denoted by $\rho_\l$
\be
\label{rl} \rho_\l \equiv \rho_{\l,DBI} + \rho_{\l,CS} \, .
\ee 
The equations of motion for the fields of the rotating soliton ansatz are obtained by extremizing the moment of inertia \eqref{deflamb} with respect to small deformations of the fields. They are presented in Appendix \ref{App:EOMrot}.

\subsection{Boundary conditions}

\label{Sec:bcrot}

\begin{table}[h]
\centering
\begin{tabular}{|c|c|c|c|}
\hline
$\xi \to 0$ & $\xi \to L \to \infty$ & $r \to 0$ & $r \to r_{\text{IR}} \to \infty $  \\ \hline
$ v + \td{\k}_2  \to 0$ & $v \to -1$ & $v\to -1$  &  $\partial_r v \to 0$     \\
$\td{\k}_1 \to 0$ & $ \td{\k}_1 \to 0$ & $ \td{\k}_1 \to -\sin{\theta} $ & $\td{\k}_1 \to 0 $     \\
$\partial_{\xi}\td{\k}_2 \to 0$ & $\td{\k}_2 \to -1$ & $\td{\k}_2 \to \cos{\theta} $ & $\partial_r\td{\k}_2 \to 0 $      \\
 $\partial_{\xi}B_{\xi} \to 0$  & $\partial_{\xi}B_{\xi}\to 0$ & $ B_{\xi}\to 0$ & $ B_{\xi} \to 0$       \\
$B_r \to 0$  & $B_r \to 0$ & $\partial_r B_r \to 0$ &  $\partial_r B_r \to 0$      \\
$\partial_{\xi}\rho - B_{\xi} \to 0$ & $ \rho \to 0 $ & $\rho  \to 0$ & $ \rho \to 0$ \\
$Q \to 0$ & $Q\to 0$ & $Q\to 0$ & $\partial_r Q\to 0$ \\
$\zeta \to 0$ & $\zeta \to 0$ & $r\zeta\to 0$ & $\zeta \to 0$ \\ \hline
\end{tabular}
\caption{Boundary conditions for the rotating soliton solution in Lorenz gauge.}
\label{tab:bcsLgs}
\end{table}

We present in Table \ref{tab:bcsLgs} the boundary conditions that are imposed on the fields of the rotating soliton ansatz \eqref{ansLs}-\eqref{ansLhrs} and \eqref{ansTs}, which obey the equations of motion \eqref{Enab01b}-\eqref{Ezans}. We discuss separately the 4 boundaries of the $(\xi,r)$ space
\begin{itemize}
\item \textbf{\underline{UV :}} In the UV limit $r \to 0$, the condition that $v, \td{\k}, B_{\xi}, \rho$ and $Q$ should vanish comes from requiring that there is no source for the gauge fields at the boundary. Moreover, the condition for $B_r$ comes from imposing the Lorenz gauge.

The condition for $\zeta$ is somewhat more subtle than the other fields. The reason is that, because the quark mass is set to 0, there is no source term for the tachyon field. The abelian phase $\zeta$ is therefore not associated with any source. At the level of the near-boundary behavior, it translates into the fact that the boundary value of $\zeta$ appears in front of the vev term for the tachyon field
\be
\label{TUVz} T(r,\xi) \underset{r\to 0}{=} \ell\Sigma(\xi)\exp{\left(i\zeta(0,\xi)\frac{\vec{\omega}\cdot \vec{x}}{\xi} + i\theta(0,\xi) \frac{x\cdot\s}{\xi}\right)} r^3(-\log{(r\L)})^{-c}\left(1 + \cdots\right)  ,
\ee
where the dots refer to terms that go to 0 near the boundary; $\Sigma(\xi)$ is proportional to the modulus of the chiral condensate in the boundary theory $\left|\left<\bar{\psi}\psi\right>\right|$. The UV condition for $\zeta$ will therefore not come from a choice of source at the boundary, but rather from the requirement that the solution be regular. The equation of motion for $\zeta$ \eqref{Ezans} has two linearly independent solutions, one of which behaves as $r^{-2}$ near the boundary and the other as $r^0$. Requiring that
\be
\label{zuv} r\zeta \underset{r\to 0}{\to} 0 \, ,
\ee
will therefore select the regular behavior.

\item \textbf{\underline{$\xi \to L\, (L \to \infty)\,:$}}  Requiring that the moment of inertia \eqref{deflamb} should be finite imposes that, as $\xi \to\infty$
\be
\label{cEDBIfnab} \xi^{3/2} \tilde{\k}_1 \to 0 \sp \xi^{3/2} \partial_{\bar{\mu}}\tilde{\k}_2 \to 0 \sp  \xi^{3/2} \partial_{\bar{\mu}}v \to 0  \, ,
\ee
\be
\label{cEDBIfab}\xi^{3/2} \left(B_{\bar{\mu}} + \frac{1}{2}\partial_{\bar{\mu}}\zeta \right) \to 0 \sp\xi^{1/2}\left(\rho+\frac{1}{2}\zeta \right) \to 0 \sp \xi^{3/2}\partial_{\bar{\mu}}Q \to 0 \, .
\ee
For regularity $\partial_{\xi}\zeta$ should go to 0 as $\xi \to \infty$ so $B_{\xi}$ should also tend to 0 from \eqref{cEDBIfab}. Then the Lorenz gauge condition in the limit where $\xi \to \infty$ reads
\be
\label{Lgcxinf} \partial_r B_r = 0 = -\frac{1}{2}\partial_r^2 \zeta \, ,
\ee
so at $\xi\to\infty $, $\zeta$ should take the form
\be
\label{hzxinf} \zeta \underset{\xi\to\infty}{\to} \zeta^{(0)} + r\zeta^{(1)} \, .
\ee
In the UV, from \eqref{cEDBIfab}
\be
\label{Uvzr} \underset{\xi\to\infty}{\text{lim}}\zeta(\xi,0)  = -2\underset{\xi\to\infty}{\text{lim}}\rho(\xi,0) = 0\, ,
\ee
because the source for $\rho$ is set to 0. We shall  further fix the residual gauge freedom \eqref{resg_Lg} such that $\zeta = 0$ in the IR. With this choice, the linear part also vanishes in  \eqref{hzxinf} and
\be
\label{zxinf} \zeta \underset{\xi\to\infty}{\to} 0 \, .
\ee
The final condition for $B_{\xi}$, $\partial_{\xi}B_{\xi} \to 0$ rather than $B_{\xi} \to 0$, is chosen to impose Lorenz gauge even at finite $L$ in the numerical solution.

\item \textbf{\underline{IR :}} As for the static soliton \cite{BaryonI}, the conditions in the IR limit $r \to r_{\text{IR}}$ are regularity conditions. In \cite{BaryonI} it was found that the resulting conditions for the gauge fields were equivalent to the conditions imposed in the hard-wall model for chiral symmetry to be broken on the IR wall
\be
\label{IRcbis} \left.\left(\mathbf{L}-\mathbf{R}\right)\right|_{r_{\text{IR}}} = 0 \sp \left.\left(\mathbf{F}^{(L)}_{\mu r}+\mathbf{F}^{(R)}_{\mu r}\right)\right|_{r_{\text{IR}}} = 0 \, .
\ee
In the following, our strategy is to assume that the IR regularity conditions for the gauge field are still equivalent to \eqref{IRcbis} in the case of the rotating soliton. This assumption will be confirmed numerically if a solution can be found with this behavior in the IR. It can also be checked analytically by studying the IR asymptotics of the equations of motion. With the ansatz \eqref{ansLs}-\eqref{ansLhrs}, \eqref{IRcbis} translates to
\be
\label{IRcanss1} \left.\k_1\right|_{\text{IR}} = 0 \sp \left.B_{\xi}\right|_{\text{IR}} = 0 \sp \left.\rho\right|_{\text{IR}} = 0 \, ,
\ee
\be
\label{IRcanss2} \left.\partial_r\k_2\right|_{\text{IR}} = 0 \sp \left.\partial_r Q\right|_{\text{IR}} = 0 \, .
\ee
Finally, the condition for $B_r$ comes from the Lorenz gauge \eqref{LgcB} and, as stated in the previous point, the residual gauge freedom \eqref{resg_Lg} is chosen such that $\zeta = 0$ in the IR.

\item \textbf{\underline{$\xi = 0\,:$}}
The boundary conditions in the limit where $\xi$ goes to 0 come from requiring that $\tilde{\mathbf{L}}$ and $\tilde{\mathbf{R}}$ are well defined vectors at $\xi = 0$ and that the field strength \eqref{F0rrs}-\eqref{Fhijrs} is a well defined 2-tensor. The additional constraint on $B_{\xi}$ is imposed by the choice of Lorenz gauge. The final condition $\zeta\underset{\xi\to 0}{\to} 0$ fixes what remained of the residual gauge freedom \eqref{resg_Lg} after setting $\zeta$ to 0 in the IR. Note that for this choice the tachyon matrix \eqref{ansTs} is well defined at $\xi = 0$.
\end{itemize}

\subsection{Numerical results : the spin-isospin spectrum}

We present in this subsection the numerical solution for the slowly rotating soliton configuration. The equations of motion written in Appendix \ref{App:EOMrot} are solved with the gradient descent method, imposing the boundary conditions of Table \ref{tab:bcsLgs}.

At linear order in $\omega$, the rotating solution is a probe on the static background. At leading order in $N_f$, the static background will correspond to the probe baryon solution presented in Section \ref{Sec:pb}. Order $\mathcal{O}\big(N_f^{-1}\big)$ corrections to the rotating solution will come from including the back-reaction on the tachyon in the static background, as discussed in Section \ref{Sec:br}. We start by presenting the leading order probe baryon solution and then discuss the back-reaction. We recall that the back-reacting solution is computed assuming no back-reaction on the color sector (metric and dilaton).

\subsubsection{Probe baryon background}

We start with the numerical results obtained for the probe baryon background. In this case the modulus of the tachyon field $\tau$ is fixed to its vacuum value, and the equations of motion  take the form presented in Appendix \ref{App:EOMrotpb}.

\begin{figure}[h]
\begin{center}
\begin{overpic}
[scale=0.5]{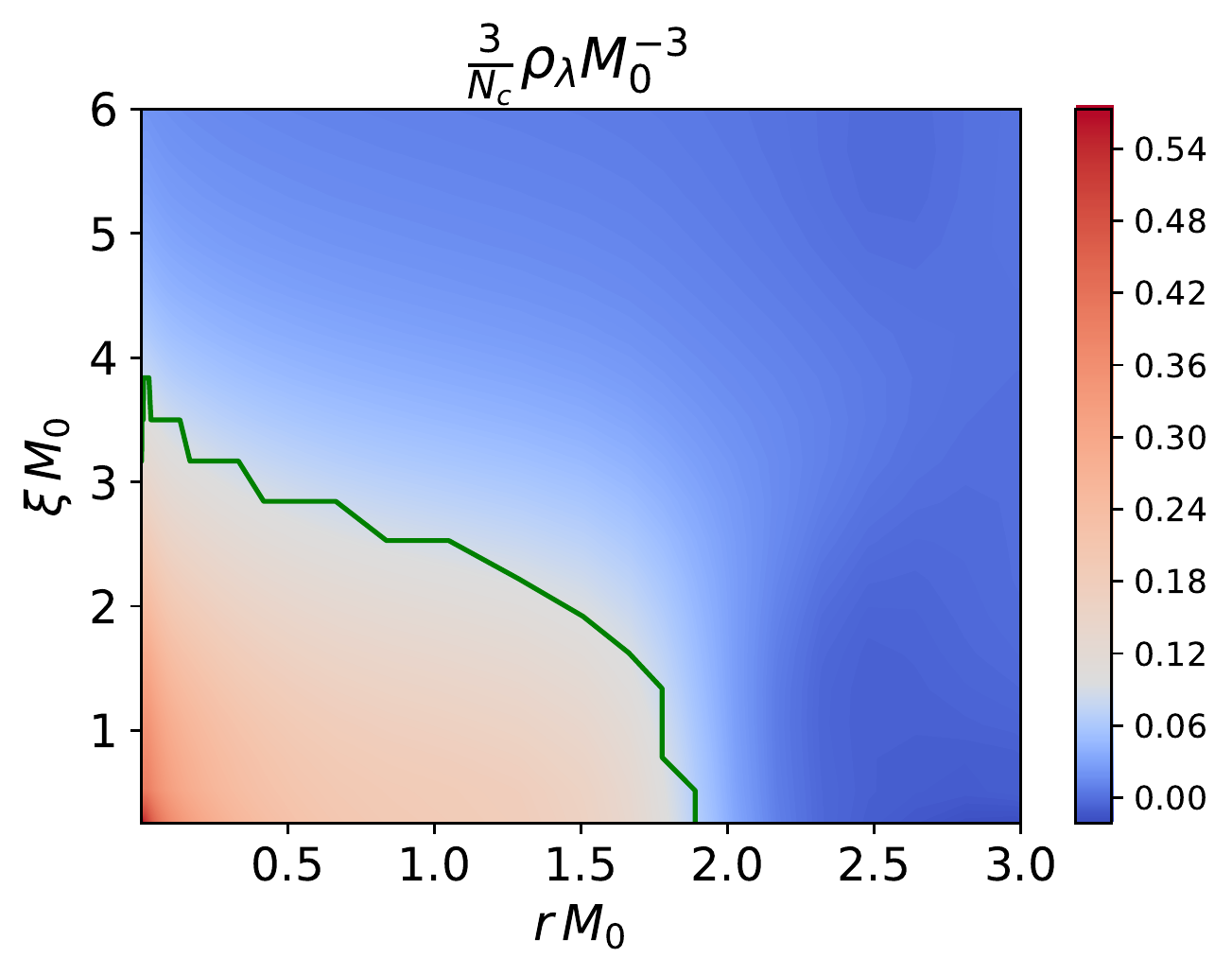}
\end{overpic}
\hspace{5mm}
\begin{overpic}
[scale=0.5]{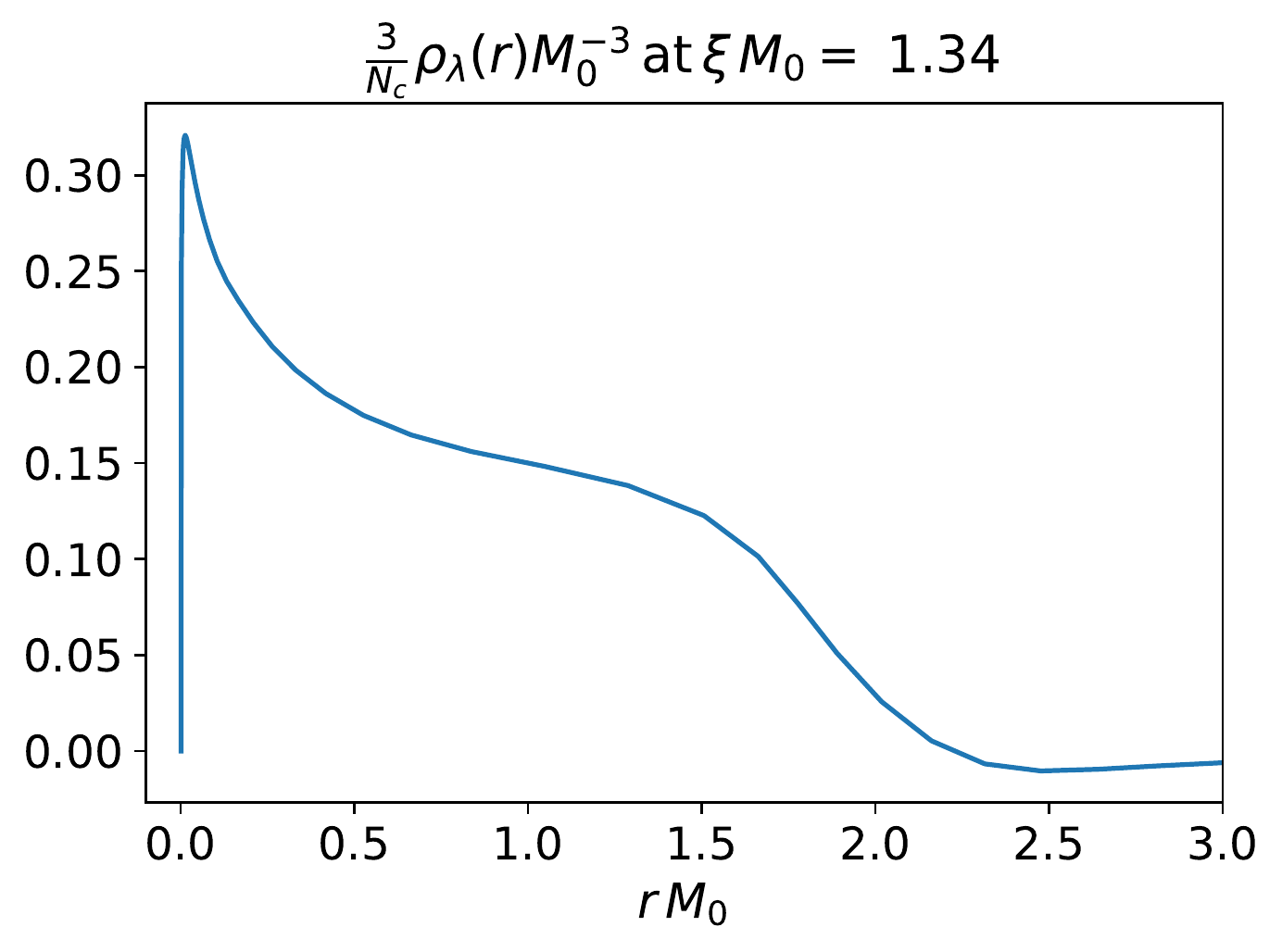}
\end{overpic}
\caption{\textbf{Left:} bulk Lagrangian density for the rotating fields \eqref{rl} in the probe baryon approximation. All quantities are expressed in units of the classical mass of the static soliton $M_0$ \eqref{M0}. The center of the soliton is located at $\xi=0$ where the density diverges as $\xi^{-1}$. The UV boundary is at $r=0$, and the green line indicates the boundary of the region over which the mean value is computed to define the relative difference in Figure \ref{fig:lrot_x}. \textbf{Right:} Same as the left figure, but at some given value of the 3-dimensional radius $\xi$. This figure makes it clear that the Lagrangian density eventually reaches 0 as $r\to 0$, as it should in absence of sources.}
\label{fig:lrot}
\end{center}
\end{figure}

The bulk Lagrangian density for the rotating fields \eqref{rl} in the $(\xi,r)$-plane is presented in Figure \ref{fig:lrot}, where all dimensionful quantities are expressed in units of the classical soliton mass $M_0$ \eqref{M0}. As for the static fields, Figure \ref{fig:lrot} shows the expected behavior for a solitonic configuration, that is the densities are confined to a region of finite extent in the bulk. The size of the lump is again of the order of $M_0^{-1}$.

Note that the density is observed to have a maximum very close to the UV boundary.
 This is associated with the flavor gauge fields of the rotating solution having the same kind of sharp behavior near the boundary. However, it is possible to check that the smooth vev-like behavior is recovered asymptotically as one goes closer to $r=0$
\be
\nonumber \mathbf{A}_\m \sim r^2 \times \text{vev} \, . 
\ee
To check this, one needs to go very close to the boundary $r=0$, so log coordinates $u = \log(r)$ are more appropriate. We have explicitly checked that the vev behavior is recovered very near the boundary.
We believe that this feature is a peculiarity of the UV behavior of our choice of V-QCD potentials, and that it can be avoided with a better choice. In particular, this behavior is not observed for the potentials of~\cite{Remes}, for which the solution is analyzed in Appendix \ref{App:OldPot}.

The numerical value for the classical moment of inertia density $\l$ in \eqref{deflamb} is obtained by integrating the Lagrangian density in Figure \ref{fig:lrot}
\be
\label{lnum} \frac{1}{\l} \simeq \frac{3}{N_c}\times60\,\text{MeV} \, .
\ee
From this result, the spin-isospin spectrum of the baryons \eqref{Es1} can be computed and compared with experimental QCD data, as shown in Table \ref{tab:bspec}. We set $N_c=3$ in the large N result to make this comparison. We recall that the estimation \eqref{M0num} for the soliton mass gives only the classical contribution, which can receive sizeable $\mathcal{O}\big(N_c^0\big)$ quantum corrections at finite $N_c$. Likewise, \eqref{lnum} is the leading order contribution to the moment of inertia in the Veneziano limit. Also, the V-QCD potentials presented in Section \ref{Sec:VQCD_intro} were not fitted to baryonic properties, but rather to QCD thermodynamics and mesonic properties. In light of these remarks, the precise numerical value for the baryon spectrum presented in Table \ref{tab:bspec} should not be taken too seriously, but rather as an indicative result. In particular, we shall not mention the numerical accuracy of the result as it is much better than the theoretical uncertainty.
\begin{table}
\centering
\begin{tabular}{|c|c|c|}
\hline
Spin & V-QCD mass & Experimental mass \\
\hline
$s = \frac{1}{2}$ & $M_N \simeq 1170 \,\text{MeV} $ & $M_N = 940 \,\text{MeV} $\\
\hline
$s = \frac{3}{2}$ & $M_\D \simeq 1260 \,\text{MeV}$ & $M_\D = 1234 \,\text{MeV}$
 \\
\hline
\end{tabular}
\caption{Baryon spin-isospin spectrum in the V-QCD model with the potentials of Section \ref{Sec:VQCD_intro}, compared with experimental data.}
\label{tab:bspec}
\end{table}

\subsubsection{Back-reacted tachyon background}

We now discuss the slowly rotating soliton solution computed on the static background that takes into account the back-reaction on the tachyon field.  The corresponding equations of motion for the rotating fields are written in Appendix \ref{App:EOMrotbr}.

\begin{figure}[h]
\begin{center}
\begin{overpic}
[scale=0.75]{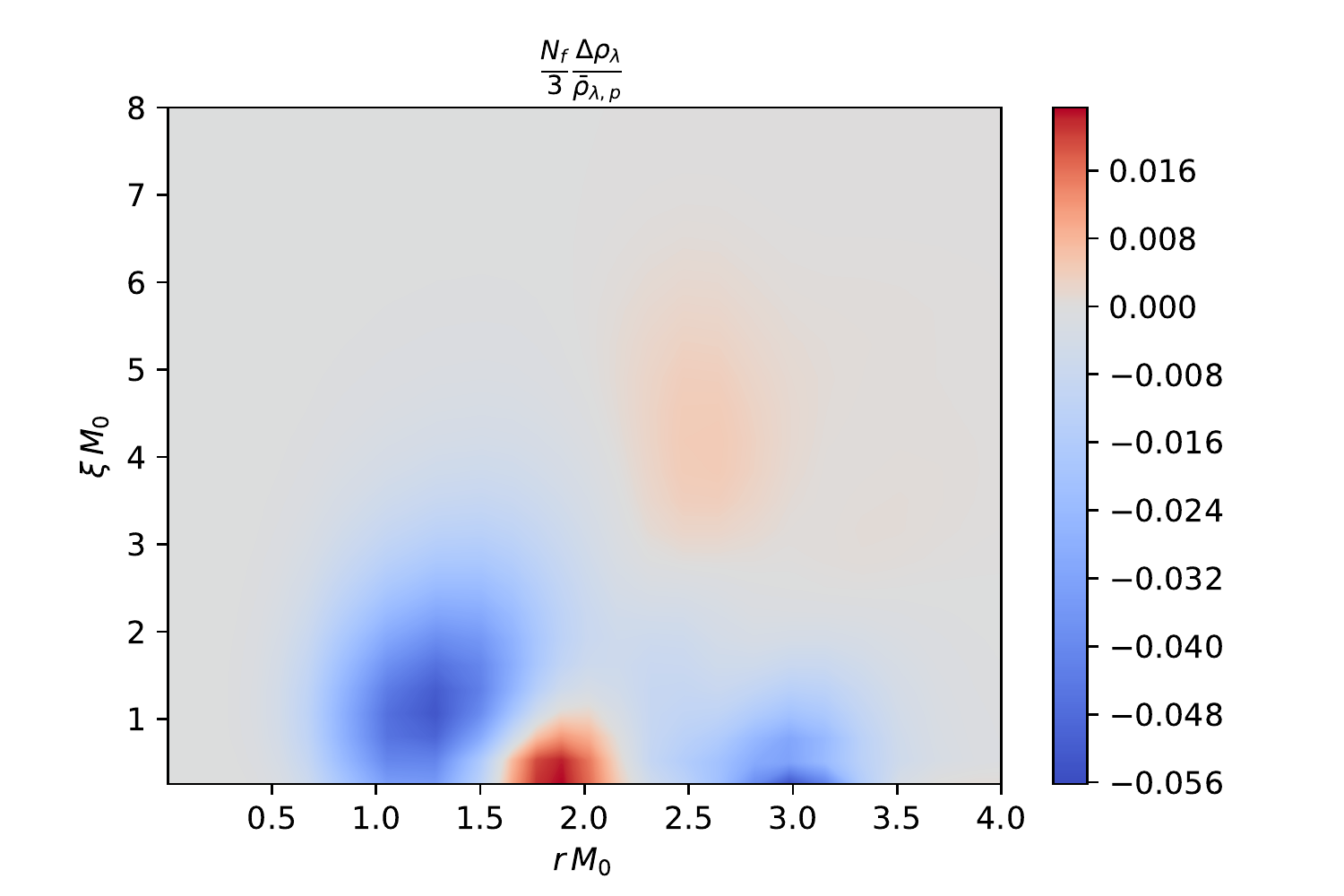}
\end{overpic}
\caption{Relative difference of bulk Lagrangian density for the slowly rotating soliton computed
on the probe baryon and the back-reacted tachyon backgrounds. The relative difference is defined as  the difference of the two densities divided by the mean value of the probe density. The mean value is taken over the area delimited by the green line in the left of Figure \ref{fig:lrot}, which is the region where the density is substantially different from zero. The ratio is multiplied by $N_f$ in order to obtain something finite in the Veneziano limit. The UV boundary is located at $r=0$ and the baryon center at $\xi = 0$. }
\label{fig:lrot_x}
\end{center}
\end{figure}

We are interested in the effect of the back-reaction on the soliton moment of inertia \eqref{deflamb}. In the large $N_f$ limit, the correction is of order $\mathcal{O}\big(N_f^{-1}\big)$
\be
\label{delM0l} \d \l = \d\tau \frac{\d \l}{\d\tau}\bigg|_{\text{probe}} + \mathcal{O}\big(N_f^{-2}\big) \, ,
\ee
where $\d\tau$ refers to the order $\mathcal{O}\big(N_f^{-1}\big)$ correction to $\tau$.
Figure \ref{fig:lrot_x} shows the relative difference between the bulk Lagrangian density for the back-reacted and probe backgrounds, when setting $N_f=3$ in the large N result. The definition of the relative difference is analogous to the static case \eqref{Drel} 
\be
\label{Drel} \D_{\text{rel}}\rho_\l \equiv \frac{\rho_{\l,\text{back-reacted}} - \rho_{\l,\text{probe}}}{\bar{\rho}_\l} \, ,
\ee
where the criterion defining the region over which the mean value $\bar{\rho}_\l$ is computed is now given by
\be
\label{cAl} \frac{3}{N_c} \rho_\l \geq M_0^2 \, .
\ee
This region is delimited by the green line in Figure \ref{fig:lrot}.

The dominant effect of the tachyon back-reaction observed in Figure \ref{fig:lrot_x} is the same as in the static case shown in Figure \ref{fig:rhoM_x}: the density decreases in the UV and is shifted towards the IR.
Even more strikingly than for the soliton mass in the static case, even at small values of $N_f$, the relative difference between the probe and back-reacted solutions is observed to be small numerically, of the order of a few percent. At the level of the soliton moment of inertia we obtain
\be
\label{rdla1} \frac{N_f}{3}\frac{\frac{1}{\l_{\text{back-reacted}}}-\frac{1}{\l_{\text{probe}}}}{\frac{1}{\l_{\text{probe}}}}\simeq 0.37\% \, .
\ee
{As for the static mass, we would like to emphasize here again that precise quantitative results such as} \eqref{rdla1} {cannot be trusted when substituting a small number of flavors in the large N result. The result in} \eqref{rdla1} {should be considered as an indication, that the moment of inertia does not seem to be affected much by the back-reaction of the baryon on the background.}


\section*{Acknowledgements}\label{ACKNOWL}
\addcontentsline{toc}{section}{Acknowledgements}

\noindent We thank D.~Arean, P. Figueras,  A.~Krikun, J.~L.~Ma\~nes,
A.~Pich,  C. Rosen, P. Sutcliffe and  P.  Yi for discussions and/or
correspondence.
This work was supported in part by CNRS contract IEA 199430. The work of MJ was supported
by an appointment to the JRG Program at the APCTP through the Science
and Technology Promotion Fund and Lottery Fund of the Korean
Government. MJ was also  supported by the Korean Local Governments --
Gyeong\-sang\-buk-do Province and Pohang City -- and by the National
Research Foundation of Korea (NRF) funded by the Korean government
(MSIT) (grant number 2021R1A2C1010834). FN would like to thank the
Asian Pacific Center for Theoretical Physics in Pohang, Korea, where
he was a visitor during a time when part of this work was being carried out.

\newpage
\appendix
\renewcommand{\theequation}{\thesection.\arabic{equation}}
\addcontentsline{toc}{section}{Appendix\label{app}}
\section*{Appendix}

\section{Conventions and Symmetry Transformations}
\label{app: conventions}

\subsection{Gauge fields}

We choose the following conventions for the $SU(N_f)$ generators $\lambda^a$
\be\label{norm gen}
(\lambda^a)^\dagger = \lambda^a \sp \tr ( \lambda^a \lambda^b ) =\frac{1}{2}\delta^{ab} \, ,
\ee
where $a,b=1,\ldots N_f^2-1$. The normalization of the $U(1)$ generator is chosen to be
\be
\label{norml0} \lambda^0 = \mathbb{I} \, .
\ee
Here $\mathbb{I}$ denotes the $N_f \times N_f$ unit matrix.
The $SU(N_f)$ generators satisfy
\be
[\lambda^a,\lambda^b]=if^{ab}_{\phantom{ab}c}\, \lambda^c \sp \tr(\lambda^a\{\lambda^b,\lambda^c\})= d^{abc} \, ,
\ee
where $f^{ab}_{\phantom{ab}c}$ are the structure constants of $SU(N_f)$ and $d^{abc}$ are the normalized anomaly Casimirs.

The gauge fields are written in terms of the generators as
\be\label{A decomp}
A_\mu= \hat A_\mu\mathbb{I} + A_\mu^a \lambda^a \, ,
\ee
where the components $\hat A_\mu$ and $A_\mu^a$ are real. In terms of differential forms, we can write, following the conventions of Appendix B in~\cite{book},
\be
F=dA -i A\wedge A \sp  D\equiv d - i A \cdot
\ee
Here $A\cdot$ stands for the action on the fields which depends on the representation of the gauge group under which they transform. The Bianchi identity can be written as
\be
DF=dF+iF\wedge A-iA\wedge F=0\ .
\ee
Moreover the derivatives of the tachyon fields are
\be
DT=dT+i \, T A_L -i A_R T \qquad \qquad DT^\dagger =dT^\dagger
-iA_L T^\dagger + i \, T^\dagger A_R \, .
\ee

\subsection{Chiral currents}

We choose conventions where the chiral currents are Hermitian. In the case of $SU(N_f)$, we write
\be
J_\mu= \frac{1}{2N_f}\hat J_\mu \mathbb{I} +J_\mu^a\lambda^a \ .
\ee
For this choice the currents should be normalized as
\be
J_{L,R}^{a\;\mu}=\tr_{\text{flavor}} \left(i\bar{q} \gamma^\mu
\frac{1\pm\gamma_5}{2}\lambda^a q \right) \, ,
\ee
\be
J^{U(1)\; \mu}_{L,R}=\tr_{\text{flavor}} \left(i\bar{q} \gamma^\mu\frac{1\pm\gamma_5}{2} q \right) \ .
\ee
For our normalization conventions then
\be
2\int d^4x\, \tr(J^\mu A_\mu) =\int d^4x \left( \hat J^{\mu} \hat A_\mu  +J^{a\,\mu} A_\mu^a\right) \ .
\ee

\subsection{Gauge Transformations}

The gauge transformations of the gauge fields and the tachyon are given as
\be
\nn A_L \to V_LA_LV^{\dagger}_L -idV_LV_L^{\dagger} \sp A_R\to V_RA_RV^{\dagger}_R -idV_RV_R^{\dagger} \, ,
\ee
\be
\label{gaugevar} F_{L}\to V_L~F_L~V_L^{\dagger}\sp F_{R}\to V_R~F_R~V_R^{\dagger} \ ,
\ee
\be
\nn T\to V_R TV_L^{\dagger}  \sp
T^\dagger\to V_L T^\dagger V_R^{\dagger} \ ,
\ee
where $(V_L,V_R)\in SU(N_f)_L\times SU(N_f)_R$.
We then consider infinitesimal gauge transformations $V_\epsilon (x)=e^{\epsilon \Lambda(x)
}\simeq 1+\epsilon\, \Lambda(x)$ such that the gauge field transforms as
$A\to A + \epsilon \delta_\Lambda A$. The transformations  \eqref{gaugevar} then imply
\be\label{gaugevar inf}
\begin{split}
&\delta_\Lambda A=-i\, D\Lambda=-i\,d\Lambda +[\Lambda,A] \, , \\
&\delta_\Lambda F =[\Lambda,F] \, ,\\
&\delta_{\Lambda_L}T=-T\Lambda_L \, , \\
&\delta_{\Lambda_R}T=\Lambda_R T \, .
\end{split}
\ee
The generators $\Lambda$ are antihermitian. Their decomposition in
the $U(1)$ and $SU(N_f)$ components is defined as
\be\label{Lambda decomp}
\Lambda= i\alpha \mathbb{I} + i\Lambda^a \lambda^a \, .
\ee
Here $\alpha$ and $\Lambda^a$ are real. Combining with (\ref{A decomp}) and (\ref{Lambda decomp}) we find
\be\label{gauge var A}
\delta \hat A_\mu =\partial_\mu \alpha\qquad \mathrm{and} \qquad \delta A_\mu^a=(D_\mu \Lambda)^a \, .
\ee

\subsection{Discrete symmetries}\label{discrete}

We specify here our conventions for the transformation of the gauge fields and the tachyon under parity and charge conjugation, following~\cite{Casero}.

\subsubsection*{Parity}

The parity transformation is
\be
\label{PP1P2} P = P_1 \cdot P_2 \, .
\ee
Here the second operator $P_2$ represents the action of parity on space
\be
\label{defP2} P_2 : (x_1,x_2,x_3) \to (-x_1,-x_2,-x_3) \, .
\ee
and $P_1$ represents the action on the flavor fields:
\be
\label{defP1} P_1 : L \leftrightarrow R \sp T \leftrightarrow T^\dagger \, .
\ee

\subsubsection*{Charge conjugation}

The charge conjugation acts on the flavor fields as follows:
\be
\label{defC}  C : L \to -R^t \sp R \to -L^t \sp T \to T^t \sp T^\dagger \to \left(T^\dagger\right)^t \, .
\ee

\subsection{Definitions for the static ansatz fields}

\label{Sec:2Dconv}

We present here the conventions and definitions for the fields of the static instanton ansatz living on the 2D subspace $(\xi, r) \equiv x^{\bar{\mu}}$. First, we choose the 2D Levi-Civita tensor as
\be
\label{defe2D} \e^{\xi r} = 1 \, .
\ee

In the ansatz \eqref{ansatzSU2iL}-\eqref{ansatzU1} for the static soliton, the fields transform under the residual gauge freedom \eqref{resg1} as follows:
\begin{itemize}
\item $\Phi$ is neutral.
\item $\phi\equiv \phi_1 + i\phi_2$ has charge 1.
\item $A_{\bar{\mu}} \equiv \Big(A_\xi, A_r\Big)$ is a gauge field.
\end{itemize}
The covariant derivatives of the complex scalars $\phi$ under the residual gauge freedom are then
\be
\label{covDiv} D_{\bar{\mu}}\phi \equiv \big( \partial_{\bar{\mu}} - iA_{\bar{\mu}} \big)\phi \, ,
\ee
with real and imaginary parts
\be
\label{covDiv_12} D_{\bar{\mu}}\phi_1 = \partial_{\bar{\mu}}\phi_1 + A_{\bar{\mu}} \phi_2 \sp D_{\bar{\mu}}\phi_2 = \partial_{\bar{\mu}}\phi_2 - A_{\bar{\mu}} \phi_1 \, .
\ee
The gauge-invariant field strength of the gauge field $A_{\bar\mu}$ is given as
\be
\label{defFLR} F_{\bar{\mu}\bar{\nu}} = \partial_{\bar{\mu}}A_{\bar{\nu}} - \partial_{\bar{\nu}}A_{\bar{\mu}} \, .
\ee

\section{Vacuum solution} \label{app:vac}
In this appendix we briefly review the solution on the gravity side
which corresponds to the  Poincar\'e-invariant vacuum of the dual
field theory. For more details, we refer the reader to \cite{spectrum}.

In the vacuum solution, the  gauge fields are set to
zero and the non-trivial fields are the 5d metric,  the dilaton
$\lambda(r)$ and, if all quark  masses are equal, the scalar tachyon
$\tau(r)$ defined below equation (\ref{v7}). The Poincar\'e invariant
solution can be put in the form:
\be \label{vac1}
ds^2 = e^{2A(r)}\left(dr^2 + \eta_{\mu\nu} dx^\mu dx^\nu\right), \quad
  \lambda = \lambda(r), \quad \tau = \tau(r),
\ee
where $r$ is the holographic radial coordinate and $x^\mu$ the
coordinates of 4d flat space-time where the dual QFT lives.

Since the gauge fields are not turned on,  the CS action does not
contribute to the field equations, and only the first two terms in
(\ref{v3}) are relevant for the solution (\ref{vac1}).

The
asymptotic form of the functions $A(r),\lambda(r),\tau(r)$   can be obtained
analytically in the UV ($r\to 0$)  and IR ($r\to +\infty$).  It is determined by the asymptotic form of the
potentials appearing in (\ref{v4}) and (\ref{v7}).

\paragraph*{UV asymptotics.}

The limit $r \to 0$ corresponds to the UV of the  field
theory. In this region,  $e^{A} \to +\infty$ and $\lambda \to 0$.  One finds (see e.g. \cite{spectrum}):
\be
\label{AUV} A(r)= -\log \left(\frac{r}{\ell}\right) + \frac{4}{9 \log (r \Lambda)} + \mathcal{O}\left(\frac{1}{\log(r \Lambda)^2}\right) \, ,
\ee
\be \label{lUV}
\lambda(r)  = -{1\over V_1}\frac{8}{9 \log (r \Lambda)} +
\mathcal{O}\left(\frac{1}{\log(r \Lambda)^2}\right)\,  ,
\ee
\begin{align}
\nn \frac{1}{\ell} \tau(r) &= m r (-\log (r \Lambda))^c \left(1 + \mathcal{O}\left(\frac{1}{\log(r \Lambda)}\right)\right) \\
\label{tUV}&\hphantom{=} + \s r^3 (-\log (r \Lambda))^{-c}  \left(1 + \mathcal{O}\left(\frac{1}{\log(r \Lambda)}\right)\right) \, .
\end{align}
Here, $\ell$ is the asymptotic AdS length,  $V_1$ is defined in equation (\ref{v12}), and  $\Lambda$, $m$ and
$\sigma$ are integration constants. In the dual field theory language,
 $\Lambda$ is the holographic
analog of the QCD scale, which measures the
breaking of conformal invariance in the UV;  $m$ corresponds to the
quark mass\footnote{We are working under the assumption that the quark
  mass  matrix is proportional to the identity.} ;
$\sigma$ is the chiral condensate. The near-boundary expansion of the
tachyon field corresponds to a   dimension-$3$ field theory operator,
with the extra logarithmic correction reproducing the mass
anomalous dimension of QCD. The exponent $c$ is fixed by the equation:
\be \label{vac2}
c = {4\over 3} \left(1 + {\kappa_1 -a_1 \over V_1} \right).
\ee

\paragraph*{IR asymptotics}
The IR  of the geometry is the region where   $e^A\to 0$ and
$\lambda \to +\infty$. In the solution exhibiting chiral symmetry
breaking, which has a non-trivial $\tau(r)$,  the tachyon also goes to
infinity in this region. With the
potential such as in (\ref{v15}), the IR  is reached as
$r\to +\infty$. In this limit, the fields behave as follows:
\be
\label{laIR} \l(r) = \ex^{\frac{3r^2}{2 R^2}+ \l_c} \left(1 + \mathcal{O}\left(r^{-2}\right)\right) \, ,
\ee
\be
\label{AIR} \ex^{A(r)} = \sqrt{\frac{r}{R}}\ex^{-\frac{r^2}{R^2}+ A_c} \left(1 + \mathcal{O}\left(r^{-2}\right)\right) \, ,
\ee
\be
\label{tauIR} \tau(r) = \tau_0 \left(\frac{r}{R}\right)^{C_\tau} \left(1 + \mathcal{O}\left(r^{-2}\right)\right) \, .
\ee
In the equations above, the constant  $C_\tau$  is fixed by the
asymptotic behavior of the potentials in equations (\ref{v10}) and
(\ref{v15}), and it must satisfy the constraint
$C_\tau >1$;  the integration constants  $R, \tau_0$ are
functions of the UV integration constants $\Lambda, m$. The constants $A_c, \lambda_c$ can be expressed in terms of the coefficients in the IR expansions of the potentials (see~\cite{Jarvinen,spectrum}).  The
quantity $R$  is the IR manifestation of the  non-perturbative scale of the
field theory.

\section{Numerical method}

\label{Num}

We detail in this appendix the numerical method that was used to solve the static soliton equations of motion. We summarize how the gradient descent method works, and specify the type of grid that was used to obtain the solution. We also present some precision tests that support the validity of the numerical solution.

\subsection{Method and grid choice}

In the gradient descent method, the solution to the equations of motion is found by rephrasing the original minimisation of the energy $M_0$ problem,  into a flow in a fictitious time, so that the limit of that flow is the solution of the original problem.
 Specifically, the limiting field configuration of the flow will minimize the soliton energy $M_0$.
  The limiting solution  is found, by starting from a reasonable initial field configuration that satisfies the boundary conditions, and then evolving the fields.
    In practice, the fictitious time evolution equations are taken to be
\be
\label{N1} \partial_\tau \Phi = \frac{\d M_0}{\d \Phi} \sp \partial_\tau\tilde{\phi}_{1,2} = -\frac{\d M_0}{\d \tilde{\phi}_{1,2}} \sp \partial_\tau A_{\bar{\mu}} = -\frac{\d M_0}{\d A_{\bar{\mu}}} \sp \partial_\tau \theta = -\frac{\d M_0}{\d \theta} \, ,
\ee
and they are solved numerically. The fields above have been defined in section \ref{31}.
Equations (\ref{N1}) must be also supplemented by the boundary conditions of table \ref{tab:bcsLg} for any value of the fictitious time.

 Note that the gradients of the action with respect to the fields are nothing but the respective equations of motion. Also, the signs in front of the gradients,  mean that $M_0$ is maximized with respect to $\Phi$, and minimized with respect to all the other fields. This choice of sign ensures that the coefficient of the Laplacian is positive in the right hand sides of \eqref{N1}. In these conditions, the ellipticity of the equations of motion guarantees that the diffusive problem has a limiting solution which is the solution we are looking for.

The numerical algorithm is then constructed by discretizing the differential equations \eqref{N1}. That is, the bulk is covered by a grid, and the derivatives are approximated by finite differences between the values of the fields at the points of this grid. We denote by $n$ the number of points in the holographic direction and by $m$ the number in the radial direction. Note that the bulk has infinite extent, but the soliton solution is confined to a finite region.  Therefore, it can be computed by considering finite cut-offs. In practice, the following  cut-offs
$$r_{\text{max}} \simeq 30 M_0^{-1} \sp  \xi_{max} \simeq 100 M_0^{-1}$$ were  found to be good enough to solve the problem reliably. Those numbers may seem large given that the baryon density is mainly confined to $rM_0$ and $\xi M_0$ of order 1, as seen on Figure \ref{fig:rho_Ni_M_v8_probe}. However, the chiral gauge fields have non-trivial power-law asymptotics\footnote{This long-range behavior of the meson cloud is due to the fact that we consider the chiral limit, and will be suppressed exponentially when introducing finite quark masses.} away from the baryon center, which can only be captured by using sufficiently large cut-offs.  The gradients associated with these long-range tails are small though, so that a few grid points are sufficient to describe them. Instead of a linear grid, we therefore considered a logarithmic grid, where most points are concentrated near the baryon center, and only a few points cover the region that separates the baryon from the cut-offs. The precise definition of the grid that we used is given by
\be
\label{N1b} r(k) = \exp{\big( k\D r + \log{(r_{min} + 1)}\big)} - 1 \sp  0 \leq k \leq n \, ,
\ee
\be
\label{N1bb} \xi(j) = \exp{\big( j\D \xi \big)} - 1 \sp 0 \leq j \leq m  \, ,
\ee
where the spatial steps are
\be
\label{N1c} \D r = \frac{1}{n} \big( \log{(r_{max}+1)} - \log{(r_{min}+1)} \big) \sp \D \xi = \frac{1}{m} \log{(\xi_{max}+1)} \, .
\ee
Note that we introduced a UV cut-off $r_{min}$ since the boundary is a singular point of the equations and the boundary conditions there must be imposed on a shifted boundary, as usual.
 As long as $r_{min} \ll r(1)$ in (\ref{N1b}), having $r_{min}$ finite does not influence the baryon solution. In practice we found that the precision of the vacuum solution for the tachyon field is very good for $r > r_{min} \simeq 0.02 M_0^{-1}$. Also, the size of the grid that was chosen to produce the numerical results presented in the text is  $(n,m)=(100,100)$. Below, we show evidence that the baryon solution has already converged well for such a grid size.

A remark is that, instead of \eqref{N1}, one can consider the diffusive problem with general diffusion coefficients, which are generically field and position-dependent
\be
\label{N2} \partial_\tau \Phi = D_\Phi(r,\xi)\frac{\d M_0}{\d \Phi} \sp \partial_\tau\tilde{\phi}_{1,2} = -D_\phi(r,\xi)\frac{\d M_0}{\d \tilde{\phi}_{1,2}} \sp \partial_\tau A_{\bar{\mu}} = -D_A(r,\xi)\frac{\d M_0}{\d A_{\bar{\mu}}} \, ,
\ee
\be
\nn \partial_\tau \theta = -D_\theta(r,\xi)\frac{\d M_0}{\d \theta} \, .
\ee
Although a uniform diffusion coefficient amounts to a redefinition of the fictitious time $\tau$, the dependence on the fields and position in the bulk actually modify the diffusive problem. A good choice for the purpose of solving \eqref{N2} numerically is to set the diffusion coefficients such that the coefficients of the fields' Laplacians are equal to 1. This choice has the advantage that the stability properties of the discretized version of \eqref{N2} are more tractable. If we ignore the first and zeroth order derivatives, requiring the absence of modes growing with $\tau$ implies the famous bound on the time step $\D\tau$ (see for example chapter 3.6 of \cite{NumBook})
\be
\label{N3} \D\tau < \frac{\D \xi^2 \D r^2}{2(\D \xi^2 + \D r^2)} \, ,
\ee
which is valid for uniform spatial grids, with steps $\D x$ in the $\xi$ direction, and $\D r$ in the $r$ direction. For the logarithmic grid that we used \eqref{N1b}-\eqref{N1bb}, \eqref{N3} would still apply with the definitions of $\D r$ and $\D \xi$ from \eqref{N1c}, since they correspond to the smallest point separation on the grid. However, the full equations of motion include first and zeroth order derivatives of the fields, so that the stability bound is more complicated than \eqref{N3}. For the grid that we used, we found that using as a bound
\be
\label{N4} \D\tau < \frac{\D r^2}{4} \, ,
\ee
resulted in stable algorithms. We do not claim any generality of this result though, and different bounds were actually observed for different grids.

\subsection{Precision tests}

\begin{figure}[h]
\begin{center}
\includegraphics[width=0.505\textwidth]{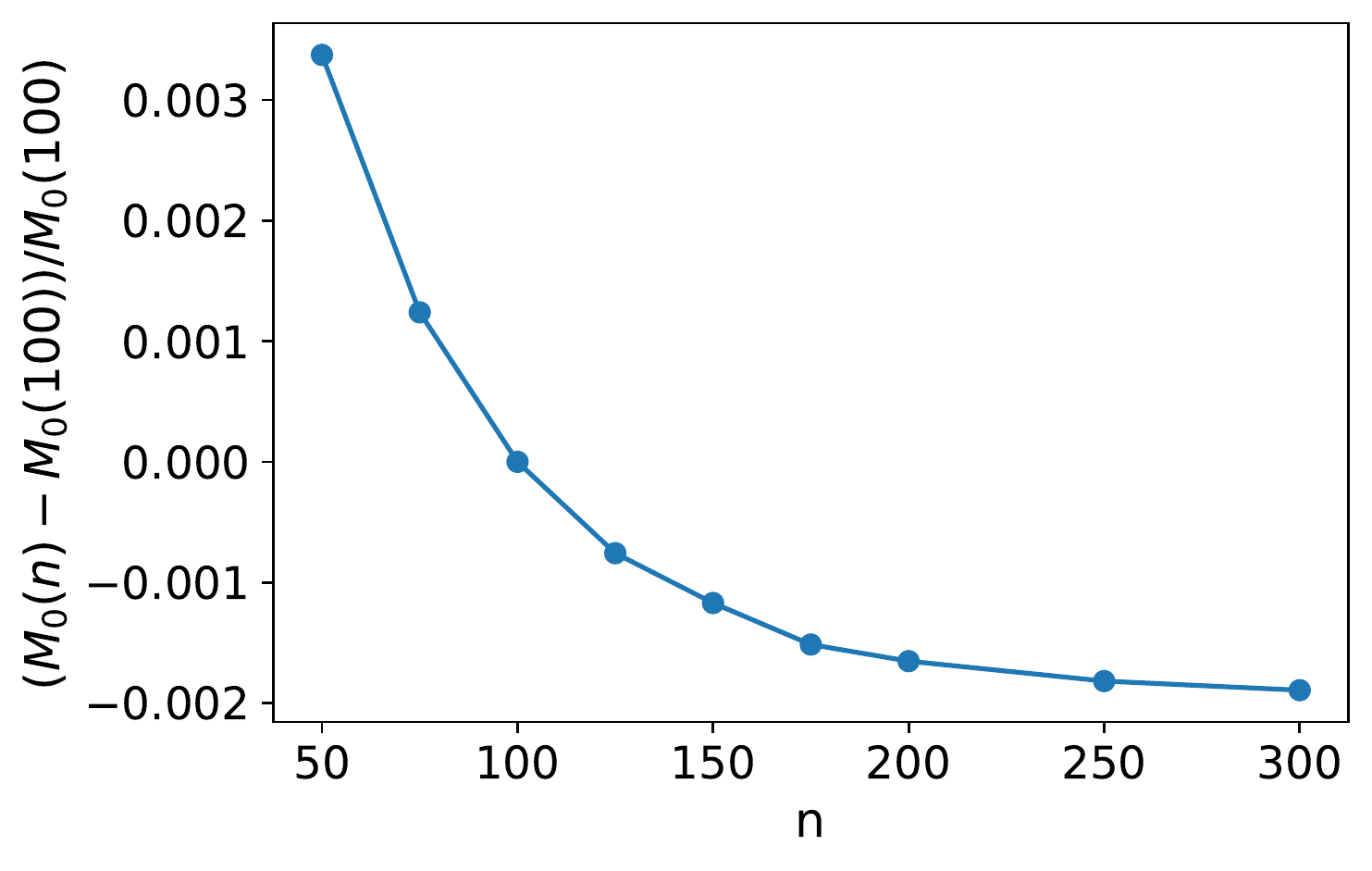}
\hspace{3mm}
\includegraphics[width=0.45\textwidth]{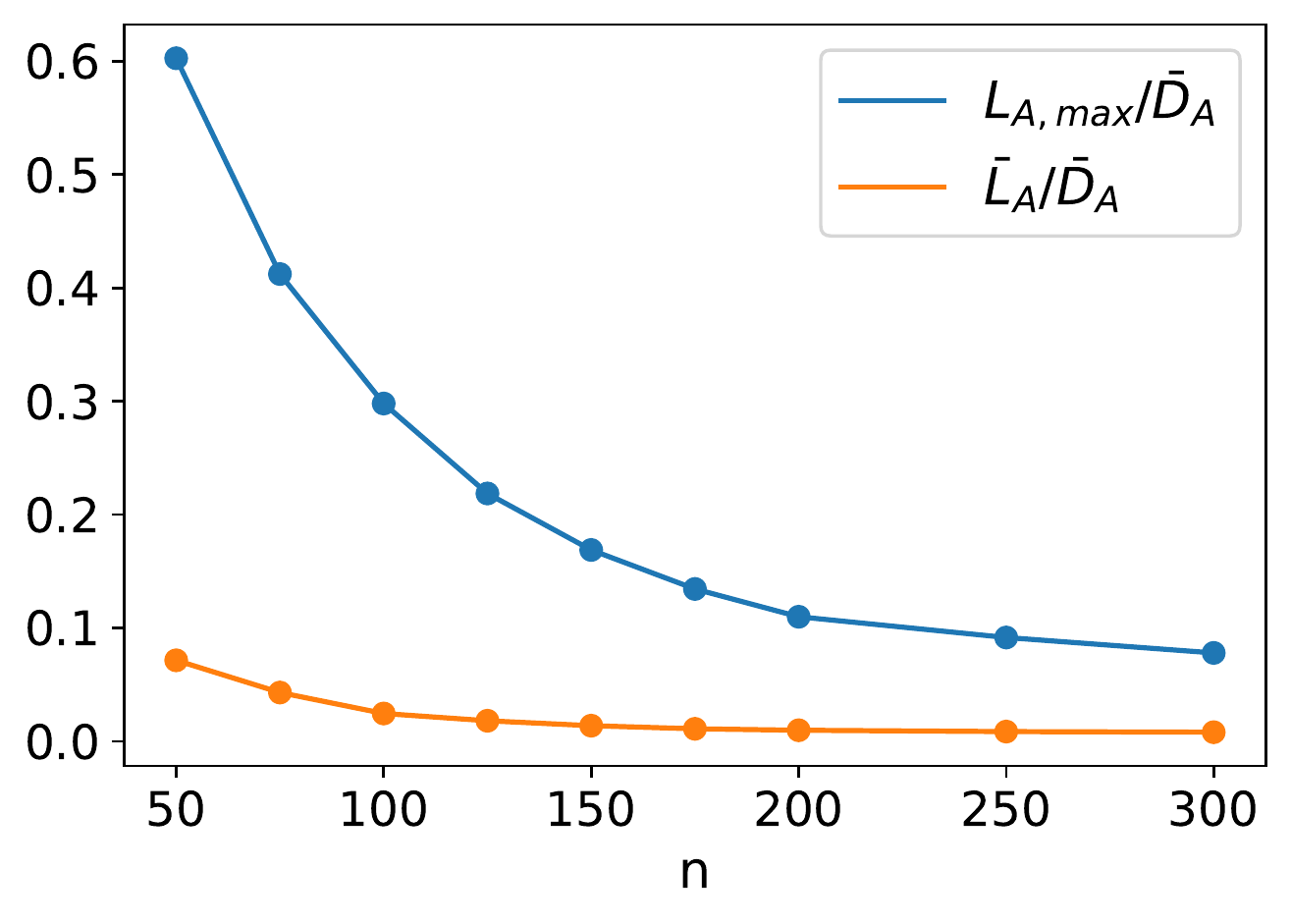}
\caption{\textbf{Left} : Relative difference of the soliton mass $M_0$ with the value at $n_0=100$, as a function of the grid size $n$. \textbf{Right} : the two indicators for the accuracy of the Lorenz gauge fixing, $i_1$ (orange) and $i_2$ (blue) from \eqref{N7}, as a function of the grid size $n$.}
\label{Fig:gs}
\end{center}
\end{figure}

We now present some tests of the numerical precision of the soliton solution computed via the gradient-descent method. We first investigate the convergence of the observables as a function of the grid size. Left figure \ref{Fig:gs} shows the evolution of the calculated soliton mass as a function of the grid size for $n=m$. It is observed that the soliton mass converges in the limit of large grid size, and the value that we derive at $n=m=100$ is already within $0.2\%$ of accuracy from the limiting value. Actually, even at $n=50$, the error on the mass is already less than one percent.

\begin{figure}[h]
\begin{center}
\includegraphics[width=0.75\textwidth]{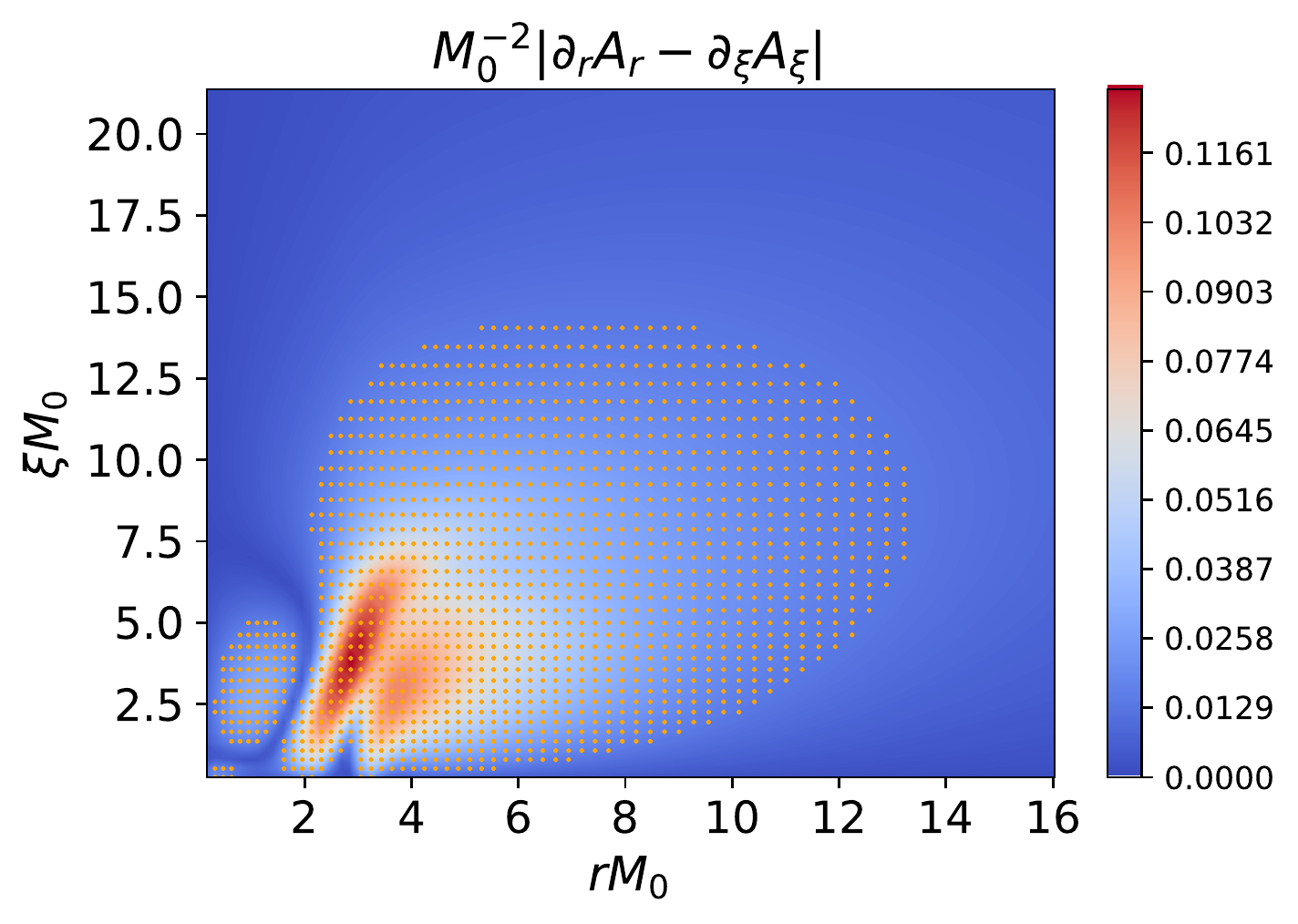}
\caption{Density plot of the quantity $D_A$ defined in \eqref{N6}, as an function of $r$ and $\xi$. All dimensionful quantities are expressed in units of the soliton mass $M_0$, and the grid size is $(n,m)=(100,100)$. The orange dots correspond to the grid points that belong to the region $\mathcal{A}_D$ \eqref{N8}, where $D_A$ is significantly different from 0.}
\label{Fig:AD}
\end{center}
\end{figure}

As another test of the precision of the solution, we also investigated the convergence for the divergence of the 2-dimensional gauge field
\be
\label{N5} L_A(r,\xi) \equiv |\partial_r A_r + \partial_\xi A_\xi| \, .
\ee
Due to the Lorenz gauge fixing \eqref{Lorenzc1}, the baryon solution should be such that $L_A$ vanishes everywhere in the bulk. To estimate what is the error on the Lorenz condition in the numerical solution, $L_A(r,\xi)$ is computed numerically and compared with the typical scale of the gradients of $A_{\bar{\mu}}$. One way of defining this typical scale, is via the combination of derivatives orthogonal to \eqref{N5}
\be
\label{N6} D_A(r,\xi) \equiv |\partial_rA_r - \partial_\xi A_\xi| \, .
\ee
The accuracy of the Lorenz gauge fixing will therefore be estimated by comparing the two quantities $L_A$ and $D_A$, where the criterion for accuracy is that $L_A$ should be much smaller than $D_A$. Because $D_A$ vanishes in some places in the bulk, calculating the ratio of $L_A$ over $D_A$ does not give a globally well-defined indicator of the precision of the Lorenz gauge. Instead, we will use two global indicators, that are defined from the maxima and mean values of $L_A$ and $D_A$
\be
\label{N7} i_1 \equiv \frac{\bar{L}_A}{\bar{D}_A} \sp i_2 \equiv \frac{L_{A,max}}{\bar{D}_A} \, .
\ee
The first indicator $i_1$ estimates the average error on the Lorenz gauge fixing over the whole solution, whereas $i_2$ indicates what is the error at the location in the bulk where it is the worst. Here, it should be clarified what we mean by the mean values $\bar{L}_A$ and $\bar{D}_A$ in \eqref{N7}. The definition is similar to what was used to define the mean of the Lagrangian density \eqref{mean}. That is, the mean value is computed over a region of the bulk where the derivatives of $A_{\bar{\mu}}$ are significantly different from zero. Specifically, the  corresponding region $\mathcal{A}_D$ is defined as
\be
\label{N8} (r,\xi) \in \mathcal{A}_D \iff D_A(r,\xi) \geq 0.1 D_{A,max} \, .
\ee
In the numerical solution, $\mathcal{A}_D$ contains a finite number of grid cells, $N_{cells}$, and the definition of the mean values over $\mathcal{A}_D$ is analogous to \eqref{mean}
\be
\label{N9} \bar{L}_A = \frac{1}{N_{\mathrm{cells}}} \sum_{i \in \mathcal{A}_D} L_A(i) \sp \bar{D}_A = \frac{1}{N_{\mathrm{cells}}} \sum_{i \in \mathcal{A}_D} D_A(i) \, .
\ee
For concreteness, Figure \ref{Fig:AD} shows the density plot of $D_A$ over the $(r,\xi)$ plane for a grid of size $(n,m)=(100,100)$, together with the grid points that belong to $\mathcal{A}_D$.

The plot of the two indicators $i_1$ and $i_2$ defined in \eqref{N7} is shown in the right of Figure \ref{Fig:gs}, as a function of the grid size. Although it is much slower than for the baryon mass, $i_1$ and $i_2$ are also found to converge, towards a value which is consistent with zero. The indicator $i_2$ is observed to be much larger than $i_1$, which means that the maximum of the gauge field divergence $L_A$ is reached at the top of a narrow peak. In particular, for the grid that we used to produce the numerical results presented in the main text $(n,m) = (100,100)$, $i_1$ and $i_2$ are found to take the following values
\be
\label{N10} i_1(100) \simeq 2.5\% \sp i_2(100) \simeq 30\% \, .
\ee
This means that on average the Lorenz condition is well obeyed within $2\%$ over most of the baryon solution, but there is a narrow region in the bulk where the error grows up to $30\%$. For the largest grid investigated that has $(n,m)=(300,300)$, the average accuracy of the Lorenz gauge fixing is given by $i_1(300)\simeq 0.8\%$, and the maximum error by $i_2(300)\simeq 8\%$.

\begin{figure}[h]
\begin{center}
\includegraphics[width=0.485\textwidth]{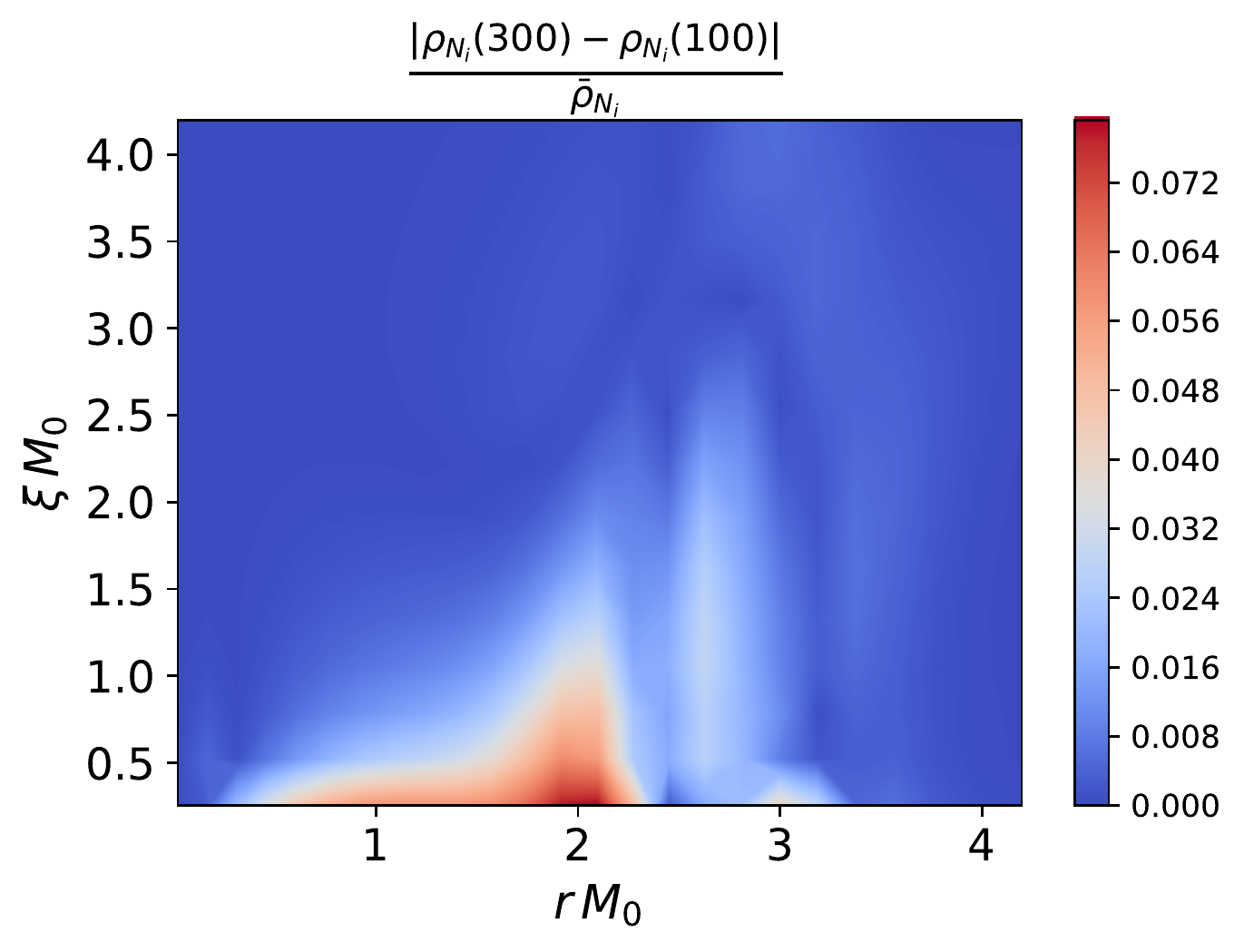}
\includegraphics[width=0.5\textwidth]{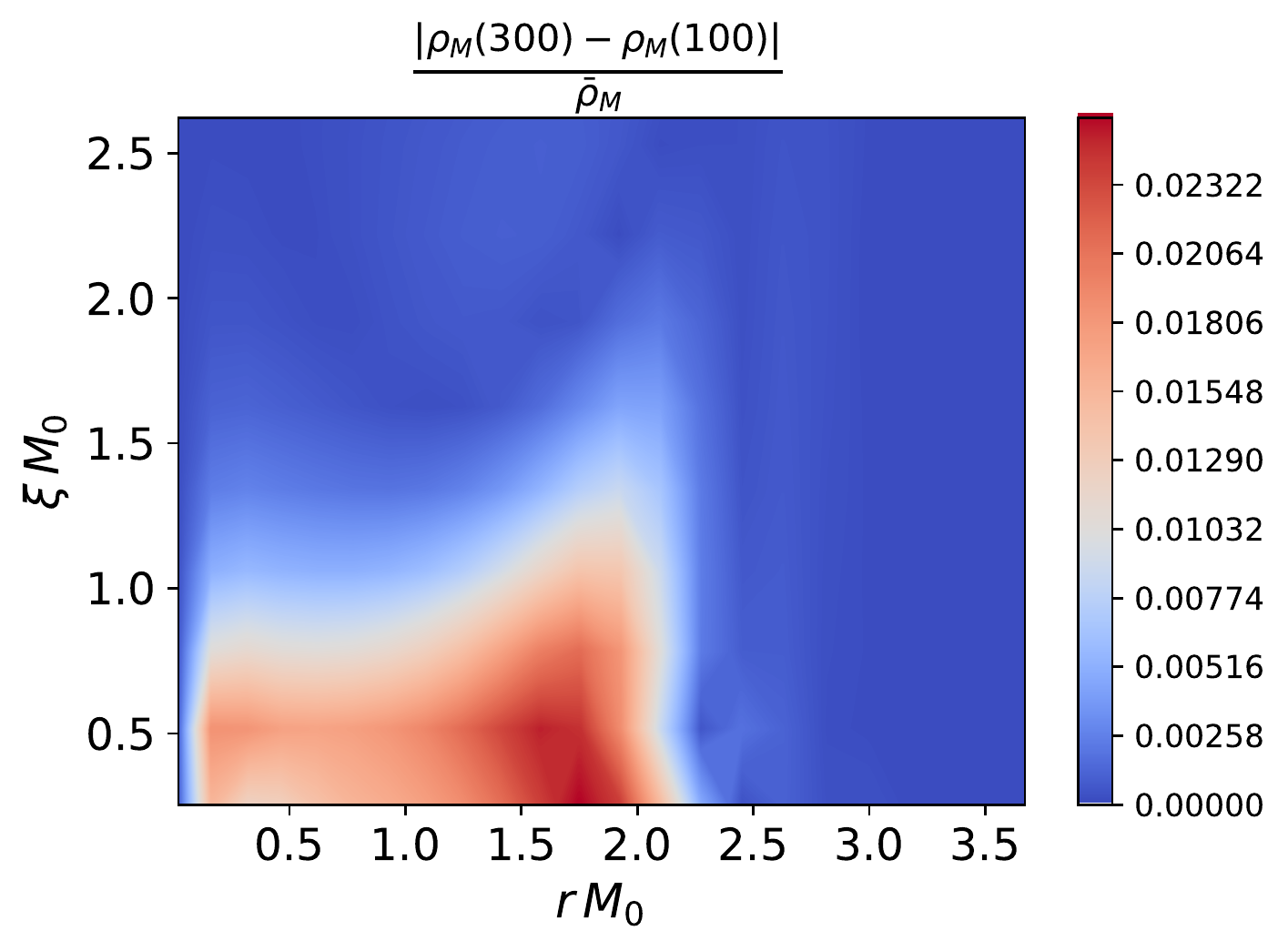}
\caption{\textbf{Left} : Relative difference of the instanton number densities computed on the grids with $(n,m)=(100,100)$ and $(n,m)=(300,300)$. \textbf{Right} : Relative difference of the bulk Lagrangian densities computed on the grids with $(n,m)=(100,100)$ and $(n,m)=(300,300)$.}
\label{Fig:Drho100v300}
\end{center}
\end{figure}

The fact that the value $i_2(100)$ is quite high, indicates that there is a small region in the bulk where the Lorenz condition is not very well obeyed for the solution on the grid that we used, with $(n,m) = (100,100)$. Since $i_2(300) \simeq 8\%$ is much smaller, another way of checking the quality of the solution at $n=100$ is to compare the numerical field configurations computed at $n=100$ with those at $n=300$. In particular, we will focus on the two bulk quantities that were analyzed in this work, that are the instanton number and Lagrangian densities \eqref{rNi} and \eqref{defrM}. The relative differences of the two types of densities between the $n=100$ and the $n=300$ solutions are shown in Figure \ref{Fig:Drho100v300}. The definition for the relative differences is the same as \eqref{mean}, where the mean of the bulk Lagrangian density is still computed over the region bounded by the green line in Figure \ref{fig:rho_Ni_M_v8_probe}. For the instanton number density, the average is computed over the region where $|\rho_{N_i}|M_0^{-4} \geq 2.8\times 10^{-3}$. The two solutions are compared on the grid with $n=100$. Figure \ref{Fig:Drho100v300} shows that the densities computed at $n=100$ and $n=300$ only differ by a few percent of the mean value, with a maximum of about $8\%$ in the case of the instanton number density. 

In addition to the grid size, we also studied the convergence of the soliton mass as a function of the cut-offs $\xi_{max}$ and $r_{max}$. As far as the radial cut-off is concerned, we found that larger $\xi_{max}$ modified the soliton mass by less than $0.01\%$. For the holographic coordinate, it was found that larger IR cut-offs $r_{max}$ do not affect the soliton mass by more than about $0.2\%$.

The conclusion of this analysis of the numerical precision is that the grid that we used with $(n,m) = (100,100)$ gives a very precise value for the soliton mass $M_0$, and a result within a few percent of accuracy for the densities. This means that the solution can be trusted at the qualitative level, and also at the quantitative level as far as $M_0$ is concerned, which is the observable that we were interested in in this work. When extracting other observables from the baryon solution, one should check in each case the properties of convergence, and adapt the grid size to the desired level of accuracy. In particular, the analysis of the densities in Figure \ref{Fig:Drho100v300} indicates that accuracies better than the percent level may require grid sizes larger than $(n,m) = (100,100)$. 

\section{Numerical solutions for a different set of potentials}

\label{App:OldPot}

In this appendix we present, for comparison, the numerical baryon solution obtained for a different set of V-QCD potentials. We focus on the leading order probe baryon case, for static and rotating baryons. The potentials are those derived in~\cite{Remes}, the set ``7a'' in Appendix~A of this reference, and can also be found in Appendix B of \cite{Ishii}. The value of the pion decay constant $f_\pi$ for these potentials is significantly smaller than the experimental value, so they are not expected to give a quantitatively good description of all baryonic properties. The potentials of~\cite{Remes} have the same UV and IR asymptotics fitted to QCD properties as the ones that were used in this work though (with parameters given in Table \ref{tab:bcsLg}), so the qualitative behavior should be the same. The purpose of this appendix is to check the previous statement by reproducing the plots of the main text for the potentials of~\cite{Remes}. We  also compute the baryon spectrum in this case and compare with Table \ref{tab:bspec}.

\subsection{Static soliton}

\begin{figure}[h]
\begin{center}
\begin{overpic}
[scale=0.45]{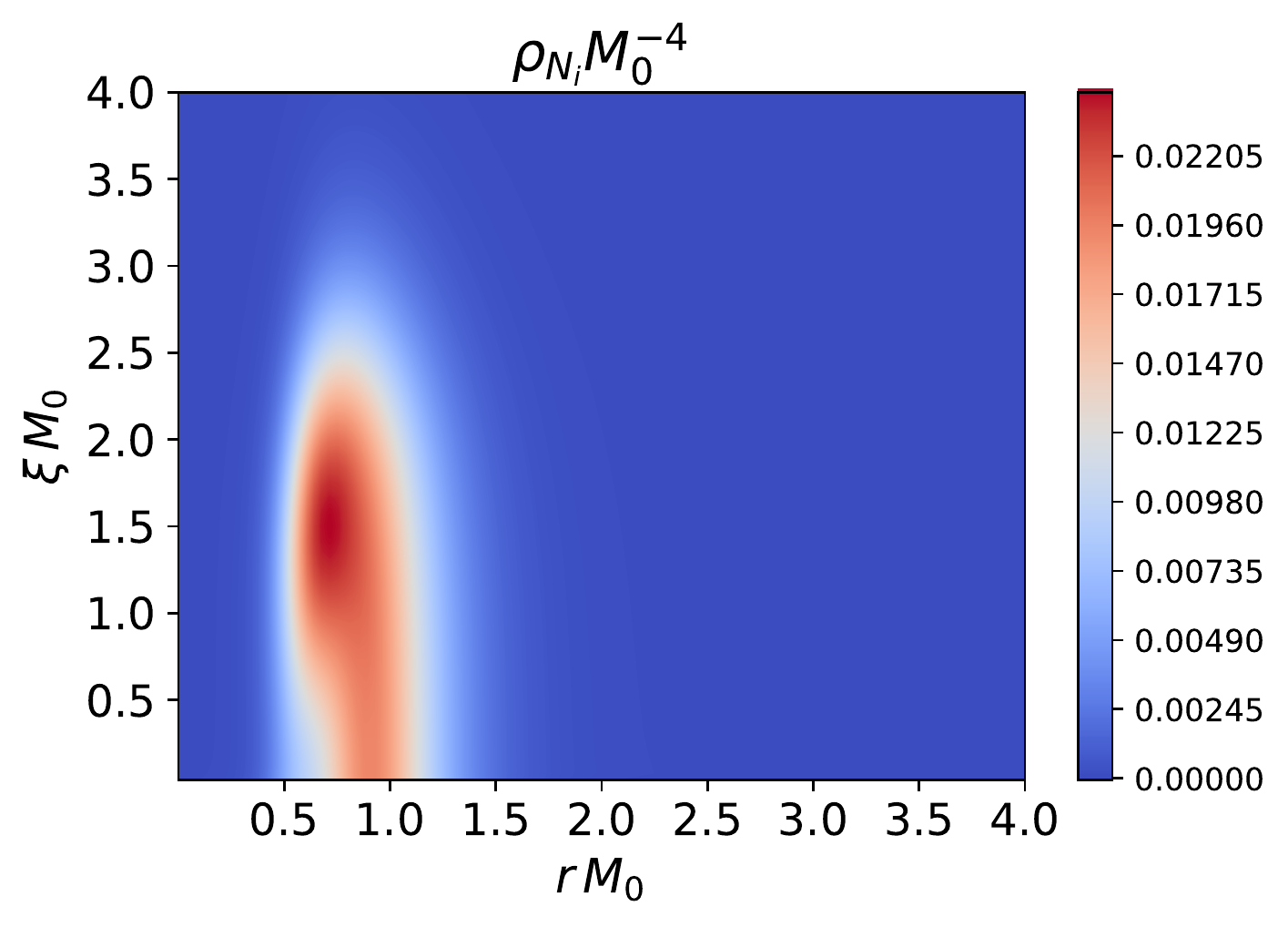}
\end{overpic}
\begin{overpic}
[scale=0.45]{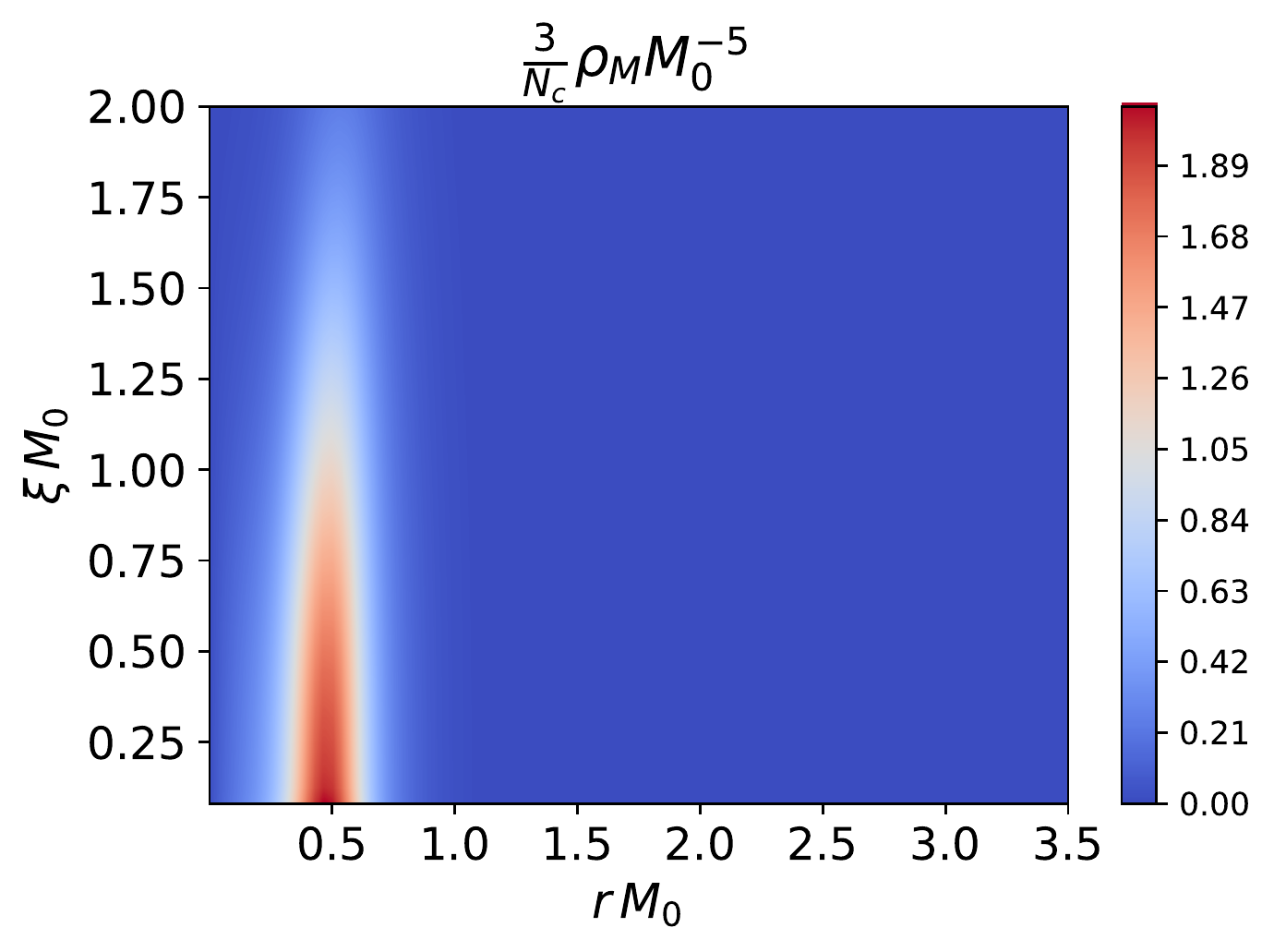}
\end{overpic}
\caption{Instanton number (left) and bulk Lagrangian (right) density for the static soliton solution in the probe baryon regime. All quantities are expressed in units of the classical mass of the soliton \eqref{M0}. The center of the soliton is located at $\xi=0$ where the density diverges as $\xi^{-1}$.}
\label{fig:rho_Ni_M_old_probe}
\end{center}
\end{figure}

The instanton number and Lagrangian density in the $(\xi,r)$-plane are presented in Figure \ref{fig:rho_Ni_M_old_probe}, where all dimensionful quantities are expressed in units of the classical soliton mass \eqref{M0}. By comparing with Figure \ref{fig:rho_Ni_M_v8_probe}, it is observed that the qualitative shape of the densities are similar. However, the densities are typically located significantly closer to the UV boundary for the potentials of~\cite{Remes}, in units of the soliton mass. Also, the extent of the densities in the holographic direction is smaller, especially for the Lagrangian density. The numerical value for the classical soliton mass $M_0$ is obtained by integrating the Lagrangian density in Figure \ref{fig:rho_Ni_M_old_probe}
\be
\label{M0num_old} M_0 \simeq \frac{N_c}{3}\times 265\, \text{MeV} \, .
\ee
This value is about 4 times smaller than for the potentials used in the main text \eqref{M0num}. Our numerical analysis indicate that this small value for $M_0$ can be traced back to the smallness of the pion decay constant for the potentials of~\cite{Remes}.

\begin{figure}[h]
\begin{center}
\begin{overpic}
[scale=0.7]{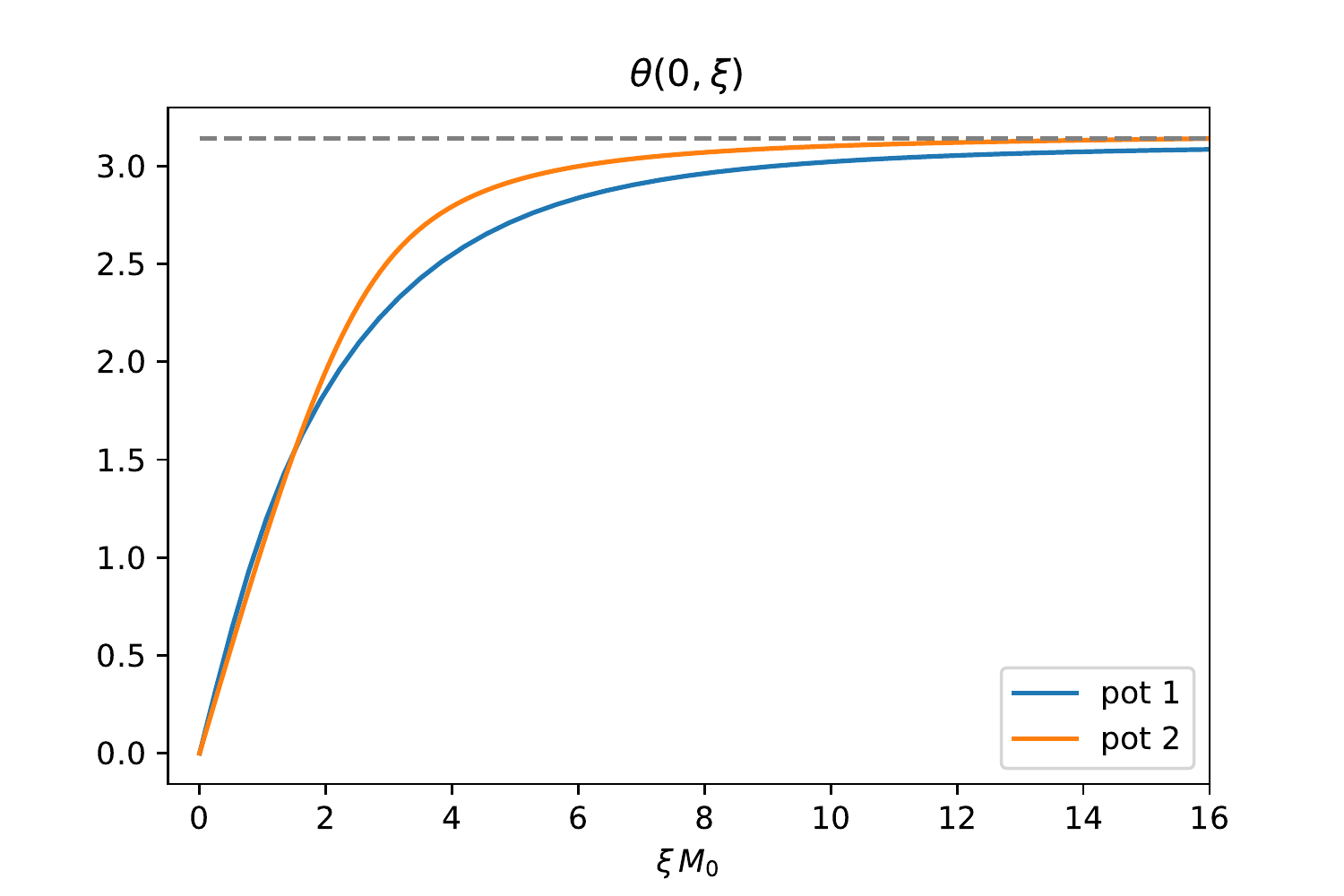}
\end{overpic}
\caption{Radial profile of the non-abelian phase of the tachyon field \eqref{TU2} at the UV boundary. The blue line corresponds to the solution for the V-QCD potentials presented in Section \ref{Sec:VQCD_intro}, and the orange line for those of \cite{Remes}. The dashed gray line indicates the asymptotic value $\pi$.}
\label{fig:thx_probe_old}
\end{center}
\end{figure}

In Figure \ref{fig:thx_probe_old}, we also plot the profile at the boundary ($r=0$) for the non-abelian phase $\theta$ of the tachyon field \eqref{TU2}, and compare with Figure \ref{fig:thx_probe_v8}. We see that the results are close up to the rescaling of the radial coordinate as a function of the soliton mass $M_0$. That being said, we notice that $\theta(\xi M_0)$ increases somewhat faster from the baryon center for the potentials of \cite{Remes} compared with those of the main text.

\subsection{Rotating soliton}

\begin{figure}[h]
\begin{center}
\begin{overpic}
[scale=0.7]{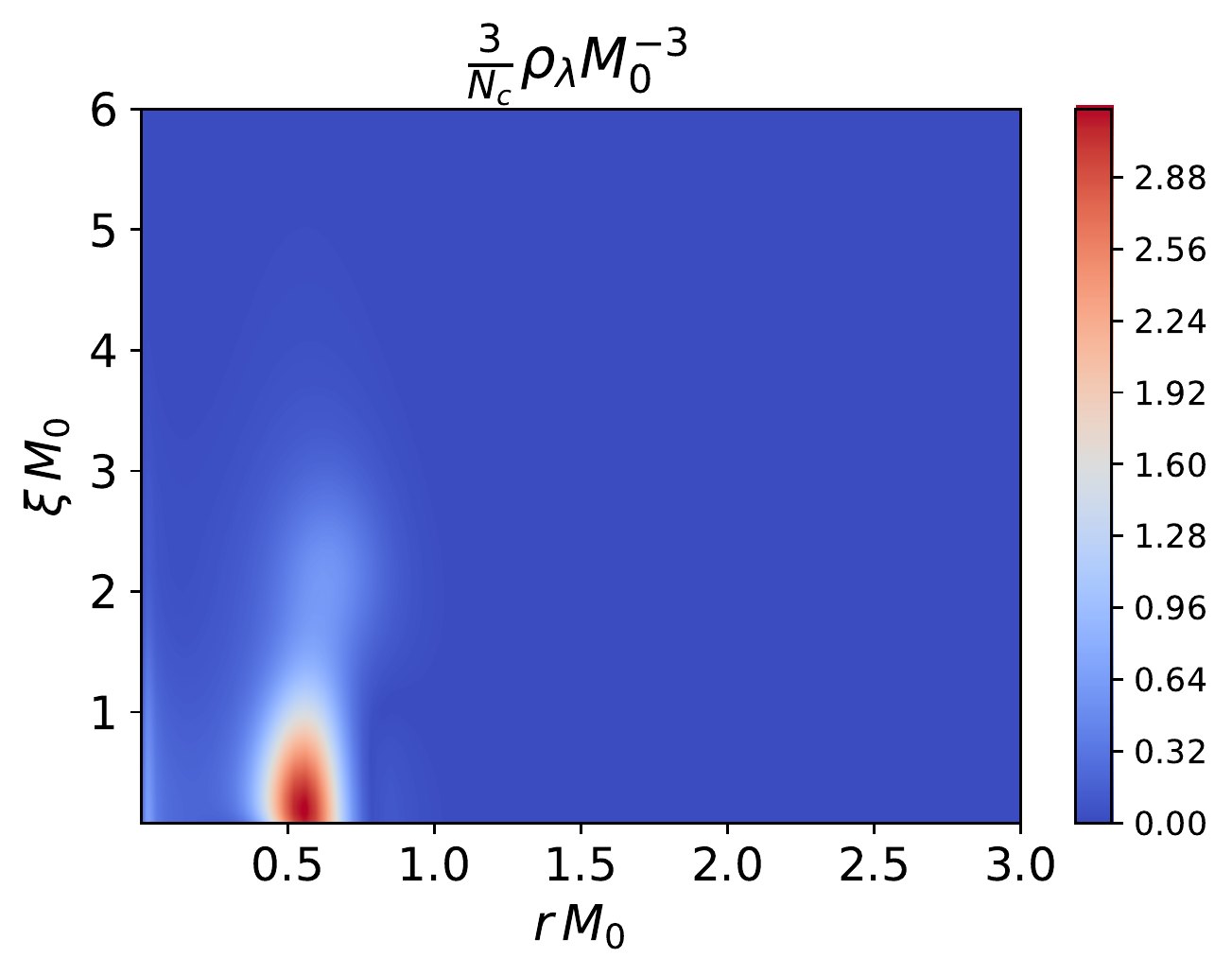}
\end{overpic}
\caption{Lagrangian density for the rotating fields in the probe baryon approximation, for the potentials of~\cite{Remes}. All quantities are expressed in units of the classical mass of the static soliton $M_0$ \eqref{M0}. The center of the soliton is located at $\xi=0$ where the density diverges as $\xi^{-1}$. The UV boundary is at $r=0$.}
\label{fig:lrot_pr_old}
\end{center}
\end{figure}

The bulk Lagrangian density for the rotating fields in the $(\xi,r)$-plane is presented in Figure \ref{fig:lrot_pr_old}, where all dimensionful quantities are expressed in units of the classical soliton mass $M_0$ \eqref{M0}. Comparing with Figure \ref{fig:lrot}, we see that the Lagrangian densities for the two sets of potentials are qualitatively different. In particular, the maximum of the density is well separated from the boundary for the potentials of~\cite{Remes}. As stated in the main text, we believe that this behavior is more generic than that of Figure \ref{fig:lrot}. Also, as was observed for the static solution, the density has a lesser extent in the holographic direction (in units of $M_0$) for the potentials of~\cite{Remes}.

The numerical value for the classical moment of inertia density $\l$ in \eqref{deflamb} is obtained by integrating the Lagrangian density in Figure \ref{fig:lrot_pr_old}
\be
\label{lnum_old} \frac{1}{\l} \simeq \frac{3}{N_c}\times 200\,\text{MeV} \, .
\ee
From this result, the spin-isospin spectrum of the baryons can be computed and compared with the result for the potentials of the main text, shown in Table \ref{tab:bspec}. This comparison is presented in Table \ref{tab:bspec_old}, setting $N_c=3$ in the large N result.
\begin{table}
\centering
\begin{tabular}{|c|c|c|c|}
\hline
Spin & Pot 2 & Pot 1& Experimental mass \\
\hline
$s = \frac{1}{2}$ & $M_N \simeq 340\,\text{MeV}$ & $M_N \simeq 1170 \,\text{MeV} $ & $M_N = 940 \,\text{MeV} $\\
\hline
$s = \frac{3}{2}$ & $M_\D \simeq 640\,\text{MeV}$ & $M_\D \simeq 1260 \,\text{MeV}$ & $M_\D = 1234 \,\text{MeV}$
 \\
\hline
\end{tabular}
\caption{Baryon spin-isospin spectrum in the V-QCD model with the potentials of~\cite{Remes} (Pot 2), compared with the potentials used in the main text (Pot 1) and experimental data.}
\label{tab:bspec_old}
\end{table}
Table \ref{tab:bspec_old} indicates that the baryon masses for the potentials of~\cite{Remes} are much smaller than experimental data, which is a consequence of the low mass of the soliton \eqref{M0num_old}. Note however that the mass difference between the nucleon and the $\D$ is much closer to experimental data than for the potentials of Section \ref{Sec:VQCD_intro}.

\section{Equations of motion for the rotating soliton}

\label{App:EOMrot}

We present in this appendix the expressions for the equations of motion for the rotating soliton ansatz fields. These equations are obtained by extremizing the moment of inertia \eqref{deflamb} with respect to variation of the fields. We start by giving the expression of the components of the field strength that are turned on by the slow rotation, as those are useful to compute the moment of inertia. We then write the equations of motion, first in the general tachyon back-reaction regime and then in the probe baryon regime.

\subsection{Field strength}

For the slowly rotating soliton ansatz of \eqref{ansLs}-\eqref{ansLhrs} and \eqref{ansTs}, the components of the field strength that are already non-zero in the static soliton solution are identical to the static solution at linear order in $\omega$. Their expression are given by equations (F.14)-(F.16) in \cite{BaryonI}. The effect of slow rotation at linear order is to source the components $F_{0r}$, $F_{0i}$, $\hat{F}_{ij}$ and $\hat{F}_{ir}$, where the superscript $(L/R)$ is implicit. These components are expressed in terms of the fields of the ansatz as
\begin{align}
\nn F_{0r} &= V(t)\left(- \left(\partial_r \k_1 + A_r \k_2 \right) \e^{abc} \omega^b \frac{x^c}{\xi} \right. \\
\label{F0rrs} & \hphantom{= V(t)} \left. + \left( \partial_r\k_2 - A_r\k_1 \right)\left( \omega^a - (\vec{\omega}.\vec{x})\frac{x^a}{\xi^2} \right) - \partial_r v\, (\vec{\omega}.\vec{x})\frac{x^a}{\xi^2}\right)\frac{\s^a}{2}V(t)^\dagger  \, ,
\end{align}
\begin{align}
\nn F_{0i} &= V(t)\left(x^a\e^{ibc}\omega^b\frac{x^c}{\xi^2}\left(-\k_1\frac{1+\phi_2}{\xi}\right) \right. \\
\nn &\hphantom{= V(t)} + \frac{x^a}{\xi} \left( \omega^i - \frac{x^i}{\xi^2}(\vec{\omega}.\vec{x}) \right) \left( \k_2 \frac{\phi_2}{\xi} + \k_1 \frac{\phi_1}{\xi} - \frac{v}{\xi} \right) \\
\nn &\hphantom{= V(t)} + \frac{\vec{\omega}.\vec{x}}{\xi}\left( \frac{x^ix^a}{\xi^2} - \d^{ia} \right) \frac{\k_2 - v\phi_2}{\xi} \\
\nn &\hphantom{= V(t)} + \e^{aki}\frac{(\vec{\omega}.\vec{x})x^k}{\xi^2}(\k_2 + v)\frac{\phi_1}{\xi} + \e^{aki}\omega^k \left( -\frac{\k_1 + \k_2\phi_1}{\xi}  \right) \\
\nn &\hphantom{= V(t)} + x^i\e^{abc}\omega^b\frac{x^c}{\xi^2}\left( \frac{\k_1 + \k_2 \phi_1}{\xi} - (\partial_{\xi}\k_1 + A_{\xi} \k_2) \right)\\
\label{F0irs} &\hphantom{= V(t)} \left.+ \frac{x^i}{\xi} \left( \omega^a - \frac{x^a}{\xi^2}(\vec{\omega}.\vec{x}) \right) \left(\partial_{\xi}\k_2 - A_{\xi}\k_1 \right) + \frac{(\vec{\omega}.\vec{x})x^ix^a}{\xi^3} (-\partial_{\xi}v) \right)\frac{\s^a}{2}V(t)^\dagger \, ,
\end{align}
\be
\label{Fhirrs} \hat{F}_{ir} = \frac{x^i(\vec{\omega}.\vec{x})}{\xi^2} \left( \partial_{\xi}B_r - \partial_r B_{\xi} \right) + \left( \omega^i - \frac{x^i(\vec{\omega}.\vec{x})}{\xi^2} \right) \frac{B_r - \partial_r\rho}{\xi} - \e^{ibc}\omega^b\frac{x^c}{\xi} \partial_r Q \, ,
\ee
\be
\label{Fhijrs} \hat{F}_{ij} = \frac{x^i\omega^j - \omega^i x^j}{\xi^2}\left( \partial_{\xi}\rho - B_{\xi} \right) + 2\frac{Q}{\xi} \e^{ijk}\omega^k + \left( \partial_{\xi}Q - \frac{Q}{\xi}\right) \omega^b \frac{x^c}{\xi} \left( \e^{jbc}\frac{x^i}{\xi} - \e^{ibc}\frac{x^j}{\xi} \right) \, .
\ee

\subsection{Equations of motion in the tachyon back-reaction regime}

\label{App:EOMrotbr}

The general equations of motion in the tachyon back-reaction regime are given by
\begin{align}
\nn & w^2 \Big[ \partial_{\xi}\big(\fxx\xi^2 \partial_{\xi}v\big) - 2\cX\big( v(1 + |\phi|^2) - (\k\phi^* + h.c.) \big) \Big]+\\
\nn & + \partial_r\Big( \xi^2 w^2\frr \partial_r v \Big)  - \partial_r\Big( \xi^2 w^2\fxr \partial_\xi v \Big) - \partial_\xi\Big( \xi^2 w^2\fxr \partial_r v \Big) = \\
\nn &\hphantom{=} \frac{2\e^{\bar{\mu}\bar{\nu}}}{\pi^2M^3}  \bigg( B_{\bar{\mu}\bar{\nu}}\left[\frac{1}{2}(f_1 + f_3)\, (|\phi|^2 - 1) + (f_1-f_3-if_2)\tilde{\phi}_1^2  \right]+ \\
\nn &\hphantom{=\frac{1}{\pi^2M^3}}  + (f_1+f_3)\xi QF_{\bar{\mu}\bar{\nu}} - 2(f_3-if_2)\td{B}_{\bar{\mu}} (D_{\bar{\nu}}\td{\phi} + h.c.)\td{\phi}_1 +\\
\label{Enab01b} &\hphantom{=\frac{1}{\pi^2M^3}}
+ \partial_{\bar{\nu}}\tau \left[ (f_1' + f_3') \left( \td{B}_{\bar{\mu}}(1-|\phi|^2) - 2\xi Q \td{A}_{\bar{\mu}} \right) + 2i f_2'\td{B}_{\bar{\mu}} \td{\phi}_1^2 \right]\bigg) \, ,
\end{align}
\begin{align}
\nn & w^2 \Big[ D_{\xi}\big(\fxx\xi^2 D_{\xi}\tilde{\k}\big) \!+\! \cX\big(2v\tilde{\phi} - \tilde{\k}(1+|\phi|^2)\big) \Big] \!+\! D_r\big( \xi^2 w^2\frr D_r \tilde{\k} \big)- \\
\nn & - D_r\big( \xi^2 w^2\fxr D_\xi \tilde{\k} \big) - D_\xi\big( \xi^2 w^2\fxr D_r \tilde{\k} \big) + h.c. \\
\nn  &\hphantom{=} = 4 \xi^2\ex^{2A}\cX\ka \tau^2\tilde{\k}+ \\
\nn &\hphantom{= =} +\frac{\e^{\bar{\mu}\bar{\nu}}}{\pi^2 M^3}\bigg( 6(f_1-f_3)D_{\bar{\mu}}\rho D_{\bar{\nu}}\td{\phi} -4i(f_3-if_2)\td{\rho}\,\td{A}_{\bar{\mu}} \partial_{\bar{\nu}}\td{\phi} +   \\
\nn &\hphantom{= =+\frac{1}{\pi^2M^3}} + \partial_{\bar{\mu}}(\xi Q)\Big( 2i(f_1 + f_3)D_{\bar{\nu}}\td{\phi} + 4(f_1+f_3-3if_2)\td{A}_{\bar{\nu}}\td{\phi} \Big) +  \\
\label{Enab02} &\hphantom{==+\frac{1}{\pi^2M^3}} + 2\partial_{\bar{\nu}}\tau \Big[ (f_1' - f_3')D_{\bar{\mu}}\rho\, \td{\phi} - (f_1'-f_3')\td{\rho} D_{\bar{\mu}}\td{\phi} \Big]\bigg) + h.c. \, ,
\end{align}
\begin{align}
\nn & w^2 \Big[-i D_{\xi}\big(\fxx\xi^2 D_{\xi}\tilde{\k}\big) -i\cX\big( 2v\tilde{\phi} -\tilde{\k}(1+|\phi|^2)\big) \Big] -\\
\nn & - i D_r\Big( \xi^2 \frr w^2 D_r \tilde{\k} \Big) + i D_r\Big( \xi^2 \fxr w^2 D_\xi \tilde{\k} \Big)+\\
\nn & + i D_\xi\Big( \xi^2 \fxr w^2 D_r \tilde{\k} \Big) + h.c. \\
\nn &=  \frac{\e^{\bar{\mu}\bar{\nu}}}{\pi^2 M^3}\bigg( D_{\bar{\mu}}\rho \left( -2i(f_1+f_3) D_{\bar{\nu}}\td{\phi} - 4(f_1-f_3-if_2)\td{A}_{\bar{\nu}}\td{\phi} \right)  + \\
\nn &\hphantom{=\frac{1}{\pi^2M^3}}
+ 2\partial_{\bar{\mu}}(\xi Q)(f_1 + f_3)D_{\bar{\nu}}\td{\phi} + 2(f_3-if_2)\td{\rho} \left( 2\td{A}_{\bar{\mu}}\partial_{\bar{\nu}}\td{\phi} -F_{\bar{\mu}\bar{\nu}}\td{\phi} \right) + \\
\label{Enab03} &\hphantom{=\frac{1}{\pi^2M^3}} + 2\partial_{\bar{\nu}}\tau \Big[ i(f_1' + f_3')\td{\rho}\, D_{\bar{\mu}}\td{\phi} -2if_2'\td{\rho}\td{A}_{\bar{\mu}}\td{\phi} + (f_1'+f_3')\partial_{\bar{\mu}}(\xi Q)\td{\phi} \Big]\bigg) + h.c. \, ,
\end{align}
\begin{align}
\nn & w^2 \Big[ \partial_{\xi}\big(\cX\left[\ex^{2A}\D_{rr}(1-\ex^{2A}\D_{\xi\xi})-\ex^{4A}\D_{\xi r}^2\right]\xi^2B_{\xi r}\big) + 2\cX\ex^{2A}\big(\D_{rr} D_r\rho - \D_{\xi r} D_\xi\rho \big) \Big] = \\
\nn &\hphantom{=} 4 \ex^{4A} \cX\D_{rr}\ka \tau^2 \xi^2 \td{B}_r - 4 \ex^{4A} \cX\D_{\xi r}\ka \tau^2 \xi^2 \td{B}_\xi -\\
\nn &\hphantom{=} - \frac{1}{2\pi^2M^3}\bigg( (f_1+f_3)(D_{\xi}\phi\, \k^* + h.c.) - 2(2f_3 - f_1)\left( D_{\xi}\td{\phi}+h.c.\right) \td{\k}_1 -  \\
\nn &\hphantom{=-\frac{1}{N_f\pi^2M^3}}
-4 (f_1-f_3-if_2)\td{A}_{\xi}\td{\phi}_1\td{\k}_2 - 8(f_1+f_3)\xi Q\partial_{\xi}\Phi +\\
\nn &\hphantom{=-\frac{1}{N_f\pi^2M^3}} + 2v(f_3-if_2)(D_{\xi}\td{\phi}+h.c.)\td{\phi}_1 -\\
\nn &\hphantom{=-\frac{1}{N_f\pi^2M^3}} - 2\partial_{\xi}\left( v\left[\frac{1}{2}(f_1 + f_3)\, (|\phi|^2 - 1) + (f_1-f_3-if_2)\td{\phi}_1^2  \right]\right) -\\
\label{Eabr} &\hphantom{=-\frac{1}{N_f\pi^2M^3}}
-\partial_\xi\tau\Big[ (f_1'+f_3')v(1-|\phi|^2)+ 2if_2'\,v\td{\phi}_1^2 -2(f_1'-f_3')\td{\k}_1\td{\phi}_1 \Big] \bigg) \, ,
\end{align}
\begin{align}
\nn & \xi^2 \partial_{r}\Big(w^2\cX\big[\ex^{2A}\D_{rr}(1-\ex^{2A}\D_{\xi\xi})-\ex^{4A}\D_{\xi r}^2\big] B_{r \xi}\Big)+\\
\nn & + 2w^2\cX\Big((1-\ex^{2A}\D_{\xi\xi})D_{\xi}\rho - \ex^{2A}\D_{\xi r}D_r\rho \Big) = \\
\nn&\hphantom{=} 4 \ex^{2A} \fxx\ka \tau^2 \xi^2\td{B}_{\xi} -4 \ex^{4A} \cX\D_{\xi r}\ka \tau^2 \xi^2\td{B}_{r} + \\
\nn &\hphantom{=}
+ \frac{1}{2\pi^2M^3}\bigg[ (f_1+f_3)(D_{r}\phi\, \k^* + h.c.) - 2(2f_3 - f_1)(D_{r}\td{\phi}+h.c.) \td{\k}_1 -  \\
\nn &\hphantom{=+ \frac{1}{N_f\pi^2M^3}}
-4 (f_1-f_3-if_2)\td{A}_{r}\td{\phi}_1\td{\k}_2 - 8(f_1+f_3)\xi Q\partial_{r}\Phi +\\
\nn&\hphantom{=+ \frac{1}{N_f\pi^2M^3}}+ 2v(f_3-if_2)(D_{r}\td{\phi}+h.c.)\td{\phi}_1- \\
\nn &\hphantom{=+ \frac{1}{N_f\pi^2M^3}}  - 2\partial_{r}\left( v\left[\frac{1}{2}(f_1 + f_3)\, (|\phi|^2 - 1) + (f_1-f_3-if_2)\td{\phi}_1^2  \right] \right) - \\
\label{Eabi1} &\hphantom{=+ \frac{1}{N_f\pi^2M^3}} -\partial_r\tau\Big[ (f_1'+f_3')v(1-|\phi|^2)+ 2if_2'\,v\td{\phi}_1^2 -2(f_1'-f_3')\td{\k}_1\td{\phi}_1 \Big] \bigg) \, ,\end{align}
\begin{align}
\nn & \partial_{r}\Big(w^2\frr D_r\rho\Big) +  \partial_{\xi}\Big(w^2\fxx D_{\xi}\rho\Big)- \\
\nn & - \partial_{r}\Big(w^2\fxr D_\xi\rho\Big) - \partial_{\xi}\Big(w^2\fxr D_r\rho\Big)\\
\nn&\hphantom{=} = 4 \ex^{2A} \fxx\ka \tau^2 \td{\rho} + \\
\nn &\hphantom{==} + \frac{\e^{\bar{\mu}\bar{\nu}}}{4 \pi^2 M^3} \bigg[ \partial_{\bar{\mu}}\Big( (f_1+f_3)(D_{\bar{\nu}}\phi\, \k^* + h.c.) - 2(2f_3 - f_1)( D_{\bar{\nu}}\td{\phi}+h.c.) \td{\k}_1- \\
\nn &\hphantom{==+\frac{1}{2N_f \pi^2 M^3}\bigg[\partial_\mu}  -4 (f_1-f_3-if_2)\td{A}_{\bar{\nu}}\td{\phi}_1\td{\k}_2 - 8(f_1+f_3)\xi Q \partial_{\bar{\nu}}\Phi \Big)- \\
\nn &\hphantom{==+\frac{1}{2N_f \pi^2 M^3}} - 2(f_3 - if_2)\Big(-F_{\bar{\mu}\bar{\nu}}\,\td{\phi}_1\td{\k}_2 + \td{A}_{\bar{\mu}}(-i\partial_{\bar{\nu}}\td{\phi}\td{\k} + h.c.)\Big)+ \\
\nn &\hphantom{==+\frac{1}{2N_f \pi^2 M^3}} +\partial_{\bar{\nu}}\tau\Big(f_1'(D_{\bar{\mu}}\phi\k^*+h.c.) -f_3'(D_{\bar{\mu}}\td{\phi}\td{\k} + h.c.) + 2if_2' \td{A}_{\bar{\mu}}\td{\phi}_1 \td{\k}_2 + \\
\label{Eabi2} &\hphantom{==+\frac{1}{2N_f \pi^2 M^3}+\partial_\nu\tau}  + 2(f_1'-f_3')\partial_{\bar{\mu}}(\td{\k}_1\td{\phi}_1) - 8(f_1'+f_3')\xi Q\partial_{\bar{\mu}}\Phi \Big)\bigg] \, ,
\end{align}
\begin{align}
\nn & \xi^2\partial_{r}\Big(w^2\frr \partial_r Q\Big) + w^2 \partial_{\xi}\Big(\fxx\xi^2\partial_{\xi}Q\Big)-\\
\nn& - \partial_{\xi}\Big(w^2\fxr\xi^2\partial_{r}Q\Big) - \partial_{r}\Big(w^2\fxr\xi^2\partial_{\xi}Q\Big)  = \\
\nn &\hphantom{=} \cX w^2\Big[2 - \ex^{2A}\xi \big( \partial_\xi\D_{\xi\xi} + \ex^{2A}\partial_r(\ex^{-2A}\D_{\xi r}) \big)  \Big]Q \, -\\
\nn &\hphantom{=} -\frac{\xi \e^{\bar{\mu}\bar{\nu}}}{4\pi^2M^3}\bigg[ -\partial_{\bar{\mu}}\Big( (f_1+f_3)(i\k^*D_{\bar{\nu}}\phi + h.c.) + 4(f_1+f_3-3if_2)\td{A}_{\bar{\nu}}\td{\phi}_1\td{\k}_1 \Big)- \\
\nn &\hphantom{=-\frac{1}{2N_f\pi^2M^3}} -8(f_1+f_3)D_{\bar{\mu}}\rho \partial_{\bar{\nu}}\Phi + (f_1+f_3)v\, F_{\bar{\mu}\bar{\nu}}  +\\
\label{Eabi3} &\hphantom{=-\frac{1}{2N_f\pi^2M^3}}
+ 2\partial_{\bar{\nu}}\tau(f_1'+f_3')\left(4\td{\rho}\partial_{\bar{\mu}}\Phi -\partial_{\bar{\mu}}(\td{\phi}_1\td{\k}_2) - v\, \td{A}_{\bar{\mu}} \right) \bigg] \, ,
\end{align}
\begin{align}
\nn & \xi^2\partial_r\left[ \ex^{4A} \cX\D_{rr}\ka\tau^2 \left( B_r + \frac{1}{2}\partial_r\zeta \right) \right]+ \\
\nn& + \ex^{2A} \ka \left[ \partial_{\xi}\left[\fxx\tau^2\xi^2\left( B_{\xi} + \frac{1}{2}\partial_{\xi}\zeta \right)\right] - 2\fxx\tau^2\left(\rho + \frac{1}{2}\zeta \right) \right]- \\
\nn & -\partial_r\left[ \ex^{4A} \cX\D_{\xi r}\ka\tau^2 \xi^2\left( B_\xi + \frac{1}{2}\partial_\xi\zeta \right) \right] - \partial_\xi\left[ \ex^{4A} \cX\D_{\xi r}\ka\tau^2 \xi^2\left( B_r + \frac{1}{2}\partial_r\zeta \right) \right] \\
\nn &= -\frac{\e^{\bar{\mu}\bar{\nu}}}{4\pi^2M^3} \bigg[ \partial_{\bar{\mu}}\Big( (f_3-if_2) v(D_{\bar{\nu}}\td{\phi}+h.c.)\td{\phi}_1  \Big) + \\
\nn &\hphantom{=-\frac{1}{2N_f\pi^2M^3}} + (f_3 - if_2)\left(-F_{\bar{\mu}\bar{\nu}}\,\td{\phi}_1\td{\k}_2 + \td{A}_{\bar{\mu}}(-i\partial_{\bar{\nu}}\td{\phi}\td{\k} + h.c.)\,\right) + \\
\nn &\hphantom{=-\frac{1}{2N_f\pi^2M^3}} +\partial_{\bar{\nu}}\tau\bigg( -\frac{1}{2}(f_1'+f_3')\partial_{\bar{\mu}}\left( v(1-|\phi|^2) \right) - if_2' \partial_{\bar{\mu}}(v\td{\phi}_1^2) - \\
\nn &\hphantom{=-\frac{1}{2N_f\pi^2M^3}+\partial_\nu\tau}  -\frac{1}{2} f_1'(D_{\bar{\mu}}\phi\k^*+h.c.) +\frac{1}{2}f_3'(D_{\bar{\mu}}\td{\phi}\td{\k} + h.c.) - if_2' \td{A}_{\bar{\mu}}\td{\phi}_1 \td{\k}_2 +\\
\label{Ezans} &\hphantom{=-\frac{1}{2N_f\pi^2M^3}+\partial_\nu\tau}  + 4(f_1'+f_3')\xi Q\partial_{\bar{\mu}}\Phi \bigg) \bigg]  \, ,
\end{align}
where
\be
\label{defX} \mathcal{X} = \sqrt{1 + \ex^{-2A}\ka\left((\partial_r\tau)^2+(\partial_{\xi}\tau)^2\right)}\, V_f(\l,\tau)\,\ex^A \, ,
\ee
and the symbol $\D_{\bar{\mu}\bar{\nu}}$ is given by
\be
\label{defDxx} \Delta_{\xi \xi} \equiv \frac{\ex^{6A} \ka \, (\partial_{\xi}\tau)^2 }{-\mathrm{det}\, \tilde{g}} = \frac{\ex^{-4A} \ka \, (\partial_{\xi}\tau)^2}{1 + \ex^{-2A}\ka\left((\partial_r\tau)^2+(\partial_{\xi}\tau)^2\right)} \, ,
\ee
\be
\label{defDxr} \Delta_{\xi r} \equiv \frac{\ex^{6A}\ka \, \partial_{\xi}\tau \partial_{r}\tau }{-\mathrm{det}\, \tilde{g}} = \frac{\ex^{-4A}\ka \, \partial_{\xi}\tau \partial_{r}\tau}{1 + \ex^{-2A}\ka\left((\partial_r\tau)^2+(\partial_{\xi}\tau)^2\right)} \, ,
\ee
\be
\label{defDrr} \Delta_{r r} \equiv \frac{\ex^{8A} (1 + \ex^{-2A} \ka \, (\partial_{\xi}\tau)^2) }{-\mathrm{det}\, \tilde{g}} = \frac{\ex^{-2A}  \, (1 + \ex^{-2A} \ka \, (\partial_{\xi}\tau)^2)}{1 + \ex^{-2A}\ka\left((\partial_r\tau)^2+(\partial_{\xi}\tau)^2\right)} \, .
\ee

\subsection{Equations of motion in the probe baryon approximation}

\label{App:EOMrotpb}

In the probe baryon approximation, the modulus of the tachyon field $\tau$ is fixed to its vacuum value. In particular, it does not depend on the radius $\xi$, so that the equations of motion are somewhat simplified
\begin{align}
\nn &\hphantom{=} hw^2 \left[ \partial_{\xi}(\xi^2 \partial_{\xi}v) - 2\left( v(1 + |\phi|^2) - (\k\phi^* + h.c.)\, \right) \right] + \partial_r\left( \xi^2 kw^2 \partial_r v \right) = \\
\nn &\hphantom{=}  \frac{2}{\pi^2M^3}  \bigg( 2B_{\xi r}\left[\frac{1}{2}(f_1(\tau) + f_3(\tau))\, (|\phi|^2 - 1) + (f_1(\tau)-f_3(\tau)-if_2(\tau))\tilde{\phi}_1^2  \right]+  \\
\nn &\hphantom{=\frac{2}{\pi^2M^3}  \bigg(}     + 2(f_1(\tau)+f_3(\tau))\xi QF_{\xi r} - 2(f_3(\tau)-if_2(\tau))\e^{\bar{\mu}\bar{\nu}}\td{B}_{\bar{\mu}} (D_{\bar{\nu}}\td{\phi} + h.c.)\td{\phi}_1 \bigg)+ \\
\label{Enab01p} &\hphantom{=} +\frac{2}{\pi^2M^3} \partial_r\tau \left[ (f_1'(\tau) + f_3'(\tau)) \left( \td{B}_{\xi}(1-|\phi|^2) - 2\xi Q \td{A}_{\xi} \right) + 2i f_2'(\tau)\td{B}_{\xi} \td{\phi}_1^2 \right] \, ,
\end{align}
\begin{align}
\nn & hw^2\left[ D_{\xi}(\xi^2 D_{\xi}\tilde{\k}) + 2v\tilde{\phi} - \tilde{\k}(1+|\phi|^2) \right] + D_r\left( \xi^2 kw^2 D_r \tilde{\k} \right) + h.c. = \\
\nn  &\hphantom{=} 4 \xi^2\ex^{2A}h\ka \tau^2\tilde{\k} +\\
\nn &+\frac{\e^{\bar{\mu}\bar{\nu}}}{\pi^2 M^3}\left[ 6(f_1-f_3)D_{\bar{\mu}}\rho D_{\bar{\nu}}\td{\phi} + \partial_{\bar{\mu}}(\xi Q)\left( 2i(f_1 + f_3)D_{\bar{\nu}}\td{\phi} + 4(f_1+f_3-3if_2)\td{A}_{\bar{\nu}}\td{\phi} \right)- \right. \\
\nn &\hphantom{+\frac{1}{\pi^2M^3}\Big[} \left. -4i(f_3-if_2)\td{\rho}\,\td{A}_{\bar{\mu}} \partial_{\bar{\nu}}\td{\phi} \right]+ \\
\label{Enab02p} &+ \frac{2}{\pi^2 M^3}\partial_r\tau \left[ (f_1' - f_3')D_{\xi}\rho\, \td{\phi} - (f_1'-f_3')\td{\rho} D_{\xi}\td{\phi} \right] + h.c. \, ,
\end{align}
\begin{align}
\nn &\hphantom{=} hw^2 \left[-i D_{\xi}(\xi^2 D_{\xi}\tilde{\k}) - 2iv\tilde{\phi} + i\tilde{\k}(1+|\phi|^2) \right] - i D_r\left( \xi^2 kw^2 D_r \tilde{\k} \right) + h.c. \\
\nn &=  \frac{\e^{\bar{\mu}\bar{\nu}}}{\pi^2 M^3}\Big[ D_{\bar{\mu}}\rho \left( -2i(f_1+f_3) D_{\bar{\nu}}\td{\phi} - 4(f_1-f_3-if_2)\td{A}_{\bar{\nu}}\td{\phi} \right)+ \\
\nn &\hphantom{=\frac{1}{\pi^2M^3}}
+ 2\partial_{\bar{\mu}}(\xi Q)(f_1 + f_3)D_{\bar{\nu}}\td{\phi} + 2(f_3-if_2)\td{\rho}\, \left( 2\td{A}_{\bar{\mu}}\partial_{\bar{\nu}}\td{\phi} -F_{\bar{\mu}\bar{\nu}}\td{\phi} \right) \Big]+ \\
\label{Enab03p} &\hphantom{=} + \frac{2}{\pi^2 M^3}\partial_r\tau \left[ i(f_1' + f_3')\td{\rho}\, D_{\xi}\td{\phi} -2if_2'\, \td{\rho}\td{A}_{\xi}\td{\phi} + (f_1'+f_3')\partial_{\xi}(\xi Q)\td{\phi} \right] + h.c. \, ,
\end{align}
\begin{align}
\nn &\hphantom{=} kw^2 \left[ \partial_{\xi}(\xi^2B_{\xi r}) + 2 D_r\rho \right] = 4 \ex^{2A} k\ka \tau^2 \xi^2 \td{B}_r -\\
\nn &\hphantom{=} - \frac{1}{2\pi^2M^3}\bigg[ (f_1+f_3)(D_{\xi}\phi\, \k^* + h.c.) - 2(2f_3 - f_1)\left( D_{\xi}\td{\phi}+h.c.\right) \td{\k}_1 -  \\
\nn &\hphantom{=\frac{1}{N_f\pi^2M^3}}
-4 (f_1-f_3-if_2)\td{A}_{\xi}\td{\phi}_1\td{\k}_2 - 8(f_1+f_3)\xi Q\partial_{\xi}\Phi + \\
\nn &\hphantom{=\frac{1}{N_f\pi^2M^3}}
+ 2v(f_3-if_2)(D_{\xi}\td{\phi}+h.c.)\td{\phi}_1 -\\
\label{Eabrp} &\hphantom{=\frac{1}{N_f\pi^2M^3}} - 2\partial_{\xi}\left( v\left[\frac{1}{2}(f_1 + f_3)\, (|\phi|^2 - 1) + (f_1-f_3-if_2)\td{\phi}_1^2  \right] \right) \bigg] \, ,
\end{align}
\begin{align}
\nn & \xi^2 \partial_{r}(kw^2 B_{r \xi}) + 2hw^2 D_{\xi}\rho  = 4 \ex^{2A} h\ka \tau^2 \xi^2\td{B}_{\xi} +  \\
\nn &\hphantom{=} + \frac{1}{2\pi^2M^3}\bigg[ (f_1+f_3)(D_{r}\phi\, \k^* + h.c.) - 2(2f_3 - f_1)\left( D_{r}\td{\phi}+h.c.\right) \td{\k}_1 -  \\
\nn &\hphantom{=\frac{1}{N_f\pi^2M^3}}
-4 (f_1-f_3-if_2)\td{A}_{r}\td{\phi}_1\td{\k}_2 - 8(f_1+f_3)\xi Q\partial_{r}\Phi +\\
\nn&\hphantom{=\frac{1}{N_f\pi^2M^3}}  + 2v(f_3-if_2)(D_{r}\td{\phi}+h.c.)\td{\phi}_1 -\\
\nn &\hphantom{=\frac{1}{N_f\pi^2M^3}}
  - 2\partial_{r}\left( v\left[\frac{1}{2}(f_1 + f_3)\, (|\phi|^2 - 1) + (f_1-f_3-if_2)\td{\phi}_1^2  \right] \right)-  \\
\label{Eabi1p} &\hphantom{=\frac{1}{N_f\pi^2M^3}}
 -\partial_r\tau\left[ (f_1'+f_3')v(1-|\phi|^2)+ 2if_2'\,v\td{\phi}_1^2 -2(f_1'-f_3')\td{\k}_1\td{\phi}_1 \right] \, \bigg] \, ,
\end{align}
\begin{align}
\nn & \partial_{r}(kw^2 D_r\rho) + hw^2 \partial_{\xi}D_{\xi}\rho  =  4 \ex^{2A} h\ka \tau^2 \td{\rho} + \\
\nn &\hphantom{=} + \frac{1}{4 \pi^2 M^3} \bigg[ \e^{\bar{\mu}\bar{\nu}}\partial_{\bar{\mu}}\Big( (f_1+f_3)(D_{\bar{\nu}}\phi\, \k^* + h.c.) - 2(2f_3 - f_1)\left( D_{\bar{\nu}}\td{\phi}+h.c.\right) \td{\k}_1 - \\
\nn &\hphantom{=+\frac{1}{2N_f \pi^2 M^3} \bigg[ \e^{\bar{\mu}\bar{\nu}}\partial_{\bar{\mu}}}
 -4 (f_1-f_3-if_2)\td{A}_{\bar{\nu}}\td{\phi}_1\td{\k}_2 - 8(f_1+f_3)\xi Q \partial_{\bar{\nu}}\Phi \Big)- \\
\nn &\hphantom{=+\frac{1}{2N_f \pi^2 M^3}}
- 2(f_3 - if_2)\e^{\bar{\mu}\bar{\nu}}\left(-F_{\bar{\mu}\bar{\nu}}\,\td{\phi}_1\td{\k}_2 + \td{A}_{\bar{\mu}}(-i\partial_{\bar{\nu}}\td{\phi}\td{\k} + h.c.)\,\right)+ \\
\nn &\hphantom{=+\frac{1}{2N_f \pi^2 M^3}}
 +\partial_r\tau\Big(f_1'(\tau)(D_{\xi}\phi\k^*+h.c.) -f_3'(\tau)(D_{\xi}\td{\phi}\td{\k} + h.c.) + 2if_2'(\tau) \td{A}_{\xi}\td{\phi}_1 \td{\k}_2 + \\
\label{Eabi2p} &\hphantom{=+\frac{1}{2N_f \pi^2 M^3}+\partial_r\tau}
+ 2(f_1'-f_3')\partial_{\xi}(\td{\k}_1\td{\phi}_1) - 8(f_1'+f_3')\xi Q\partial_{\xi}\Phi\Big) \bigg] \, ,
\end{align}
\begin{align}
\label{Eabi3p} & \xi^2\partial_{r}(kw^2 \partial_r Q) + hw^2 \left[\partial_{\xi}(\xi^2\partial_{\xi}Q) - 2Q\right] = \\
\nn &\hphantom{=} -\frac{\xi}{4\pi^2M^3}\bigg[ -\e^{\bar{\mu}\bar{\nu}}\partial_{\bar{\mu}}\left( (f_1+f_3)(i\k^*D_{\bar{\nu}}\phi + h.c.) + 4(f_1+f_3-3if_2)\td{A}_{\bar{\nu}}\td{\phi}_1\td{\k}_1 \right) - \\
\nn &\hphantom{=-\frac{\xi}{2N_f\pi^2M^3}}
-8(f_1+f_3)\e^{\bar{\mu}\bar{\nu}}D_{\bar{\mu}}\rho \partial_{\bar{\nu}}\Phi + (f_1+f_3)v\, \e^{\bar{\mu}\bar{\nu}}F_{\bar{\mu}\bar{\nu}} +\\
\nn&\hphantom{=-\frac{\xi}{2N_f\pi^2M^3}}
+ 2\partial_r\tau(f_1'+f_3')\left(4\td{\rho}\partial_{\xi}\Phi -\partial_{\xi}(\td{\phi}_1\td{\k}_2) - v\, \td{A}_{\xi} \right)\, \bigg] \, ,
\end{align}
\begin{align}
\nn & \xi^2\partial_r\left[ \ex^{2A} kw^2\ka\tau^2 \left( B_r + \frac{1}{2}\partial_r\zeta \right) \right]+\\
\nn &+ \ex^{2A} hw^2\ka\tau^2 \left[ \partial_{\xi}\left[\xi^2\left( B_{\xi} + \frac{1}{2}\partial_{\xi}\zeta \right)\right] - 2\left(\rho + \frac{1}{2}\zeta \right) \right] \\
\nn &= -\frac{1}{4\pi^2M^3} \bigg[ \e^{\bar{\mu}\bar{\nu}}\partial_{\bar{\mu}}\left( (f_3-if_2) v(D_{\bar{\nu}}\td{\phi}+h.c.)\td{\phi}_1  \right)+ \\
\nn &\hphantom{=-\frac{1}{2N_f\pi^2M^3}}
+ (f_3 - if_2)\e^{\bar{\mu}\bar{\nu}}\left(-F_{\bar{\mu}\bar{\nu}}\,\td{\phi}_1\td{\k}_2 + \td{A}_{\bar{\mu}}(-i\partial_{\bar{\nu}}\td{\phi}\td{\k} + h.c.)\,\right)+ \\
\nn &\hphantom{=-\frac{1}{2N_f\pi^2M^3}}
 +\partial_r\tau\bigg( -\frac{1}{2}(f_1'+f_3')\partial_{\xi}\left( v(1-|\phi|^2) \right) - if_2' \partial_{\xi}(v\td{\phi}_1^2)-  \\
\nn &\hphantom{=-\frac{1}{2N_f\pi^2M^3}+\partial_r\tau}
-\frac{1}{2} f_1'(\tau)(D_{\xi}\phi\k^*+h.c.) +\frac{1}{2}f_3'(\tau)(D_{\xi}\td{\phi}\td{\k} + h.c.)-\\
\label{Ezansp} &\hphantom{=-\frac{1}{2N_f\pi^2M^3}+\partial_r\tau}
- if_2'(\tau) \td{A}_{\xi}\td{\phi}_1 \td{\k}_2\,  + 4(f_1'+f_3')\xi Q\partial_{\xi}\Phi \bigg) \bigg]  \, ,
\end{align}
where
\be
\label{defhk} k(r) = \frac{\ex^A}{\sqrt{1+\ex^{-2A}\ka (\partial_r\tau)^2}}V_f(\la,\tau^2) \sp h(r) = \ex^A \sqrt{1+\ex^{-2A}\ka  (\partial_r\tau)^2} V_f(\la,\tau^2) \,  .
\ee

\section{Quantization of the rigid rotor}

\label{App:QI}

We proceed in this subsection to the quantization of the classical Lagrangian \eqref{Lrot} along the lines of \cite{Panico08}. For this purpose, it is more convenient to reintroduce the SU(2) matrix $V(t)$ from the definition of $\vec{\omega}$ in \eqref{defo}. It gives
\be
\label{Lrotu} L_{\text{rot}} = -M_0 + 2\l \sum_{a=0}^3 \dot{u}_a^2 \, ,
\ee
where we parametrized  the SU(2)  matrix $V(t)$ as
\be
\label{parVu} V(t) = u_0 \mathbb{I}_2 + i u_i \s^i \, ,
\ee
where the $u_a$'s parametrize the 3-sphere\footnote{The $u_a$'s actually parametrize $S^3/\mathbb{Z}_2$ because the collective coordinates live in $SU(2)_V/\mathbb{Z}_2$ . This means that $u_a$ and $-u_a$ correspond to the same point . In the $q^\a$ coordinates, the identification is between $(y,\theta_1,\theta_2)$ and $(y,\theta_1+\pi,\theta_2+\pi)$ .} $S^3$
\be
\label{condu} \sum_a u_a^2 = 1 \, .
\ee
$S^3$ can be alternatively described by 3 unconstrained coordinates \cite{Panico08} $q^\a \equiv (y,\theta_1,\theta_2)$ in the domains
\be
\label{domyt} y\in [-1,1] \sp \theta_1, \theta_2\in [0,2\pi) \, ,
\ee
which are related to the $u_a$'s as
\be
\label{defyt} u_1 + iu_2 \equiv z_1 = \sqrt{\frac{1-y}{2}} \ex^{i\theta_1} \sp u_0 + iu_3 \equiv z_2 = \sqrt{\frac{1+y}{2}}\ex^{i\theta_2}  \, .
\ee
In terms of these coordinates, the Lagrangian \eqref{Lrotu} is rewritten as
\be
\label{Lrotyt} L_{\text{rot}} = -M_0 + 2\l\, G_{\a\b} \dot{q}^\a \dot{q}^\b \, ,
\ee
where G is the metric of $S^3$, which in the $q^\a$ coordinates reads
\be
\label{Gq} G_{\a\b} \mathrm{d}q^\a\mathrm{d}q^\b = \frac{1}{4}\frac{1}{1-y^2}\mathrm{d}y^2 + \frac{1-y}{2}\mathrm{d}\theta_1^2 + \frac{1+y}{2}\mathrm{d}\theta_2^2 \, .
\ee
The momentum conjugate to $q^\a$ is then
\be
\label{defp} p_\a \equiv \frac{\partial L}{\partial \dot{q}^\a} = 4\l G_{\a\b}\dot{q}^\b \, ,
\ee
and the classical Hamiltonian
\be
\label{Hcl} H_c = M_0 + \frac{1}{8\l} G^{\a\b} p_\a p_\b \, .
\ee

The quantum Hamiltonian operator is found by applying the quantization rules
\be
\label{Hq} H_q = M_0 - \frac{1}{8\l}\frac{1}{\sqrt{G}} \partial_\a\left(\sqrt{G}\, G^{\a\b}\partial_\b  \right) = M_0 -\frac{1}{8\l} \nabla_\a\nabla^\a \, ,
\ee
which is hermitian \cite{Panico08} with respect to the scalar product
\be
\label{sp} \left<A|B \right> \equiv \int \mathrm{d}q^3 \sqrt{G}\, f_A(q)^* f_B(q) \, .
\ee

\paragraph*{Spin and Isospin operators} The classical Hamiltonian \eqref{Hcl} is invariant under an $SO(4)$ rotation of the momentum $p^\a$. Because $SO(4) \simeq SU(2)\times SU(2)$, this symmetry can be mapped to two SU(2) symmetries which are
\begin{itemize}
\item (Time-independent) isospin rotation of the fields, which acts on $V(t)$ as
\be
\label{UI} V(t) \to W V(t) \sp W\in SU(2) \, .
\ee
\item 3D (time-independent) spatial rotation, which acts on $V(t)$ as\footnote{Under a spatial rotation $x_i \s^i \to R^\dagger x_i\s^i R $, so the fields of \eqref{ansLs}-\eqref{ansL0s} and \eqref{ansTs} are invariant if $V(t)$ transforms as in \eqref{US} . }
\be
\label{US} V(t) \to V(t) R \sp R \in SU(2) \, .
\ee
\end{itemize}
After quantization, there should be an isospin operator $I^i$ that generates the symmetry of \eqref{UI} and a spin operator $S^i$ that generates the symmetry of \eqref{US}
\be
\label{ISop} [I^i,V] = \frac{\s^i}{2} V \sp [S^i,V] = V \frac{\s^i}{2} \, ,
\ee
where the multiplication should be understood as the action on the wave function. The  expression for the spin and isospin operators can be derived explicitly
\begin{eqnarray}
\label{S}
\left\{
\begin{array}{ll}
S^3 = -\frac{i}{2}(\partial_{\theta_1} + \partial_{\theta_2}) \, , \\
S^+ = \frac{1}{\sqrt{2}}\ex^{i(\theta_1+\theta_2)}\left( i\sqrt{1-y^2}\partial_y + \frac{1}{2}\sqrt{\frac{1+y}{1-y}}\partial_{\theta_1} - \frac{1}{2}\sqrt{\frac{1-y}{1+y}}\partial_{\theta_2} \right)\, , \\
S^- = \frac{1}{\sqrt{2}}\ex^{-i(\theta_1+\theta_2)}\left( i\sqrt{1-y^2}\partial_y - \frac{1}{2}\sqrt{\frac{1+y}{1-y}}\partial_{\theta_1} + \frac{1}{2}\sqrt{\frac{1-y}{1+y}}\partial_{\theta_2}\right)\, ,
\end{array}
\right.
\end{eqnarray}
\begin{eqnarray}
\label{I}
\left\{
\begin{array}{ll}
I^3 = -\frac{i}{2}(\partial_{\theta_1} - \partial_{\theta_2})\, , \\
I^+ = -\frac{1}{\sqrt{2}}\ex^{i(\theta_1-\theta_2)}\left( i\sqrt{1-y^2}\partial_y + \frac{1}{2}\sqrt{\frac{1+y}{1-y}}\partial_{\theta_1} + \frac{1}{2}\sqrt{\frac{1-y}{1+y}}\partial_{\theta_2} \right)\, , \\
I^- = -\frac{1}{\sqrt{2}}\ex^{-i(\theta_1-\theta_2)}\left( i\sqrt{1-y^2}\partial_y - \frac{1}{2}\sqrt{\frac{1+y}{1-y}}\partial_{\theta_1} - \frac{1}{2}\sqrt{\frac{1-y}{1+y}}\partial_{\theta_2}\right)\, ,
\end{array}
\right.
\end{eqnarray}
where the raising and lowering are defined as usual
\be
\label{defpm} S^{\pm} = \frac{1}{\sqrt{2}}(S^1\pm i S^2) \sp I^\pm = \frac{1}{\sqrt{2}}(I^1\pm i I^2) \, .
\ee
$S^i$ and $I^i$ are checked to be hermitian with the scalar product of  \eqref{sp}.

The Hamiltonian \eqref{Hq} then takes a simple form in terms of the spin and isospin operators
\be
\label{HqSI} H_q = M_0 +\frac{1}{2\l}S^2 = M_0 + \frac{1}{2\l} I^2 \, ,
\ee
which makes it clear that the eigenstates of $H_q$ have same spin and isospin and the eigenvalues are given by
\be
\label{Es} E_s = M_0 + \frac{1}{2\l}s(s+1) \, ,
\ee
where $s$ refers to the spin. In particular the nucleon states correspond to $s=1/2$ and their wavefunctions are easily found to be
\begin{eqnarray}
\label{npwf}
\begin{array}{ll}
\left|p \uparrow\right> = \frac{1}{\pi}z_1\, , \,\,\,\,\, & \left|n \uparrow\right>  = \frac{i}{\pi}z_2 \, , \\
\left|p \downarrow\right> = -\frac{i}{\pi}\bar{z}_2  \, , \,\,\,\,\, & \left|n \downarrow\right> = -\frac{1}{\pi}\bar{z}_1 \, ,
\end{array}
\end{eqnarray}
where $z_1$ and $z_2$ where defined in  \eqref{defyt}. The next level $s=3/2$ corresponds to the isobar $\D$ states.

\newpage

\end{document}